\documentclass[rmp,aps]{revtex4-2}

\usepackage[english]{babel}
\usepackage[utf8x]{inputenc}
\usepackage[T1]{fontenc}
\usepackage{times}
\usepackage[a4paper,top=3cm,bottom=2cm,left=3cm,right=3cm,marginparwidth=1.75cm]{geometry}
\usepackage{amsmath, amssymb}
\usepackage{graphicx}
\usepackage{bm}
\usepackage[colorinlistoftodos]{todonotes}
\usepackage[colorlinks=true, allcolors=blue]{hyperref}
\usepackage{paralist}
\usepackage{enumerate,xstring, hyperref}

\usepackage{aas_macros}
\usepackage{tabularx}

\usepackage{subfigure}
\usepackage{subfiles}

\newcommand{\red}{\textcolor{black}}

\newcommand{\add}{\textcolor{black}}

\newcommand{\sm}{\sigma_\mathrm{T}/m_\mathrm{DM}}
\newcommand{\s}{\sigma_\mathrm{T}}
\newcommand{\sV}{\sigma_\mathrm{V}}
\newcommand{\kms}{\,\mathrm{km\,s^{-1}}}
\newcommand{\cmg}{\,\mathrm{cm^2 \, g^{-1}}}
\newcommand{\gcm}{\,\mathrm{g \, cm^{-2}}}
\newcommand{\kpc}{\,\mathrm{kpc}}
\newcommand{\msun}{\,\mathrm{M}_\odot}
\newcommand{\mpc}{\,\mathrm{Mpc}}
\newcommand{\GeVcm}{\,\mathrm{GeV \, cm^{-3}}}
\newcommand{\Msunkpc}{\,\mathrm{M_\odot \, kpc^{-3}}}
\newcommand{\hsi}{h_\mathrm{SI}}

\begin{document}

\preprint{APS/123-QED}
\title{\textbf{Astrophysical Tests of Dark Matter Self-Interactions}}
\author{Susmita Adhikari}
\affiliation{ Department of Physics, Indian Institute of Science Education and Research, Homi Bhaba Road, Pashan, Pune 411008, India}%
\affiliation{Department of Astronomy and Astrophysics, The University of Chicago, Chicago, IL, 60637}%
\email{Contact author: susmita@iiserpune.ac.in}

\author{Arka Banerjee}
\affiliation{ Department of Physics, Indian Institute of Science Education and Research, Homi Bhaba Road, Pashan, Pune 411008, India}%

\author{Kimberly K. Boddy}
\affiliation{ Department of Physics, University of Texas at Austin, Austin, TX, 78712, USA}%

\author{Francis-Yan Cyr-Racine}
\affiliation{ Department of Physics and Astronomy, University of New Mexico, Albuquerque, New Mexico 87106, USA}%

\author{Harry Desmond}
\affiliation{ Astrophysics, University of Oxford, Denys Wilkinson Building, Keble Road, Oxford OX1 3RH, UK}

\affiliation{  McWilliams Center for Cosmology, Department of Physics, Carnegie Mellon University, 5000 Forbes Ave, Pittsburgh, PA 15213}
\affiliation{  Institute of Cosmology $\&$ Gravitation, University of Portsmouth, Dennis Sciama Building, Burnaby Road, Portsmouth, PO1 3FX, UK}
\email{Contact author: harry.desmond@port.ac.uk}

\author{Cora Dvorkin}
\affiliation{ Harvard University, Department of Physics, Cambridge, Massachusetts, 02138, U.S.A.}%

\author{Bhuvnesh Jain}
\affiliation{Center for Particle Cosmology, Department of Physics and Astronomy,
University of Pennsylvania, 209 S. 33rd St., Philadelphia, PA 19104, USA}%

\author{Felix Kahlhoefer}
\affiliation{ Institute for Theoretical Particle Physics (TTP), Karlsruhe Institute of Technology (KIT)}%

\author{Manoj Kaplinghat}
\affiliation{Department of Physics and Astronomy, University of California, Irvine, California 92697-4575, USA}%

\author{Anna Nierenberg}
\affiliation{Department of Physics, University of California, Merced,
5200 North Lake Rd., Merced CA 95343}%

\author{Annika H. G. Peter}
\affiliation{  CCAPP, Department of Physics, and Department of Astronomy, The Ohio State University, 191 W. Woodruff Ave., Columbus, OH 43210}%
\author{Andrew Robertson}
\affiliation{ Jet Propulsion Laboratory, California Institute of Technology, 4800 Oak Grove Drive, Pasadena, CA 91109, USA}%
\author{Jeremy Sakstein}
\affiliation{ Department of Physics \& Astronomy, University of Hawaii, Watanabe Hall, 2505 Correa Road, Honolulu, HI 96822, USA}%

\author{Jes\'{u}s Zavala}
\affiliation{Center for Astrophysics and Cosmology, Science Institute, University of Iceland, Dunhagi 5, 107 Reykjavik, Iceland}%


\begin{abstract}

Self-interacting dark matter (SIDM) arises generically in scenarios for physics beyond the Standard Model that have dark sectors with light mediators or strong dynamics. The self-interactions allow energy and momentum transport through halos, altering their structure and dynamics relative to those produced by collisionless dark matter. SIDM models provide a promising way to explain the diversity of galactic rotation curves, and they form a predictive and versatile framework for interpreting astrophysical phenomena related to dark matter.

This review provides a comprehensive explanation of the physical effects of dark matter self-interactions in objects ranging from galactic satellites (dark and luminous) to clusters of galaxies and the large-scale structure. The second major part describes the methods used to constrain SIDM models including current constraints, with the aim of advancing tests with upcoming galaxy surveys. This part also provides a detailed review of the unresolved small-scale structure formation issues and concrete ways to test simple SIDM models.
The review is rounded off by a discussion of the theoretical motivation for self-interactions, degeneracies with baryonic and gravitational effects, extensions to the single-component elastic-interactions SIDM framework, and future observational and theoretical prospects.

\end{abstract}

\maketitle

\tableofcontents

\section{Introduction}

Ideas for physics beyond the Standard Model (SM) often predict new stable particles, which could be part or all of the dark matter (DM) in the Universe.
Two well-known examples of such ideas introduce new symmetries, Supersymmetry and the Peccei-Quinn symmetry, to alleviate problems with the SM. The predicted DM particles, neutralinos and axions, are canonical examples of the Cold Dark Matter (CDM) model wherein DM is born non-relativistic and only its gravitational interaction is relevant for structure formation. CDM models yield predictions for the evolution of the homogeneous and inhomogeneous evolution of the Universe that agree very well with observations on large scales \cite{Aghanim:2018eyx,Chang:2018rxd}.

Collider and deep underground searches have steadily ruled out large swaths of the parameter space of
\add{weakly interacting massive particles (WIMPs), of which neutralinos are an example. Axions and axion-like particles (ALPs), while not currently strongly constrained, are coming under increased experimental scrutiny (see e.g. \cite{AxionLimits} for a compilation of relevant experiments and limits).}
As these searches progress, another front has opened up. In addition to new symmetries, physics beyond the SM could include new forces among DM particles. This is a generic consequence of a dark sector. Dark sectors play an important role in current model-building efforts in particle physics. The new particles in the dark sector do not have to be heavy but are very weakly coupled to the SM. Dark sectors also allow for qualitatively new production mechanisms for DM, such as freeze-in \cite{Hall:2009bx}. The SIDM production mechanisms offer concrete benchmark models that may be tested through laboratory experiments or astrophysical probes (e.g. \cite{Bringmann:2016din, Huo:2017vef, Hambye, Aboubrahim:2020lnr}).

A generic consequence of a new force is the ability for DM particles to scatter off of each other. If the force mediator is light, then the resulting dark sector interactions could leave observable signatures on galactic and sub-galactic scales with a cross-section that is generically velocity-dependent (e.g., Refs.~\cite{Feng:2009mn,Feng:2009hw,Buckley:2009in,Loeb:2010gj,Tulin:2012wi}). However, a light mediator is not a necessary ingredient of DM models with a large self-interaction cross section that is velocity-dependent (e.g., Refs.~\cite{Chu:2018faw,Chu:2018fzy,Tsai:2020vpi,Laha_1}).
In this review, we discuss what happens if DM is assumed to have interactions with itself (self-interactions) beyond gravitational interactions.
If dark-matter particles have a non-trivial probability of interacting on $\sim$~Gyr timescales, this will allow energy and momentum to flow from one part of the dark matter halo to another beyond what is enabled by gravity.
As we will highlight in this review, the introduction of DM scattering has profound implications for the DM distribution within individual halos and in the hierarchical assembly of structure on non-linear scales.  Furthermore, the types of particle physics models that admit strong DM self-scattering could also lead to imprints on the DM power spectrum. Thus,  self-interacting dark matter (SIDM) phenomenology includes deviations from CDM on scales of individual DM halos and subhalos, as well as their population statistics.

The initial interest in SIDM models began more than twenty years ago as a response to two puzzles in observational astronomy: the shape of \red{the rising part of the} rotation curves of spiral galaxies (the ``cusp/core problem'' \cite{1997MNRAS.290..533D,2000AJ....119.1579V,2001ApJ...552L..23D,kuzio2008}) and the number of observed satellites of the Milky Way (MW; the ``missing satellites problem'' \cite{Klypin:1999uc,Moore:1999nt}).  The term ``self-interacting dark matter" had been coined almost a decade earlier  \cite{Carlson:1992fn}, but it took these astrophysical puzzles to put dark-matter models with strong self-scattering firmly in the view of particle physicists and astronomers.

\red{
In the absence of strong feedback from star formation (e.g., \cite{Governato:2009bg}), DM halos are expected to be ``cusped'', with central density profiles $\rho \propto r^{-1}$ \cite{Navarro:1995iw}.  However, the observationally inferred rotation curves of low-surface-brightness (LSB) spiral galaxies almost universally indicate shallower DM density profiles (e.g., \cite{kuzio2008,oh2011}): the \emph{cusp/core problem}. SIDM is a way to produce isothermal, flat cores ($\rho \sim \hbox{ const}$ at small radii) in the halos hosting spiral and dwarf spheroidal galaxies \cite{Spergel:1999mh}.
New observations over the past 20 years have confirmed that the DM densities of many low surface brightness and dwarf spheroidal galaxies are intriguingly low.}

\add{The missing satellites problem pertains to the most striking prediction of the CDM paradigm: a hierarchy of DM halos down to around Earth-mass~\cite{Hofmann:2001bi,Green:2003un,Profumo:2006bv,diemand2006}. The halo mass function
scales approximately as $dN/dM \propto M^{-1.9}$ \cite{gao2004,diemand2007,springel2008}, which means there are many more low-mass halos than higher mass ones. Using galaxies to trace the halo mass function (effective down to a mass scale below which gas cannot cool and form stars;~\cite{Wechsler:2018pic,Barkana:1999apa,Bullock:2000wn,Benson:2001at,somerville2002,okamoto2008,Chen:2014cxa,Smith:2015vpa}) revealed circa 2000 many fewer halos than expected in CDM.
This provided motivation for SIDM, which suppresses
the abundance of subhalos~\cite{Spergel:1999mh}.  Interactions between DM particles in the host halo and a subhalo can kick particles out of the subhalo, which, combined with the increased tidal disruption of cored subhalos compared to cuspy subhalos, reduces the number of MW satellites.
Since the advent of wide-field digital sky surveys, dozens of faint dwarf galaxies have been identified as satellites of the MW \cite{Willman:2005cd,Drlica-Wagner:2015ufc,Kim:2015xoa,Laevens:2015kla,Caldwell:2016hrl,Homma:2016fzg,Torrealba:2018svf}.  The abundance of satellites is consistent with CDM predictions under reasonable and simulation-tested models for star formation before and after reionization \cite{Bullock:2000wn,Benson:2001at,somerville2002,tollerud2008,Koposov:2009ru,hargis2014,Jethwa:2016gra,Kim:2017iwr,newton2017,Bose:2018vpe,Nadler:2018iux,graus2018}.
Thus the missing satellites problem has now largely gone away, and the increasing numbers of confirmed satellites of the MW in fact afford
tight constraints on novel DM physics \cite{Drlica-Wagner:2019xan,DES:2020fxi}.
Furthermore, it was found that evaporation (mass loss) due to scattering events is ineffective at altering the subhalo mass function in constant cross-section SIDM models unless that cross section is so high as to have already been ruled out \cite{Vogelsberger:2012ku,Dooley:2016ajo}. Thus, to leading order, SIDM and CDM make the same preductions for the census of satellite galaxies.}

\add{Since 2000 further puzzles have however emerged, providing further motivation for studying SIDM models. One is the the ``too-big-to-fail'' problem, that CDM predicts the existence of subhalos that should be too big to fail to form stars but yet are not observed \cite{TBTF, BoylanKolchin:2011dk}. Another has come to be known as the ``diversity of rotation curves'' problem, that there is a high degree of galaxy-to-galaxy scatter in the shapes of rotation curves despite the universality of the Navarro--Frenk--White (NFW) \cite{1997ApJ...490..493N} profile in CDM simulations (e.g., \cite{deBlok_1,Simon:2004sr,Adams:2014bda,KuziodeNaray:2009oon}). It is unclear if the physics of star formation can create the diversity of density profiles observed, although it appears that SIDM could accommodate the range of observed rotation curves \cite{Oman,Creasey,Valli:2017ktb,Kamada,Robertson:2017mgj,Ren:2018jpt,Zavala:2019sjk}.
This diversity is most apparent in galaxies with rotation velocities of about $70-100 \kms$, but there is some evidence for deviation from the standard CDM cuspy halo profile even on cluster scales \cite{Newman++13a,Newman++13b,Newman++15}. Thus, instead of focusing on a narrow mass range, SIDM models should be challenged with explaining the density profiles of halos ranging from dwarf galaxy to cluster masses~\cite{Kaplinghat:2015aga}. Additional puzzling observations from the perspective of CDM involve ultra-diffuse galaxies and strong lensing perturbers, as we will discuss extensively.}

In this review, we focus on the astrophysical phenomenology of SIDM across the range of scales from unresolved subhalos to clusters of galaxies. In  Sec. \ref{sec:theory}, we provide the theoretical background to SIDM and the ways in which it is modelled.
Sec. \ref{sec:pheno} provides an exhaustive account of the physical effects of self-interactions in galaxies, clusters and large-scale structure. Sec. \ref{sec:observations} describes the ways in which these effects may be searched for, including hints of the existence of self-interactions and the discovery potential of near-term astronomical facilities. We focus predominantly on elastic interactions because
these typically dominate over inelastic interactions in particle models, and most simulation work has been done in that context. We provide a summary of the current constraints on the elastic self-interaction cross section at the end of this section.
In Sec. \ref{sec:particle} we connect to specific particle models and describe ways in which more complex interactions and effects may extend the SIDM paradigm. We close with our view on the theoretical and observational path forward to characterizing DM self-scattering in the next decade, highlighting important issues that need more work and promising future tests (Sec.~\ref{sec:future}).

\textcolor{black}{
The present review is complementary to a recent treatise by Tulin \& Yu \cite{Tulin:2017ara} on DM self-interactions. Together, these reviews provide a more complete picture of the theory, phenomenology, and current and future observational tests of SIDM models.
The present review also provides a self-contained summary of the small-scale structure challenges, but readers will benefit from referring to the influential review on this topic in Ref.~\cite{Bullock:2017xww}.
}

\red{
This review forms the second arm of the Novel Probes Project, an initiative to nurture the burgeoning field of astrophysical tests of the dark sector. The first describes astrophysical tests of extended theories of gravity (Baker \textit{et al.} \cite{Baker}).
The website of the Novel Probes Project is \url{https://www.novelprobes.org}.
}

\section{Theoretical considerations: from micro to macro}\label{sec:theory}

We will be interested in signatures of the microphysical DM scattering process on macrophysical scales, typical ${\cal O}(100)~\rm pc$ or larger.
In this section, we define various quantitative measures of the interaction strength on the microscopic scale---the differential cross section, the total cross section, and the momentum and viscosity cross section---and outline the typical range of values of the cross section for which the effects of these interactions become interesting from the perspective of structure formation. In the second part, we focus on how the effects of the microphysical interactions are modeled in simulations of structure formation. Typically, these $N$-body simulations follow the evolution of particles with masses many times $M_\odot$, so it is important to clarify the connections between the micophysical cross sections, and the modification of trajectories of the macrophysical particles in simulations. We present various methods that have been used in the literature to perform this mapping, as well as a range of tests which have been performed to validate these implementations.

\subsection{What do we mean by ``cross section''?}
\label{sec:velocity_dependence}

In this section, we tackle several issues related to the definition of a cross section and show that the microscopic definition is not the most relevant one to describe macrophysical phenomena.
We show what range of cross sections is astrophysically interesting and what types of microphysical models give cross sections in this range.

Historically, most of the discussion of SIDM was framed in the context of velocity-independent ``hard-sphere" scattering
\cite{Spergel:1999mh,Tulin:2017ara} where the outgoing momentum direction is random in the center-of-mass frame. This simple setup can be parameterized by one parameter, the total self-scattering cross section $\sigma$.

We typically, however, discuss the quantity $\sigma/m_\text{DM}$, the DM--DM cross section per unit DM particle mass $m_\text{DM}$.
For a DM particle moving at velocity $v_0$ through a background of stationary DM particles with a number density $n$, the rate at which that particle scatters with background particles is
\begin{equation}
R = \sigma \, n \, v_0 = \frac{\sigma}{m_\text{DM}} \rho \, v_0 \; ,
\label{eq:R}
\end{equation}
where $\rho$ is the mass density (henceforth ``density'') of the background particles, such that $n = \rho/m_\text{DM}$.
The total probability for the moving particle to scatter is therefore given by
\begin{equation}
p = 1 - \exp\left(- \frac{\sigma}{m_\text{DM}} \Sigma\right) \; ,
\label{eq:p}
\end{equation}
where $\Sigma \equiv \int v_0 \, \rho \, \mathrm{d}t = \int \rho \, \mathrm{d}x$ is the integrated background density through which the particle passes.
Placing this scenario in the context of virialized halos, the integrated background density is given by the column density of the halo multiplied with the number of times the DM particle passes through the halo. The former quantity is $\Sigma \sim 1\gcm$ on cluster scales (halo mass $\sim 10^{15} M_\odot$, virial velocity $\sim 10^3\hbox{ km}/\hbox{s}$) and somewhat smaller for smaller halos \cite{Lin:2015fza}, whereas the latter quantity can be estimated by the ratio of the Hubble time and the dynamical time at the virial radius and is found to be of order one.
Thus, we must have $\sigma/m_\text{DM} \sim 1\cmg$ in order for scatters to be frequent enough to induce large astrophysical effects.

\red{In realistic SIDM models, the cross section will generically change with the  relative velocity $v$ of the colliding DM particles. However, constant cross section models are also possible in the case of DM with a finite size (the ``black disk'' scattering limit), which can arise if DM is a composite particle formed by a QCD-like confining force, like neutron--neutron scattering~\cite{Buckley:2012ky}. More prosaically, we expect velocity dependence if two mass scales are relevant for the scattering process, for example, a DM and a mediator mass, or the dark proton and electron masses in the case of atomic DM.}

The best-known example \red{of velocity-dependence} is a long-range Yukawa interaction, arising from the exchange of a light scalar or vector mediator.
In the massless limit, one obtains the well-known cross section for Rutherford scattering for the case of distinguishable particles and M{\o}ller scattering for indistinguishable particles, both of which scale as $v^{-4}$.
If the mediating particle is 
not massless, {the differential cross section in the Born approximation for the scattering of identical fermions}
 is given by~\cite{Girmohanta:2022dog},
\begin{equation}
  \frac{\text{d}\sigma}{\text{d}\cos\theta} = \frac{\sigma_0}{2} \left(\frac{1}{\left(1+\xi^2\sin^2\frac{\theta}{2}\right)^2}
  + \frac{1}{\left(1+\xi^2\cos^2\frac{\theta}{2}\right)^2}
  -\frac{1}{\left(1+\xi^2\sin^2\frac{\theta}{2}\right) \left(1+\xi^2\cos^2\frac{\theta}{2}\right)}
  \right)\; ,\label{eq:longrange}
\end{equation}
where $\theta$ is the scattering angle  in the center-of-mass frame, \add{$\sigma_0=4\pi\alpha_{\rm DM}^2 m_{\rm DM}^2 / m_{\rm med}^4$ denotes the total cross section in the limit $v \to 0$}, $\alpha_{\rm DM}$ is the strength of the Yukawa potential, $m_{\rm DM}$ and $m_{\rm med}$ are the DM and mediator masses respectively,  \red{$\xi \equiv v/w$} with $w = m_\text{med}/m_\text{DM}$ is the ratio of mediator mass to DM mass. Scattering is velocity-independent for $v \ll w$ and is suppressed as $v^{-4}$ for large velocities ($v\gg w$).

{A somewhat different expression is found for the case of particle-antiparticle scattering, since the $u$-channel contribution is absent~\cite{Feng:2009hw}.
If DM particles are different from their antiparticles, the total effect of DM self-interactions will be the averaged sum of particle-particle and particle-antiparticle scattering~\cite{Yang:2022hkm}. As a simplification, the resulting self-interactions have often been modelled by the first term in Eq.~\ref{eq:longrange}, which is the t-channel contribution;} see Ref.~\cite{Girmohanta:2022dog} for a discussion of how this impacts the transfer and viscosity cross sections discussed later in this section.

Equation~\eqref{eq:longrange} holds under the Born approximation, in which the strength of the Yukawa potential $\alpha_\textrm{DM} \ll {w}$. 
{For $w \ll \alpha_\textrm{DM}$, on the other hand,
the differential cross section is no longer perturbative in $\alpha_\textrm{DM}$ and must be calculated by solving the nonrelativistic Sch\"{o}dinger equation for the Yukawa potential~\cite{Buckley:2009in,Tulin:2012wi,Tulin:2013teo}. Ref.~\cite{Loeb:2010gj} instead uses an analytic expression for the cross-section from Ref.~\cite{Feng:2009hw}.
In the low-velocity regime of $v \lesssim w$, the presence of a bound-state spectrum permits a resonant enhancement of the DM self-scattering cross section, which requires numerical solutions, whereas in the semi-classical regime $v \gg w$ one can obtain an approximate analytical solution~\cite{Colquhoun:2020adl}. In both cases it is possible to obtain large self-interactions from a weakly-coupled theory.}

{Of course, SIDM can also arise in strongly-coupled theories with $\alpha_\textrm{DM} \sim 1$. In this case DM particles may form permanent bound states~\cite{Hochberg:2014dra,Hochberg:2015vrg} analogous to the hadrons of QCD.}
While more complicated to study, these models are also expected to exhibit velocity-dependent self-interactions~\cite{Cline:2013pca,Boddy:2014yra,Boddy:2016bbu}.

\red{
If the scattering cross section exhibits a velocity dependence, it is useful to define the velocity-averaged cross section
\begin{equation}
  \langle \sigma \rangle = \frac{\int f_{\rm rel}(v_{\rm rel}) \, \sigma(v_{\rm rel})\, v_{\rm rel}^n \mathrm{d} v_{\rm rel} }{ \int f_{\rm rel}(v_{\rm rel}) \, v_{\rm rel}^n \mathrm{d} v_{\rm rel} }\; ,
  \label{eq:sigma-avg}
\end{equation}
where $f_{\rm rel}(v_{\rm rel})$ denotes the one-dimensional distribution of relative velocities in the system under consideration, and the integral is weighted by a factor of $v^n$.
The effect of a velocity dependence is then approximately captured by replacing $\sigma \to \langle \sigma \rangle$ in the discussion above for an appropriate choice of $n$ (discussed below).}
\red{
The single-particle velocity distribution $f(v)$ is typically taken to be a Maxwell--Boltzmann distribution, which is a good approximation within the thermalized core but may not work as well in the outskirts of the halo:
\begin{equation}
  f(v) \propto v^2 \exp \left(- \frac{v^2}{2\sigma_v^2} \right)
\end{equation}
with the one-dimensional velocity dispersion $\sigma_v$ of the order of the virial velocity. The distribution of relative velocities is then given by a Maxwell--Boltzmann distribution with velocity dispersion $\sqrt{2} \sigma_v$.
For an even rougher estimate, it is often sufficient to approximate $\langle \sigma \rangle \approx \sigma(v=\langle v_{\rm rel}\rangle)$ where
\add{$\langle v_{\rm rel}\rangle = 4/\sqrt{\pi} \sigma_v$} for a Maxwell--Boltzmann distribution.
}

\red{
For a velocity weighting of $n=1$ in Eq.~\eqref{eq:sigma-avg}, the substitution $\sigma \to \langle\sigma\rangle$ appropriately captures the average rate that self interactions occur within the halo.
One of the main features of SIDM, however, is that the self-scattering of DM particles permits heat flow from one region of the halo to another.
In order to capture this effect, a different value of $n$ is needed.
Under the assumption of a Maxwell--Boltzmann equilibrium distribution, the amount of momentum and energy transferred in a single collision can be used to obtain momentum- and energy-transfer rates, which are proportional to Eq.~\eqref{eq:sigma-avg} for $n=2$ and $n=3$, respectively~\cite{Colquhoun:2020adl}, reflecting the fact that the transferred momentum is proportional to $v$ and the transferred energy to $v^2$.
A more robust treatment of the velocity dependence exists in the fluid regime, in which the DM distribution function experiences a small departure from equilibrium.
Integrating the non-relativistic collisional Boltzmann equation at linear order~\cite{pitaevskii2012physical,chapman1990mathematical} yields an energy flux that is proportional to Eq.~\eqref{eq:sigma-avg} for $n=5$~\cite{Outmezguine:2022bhq}.
}

From a phenomenological perspective, velocity-dependent self-interactions are interesting because they strongly modify the relative importance of different types of astrophysical systems.
If, as in the example above, scattering is suppressed for large velocities, the sensitivity of very massive systems such as galaxy clusters (where relative velocities are typically large) is reduced compared to less massive systems like galaxies, because $\langle \sigma \rangle$ is predicted to be much smaller for the former than for the latter.
Velocity-dependent interactions have, therefore, frequently been considered to evade the strong constraints on $\sigma$ that arise from observations of galaxy clusters~\cite{Buckley:2009in,Loeb:2010gj,Kaplinghat:2015aga}, which will be discussed in more detail below.

Equation~\eqref{eq:longrange} exhibits another interesting property: the velocity dependence is accompanied by an angular dependence.
While scattering is isotropic for $v \ll w$, the differential cross section is strongly peaked towards $\theta \to 0,\pi$ for larger velocities.
The combination of a velocity dependence and an angular dependence is in fact very generic, as they often arise simultaneously from a fundamental dependence on the momentum transfer $q$, which is given by $q^2 = 2 \, m_\textrm{DM}^2 \, v^2 (1 - \cos \theta)$.
In other words, if scattering with large velocity is suppressed, there is usually also a suppression of large scattering angles.

If DM self-scattering is not isotropic, it is no longer clear that the total cross section $\sigma = \int (\mathrm{d}\sigma / \mathrm{d}\theta) \mathrm{d}\theta$ is the relevant quantity for calculating effects in astrophysical systems.
In order for the phase space density of DM to change on account of scattering, we must consider how scattering affects the exchange of momentum and energy between particles.
In fact, the more traditional quantity to use is the momentum transfer cross section
\begin{equation}
  \s = \int \frac{\mathrm{d}\sigma}{\mathrm{d}\theta} (1 - \cos \theta) \mathrm{d}\theta \; ,
\end{equation}
which down-weights forward-angle scattering (which does not appreciably change the momentum of the scattering particles) and up-weights backward-angle scattering (which yields the largest momentum transfer).
In analogy to $\langle \sigma \rangle$, one can then define the velocity-averaged momentum transfer cross section $\langle \s \rangle$ \cite{Tulin:2012wi,Tulin:2013teo}.

This definition of the momentum transfer cross section is, however, not self-consistent for the scattering of identical particles, as there should be no physical difference between $\theta = 0$ and $\theta = \pi$.
To address this issue, a modified definition is~\cite{Kahlhoefer:2013dca}.
\begin{equation}
  \tilde{\sigma}_\mathrm{T} = \int \frac{\mathrm{d}\sigma}{\mathrm{d}\theta} (1 - |\cos \theta|) \mathrm{d}\theta \; .
\end{equation}
Indeed, explicit numerical simulations show this latter definition is more appropriate for a differential cross section of the form of Eq.~\eqref{eq:longrange}~\cite{Robertson:2016qef}.%
\footnote{This conclusion may be different for the scattering of non-identical particles, e.g.\ for particle-antiparticle scattering~\cite{Agrawal:2016quu}.}
Alternatively, Refs.~\cite{Tulin:2013teo,Boddy:2016bbu} suggest considering the viscosity cross section
\begin{equation}
  \sV = \int \frac{\mathrm{d}\sigma}{\mathrm{d}\theta} \sin^2 \theta \mathrm{d}\theta \; ,
\end{equation}
which weighs the cross section by the fractional transverse energy transfer. Simulations show that $\sigma_V$ is also more appropriate than $\sigma_T$ for Rutherford scattering in Eq.~\eqref{eq:longrange}, as well as for M{\o}ller scattering~\cite{Yang:2022hkm}. Both the modified momentum transfer cross section and the viscosity cross section exhibit the appropriate symmetry: they suppress forward- and backward-angle elastic scattering, which are identical processes for identical particles.

For numerical simulations of isotropic systems, such as isolated DM halos, it has been shown that it is not necessary to explicitly include the angular dependence of the scattering in the simulation, as long as the velocity dependence is correctly included \cite{Robertson:2016qef}.
This is also true when considering average properties of entire populations, such as stacked weak lensing profiles or stacked satellite counts around galaxy clusters \cite{Banerjee:2019bjp}.
For systems with a preferred direction, such as merging galaxy clusters, however, the detailed angular dependence does become relevant~\cite{Robertson:2016qef}.
In other words, it is no longer possible to capture all features of DM self-interactions from the momentum transfer cross sections (or other variants of the cross section), since the angular information has been lost in the integration over angles.
We will return to the effects of an angular dependence in Sec.~\ref{sec:drag}.

In summary, what we mean by ``the cross section'' depends on the combination of the type of microphysical model and the astrophysical application.
Our guiding principle is capturing the flow of particles in momentum and energy space arising from interactions.
For all astrophysical applications, though, we expect the relevant cross section to be of order $\sigma/m_\text{DM} \sim 1~\cmg$ in order for there to be a significant effect for halos, which typically have DM column densities of order $\Sigma \sim 1\gcm$.

\subsection{Modeling the effective macroscopic phenomena arising from microscopic interactions}

For most systems of interest in an SIDM cosmology, cross sections $\mathcal O(1) \cmg$ imply that they fall in an interesting regime that is an intermediary between collisionless dynamics---simulated using $N$-body methods---and fully collisional hydrodynamics typically simulated with either smoothed particle hydrodynamics (e.g. Ref.~\cite{Springel2010}) or grid-based methods (e.g. Ref.~\cite{Teyssier2015}). To see this, we note that the mean free path ---the average distance a particle travels between scattering---is
\begin{equation}
\lambda = \frac{v_0}{R} = \frac{1}{\frac{\sigma}{m_\text{DM}} \rho} \; ,
\label{eq:mean_free_path}
\end{equation}
which, evaluated for the DM-density in the solar-neighborhood ($\sim 0.4 \GeVcm \approx 10^7 \Msunkpc$ \cite{Read2014}), is $\sim 0.5 \mpc$ for $\sigma / m_\text{DM} = 1 \cmg$. Thus, SIDM scattering is neither frequent enough that the distribution of particles can be described by fluid elements with isotropic pressure, nor infrequent enough to be ignored. For spherically symmetric systems, the effects of these infrequent interactions can be modelled by considering energy transport, where the thermal conductivity depends on $\lambda$ (cf. Refs.~\cite{Balberg2002,Koda2011,Shapiro2014,Pollack:2014rja,nishikawa2019,Outmezguine:2022bhq,Yang:2022hkm,Yang:2022zkd}). However, most work on astrophysical SIDM phenomena is performed in the context of numerical simulations.  The solution has been to use standard $N$-body methods, with the addition of Monte Carlo scattering to account for self-interactions, which we now describe further.

The generalization of equation~\eqref{eq:R} for the interaction rate to the case of a non-stationary background is
\begin{equation}
R(\vec{r},\vec{v}) = \int f_v(\vec{r},\vec{v}') \frac{\sigma}{m_\text{DM}} \rho(\vec{r}) \, |\vec{v}-\vec{v}'| \, \text{d}^3\vec{v}' \; ,
\label{eq:R_integral}
\end{equation}
where $f_v$ is the velocity distribution function, here normalized such that $\int f_v(\vec{r},\vec{v}) \, \text{d}^3\vec{v} = 1$. In order to determine whether a particle in an $N$-body simulation should scatter, the {local values of $f_v$ and $\rho$ must first be estimated at the position of the particle, and the rate $R$ multiplied by the length of the particle's time-step to produce a probability for scattering.} In most implementations, the value of the interaction probability computed in this way is then compared to a random number generated from a uniform distribution between $0$ and $1$, and if the computed probability exceeds the random number, a scatter takes place.

For velocity-independent elastic scattering, especially, treating the scattering of simulation particles (which may represent $\sim 10^{60}$ or more ``real'' DM particles) in the same fashion as microphysical scattering is a sensible assumption.  In other words, one can visualize the scattering of simulation particles as a scaled up version of the scattering of real particles.  One reason for this is that the astrophysically relevant quantity is the cross section per unit mass, rather than the cross section itself. Second, it can be shown that the actual evolution of the fine-grained phase space density of microphysical DM particles is mimicked by that of the coarse-grained phase space density (the phase-space density averaged over many particles) for macroscopic collections of particles. This suggests that the overall flow of energy and momentum is correctly captured by this procedure. It is important to remember that the transfer of energy and momentum among particles motivated the discussion of the different types of cross sections ($\sigma_{\rm T}$ and $\sigma_{\rm V}$) in the previous section. The choices presented in that section were also a result of the underlying assumption that the interactions of simulation particles are completely analogous to the scattering of microscopic particles. More complicated self-interaction models, e.g. those with non-trivial angular or velocity dependence in the differential cross section, have also been simulated in an exactly analogous manner, but with appropriate use of the momentum transfer cross section or the viscosity cross section.

In the context of velocity-independent elastic scattering, a number of early SIDM simulations first calculated the probability for a simulation particle to scatter, and then chose a neighboring partner with which to scatter \cite{Kochanek2000}. Most recent simulations have implemented scattering on a pair-by-pair basis where the probability for individual pairs of nearby particles to scatter is calculated, and a random number drawn for each pair to see if they do \cite{Vogelsberger:2012ku,Rocha13,Robertson17BC,Banerjee:2019bjp}.\footnote{A notable exception are Refs.~\cite{Fischer:2020uxh,Fischer:2021qux,Fischer:2022rko}, which consider the case of very frequent self-interactions leading to an effective drag force. However, also in this case it is necessary to consider pairs of simulation particles in order to re-add the energy lost due to the
drag force and ensure overall energy conservation.} This ensures not just that particles scatter at the correct rate, but that the particles they scatter with correctly sample the local velocity distribution. The probability of two nearby particles, $i$ and $j$, scattering within the next time step, $\Delta t$, is given by
\begin{equation}
\label{eq:P_ij}
P_{ij}= \frac{\sigma}{m_\text{DM}} \rho_{ij} |\vec{v}_i - \vec{v}_j| \Delta t\; ,
\end{equation}
where $\rho_{ij}$ is the contribution of particle $j$ to the density estimate at the location of particle $i$. Modern implementations of SIDM within $N$-body simulations have differed primarily in the form of $\rho_{ij}$. For example:
\begin{eqnarray}
\label{eq:rho_ij_vog}
\mathrm{Vogelsberger}~\cite{Vogelsberger:2012ku} &:&   \rho_{ij} =  m_\mathrm{p} W(r_{ij},\hsi) \\
\label{eq:rho_ij_rocha}
\mathrm{Rocha}~\cite{Rocha13} &:& \rho_{ij} =  m_\mathrm{p} \int W(|\vec{x}|,\hsi) W(|\vec{x} + \vec{r}_{ij}|,\hsi) \, \text{d}^3\vec{x}\\
\label{eq:rho_ij_rob}
\mathrm{Robertson}~\cite{Robertson17BC} &:& \rho_{ij} =  \begin{cases}
    m_\mathrm{p} / \frac{4}{3} \pi \hsi^3,& r_{ij} \leq \hsi\\
    0,              & r_{ij} > \hsi
\end{cases}
\end{eqnarray}

where $m_p$ is the mass of the simulation particle, $h_{SI}$ is a smoothing length enclosing $k$ nearest neighbours of the particle in \cite{Vogelsberger:2012ku} and \cite{Robertson17BC}, whereas it is defined as a length scale that defines a smoothing kernel of particles to evaluate an overlap fraction when two particles come close to each other in \cite{Rocha13}, $W(r,h)$ is  given by,
\begin{equation}
W(r,h) = \frac{8}{\pi h^3} \begin{cases}
    1 - 6 (\frac{r}{h})^2 + 6 (\frac{r}{h})^3,& 0 \leq \frac{r}{h} \leq 1/2 \\
    2(1-\frac{r}{h})^3,              & 1/2 < \frac{r}{h} \leq 1 \\
    0, & 1 < \frac{r}{h} \: .
\end{cases}
\end{equation}

The other main difference between recent SIDM implementations has been the choice of $\hsi$. Ref.~\cite{Vogelsberger:2012ku} used a variable $\hsi$ that adapted to the local density to keep the number of neighbours $\sim 38$, while both Refs. \cite{Rocha13} and \cite{Robertson17BC} used $\hsi$ that was fixed for all particles. Ref. \cite{Rocha13} found that $\hsi \gtrsim 0.2 \, (\rho / m_\mathrm{p})^{-1/3} $ was required for scattering to be correctly implemented (i.e. $\hsi$ must be larger than 20\% of the mean inter-particle separation), and set $\hsi$ such that this was the case at densities a few times lower than the lowest densities for which self-interactions should be significant. However, Ref. \cite{Robertson17BC} (who used a smaller $\hsi$, similar to the Plummer-equivalent gravitational softening length) demonstrated that scattering can be correctly implemented when $\hsi \ll 0.2 \, (\rho / m_\mathrm{p})^{-1/3}$, and that the result in Ref.~\cite{Rocha13} was likely a case of pairwise scattering probabilities saturating above 1.\footnote{For more details see Appendix A in Ref. \cite{Robertson17BC}.} Ref. \cite{Robertson17BC} also compared their fixed $\hsi$ approach to that adopted by Ref. \cite{Vogelsberger:2012ku} and found the results to be very similar.

While a detailed study comparing the differences between SIDM implementations has not been performed, large differences are not expected. Given the inherently stochastic nature of SIDM interactions, whether the interaction probability is smooth (equations \eqref{eq:rho_ij_vog} and  \eqref{eq:rho_ij_rocha}) or discontinuous (equation \eqref{eq:rho_ij_rob}) with respect to the particle separation, and exactly how that probability varies with separation should be unimportant. Also, large $\hsi$ results in many neighbors being found, but the probability of scattering from a particular one being small (in all cases $\rho_{ij} \propto 1/\hsi^3$), while reducing $\hsi$ essentially moves the stochasticity from the drawing of an unlikely random number, to the unlikely event of finding a particle within the search region.

\begin{figure}
\centering
\includegraphics[width=0.8\textwidth]{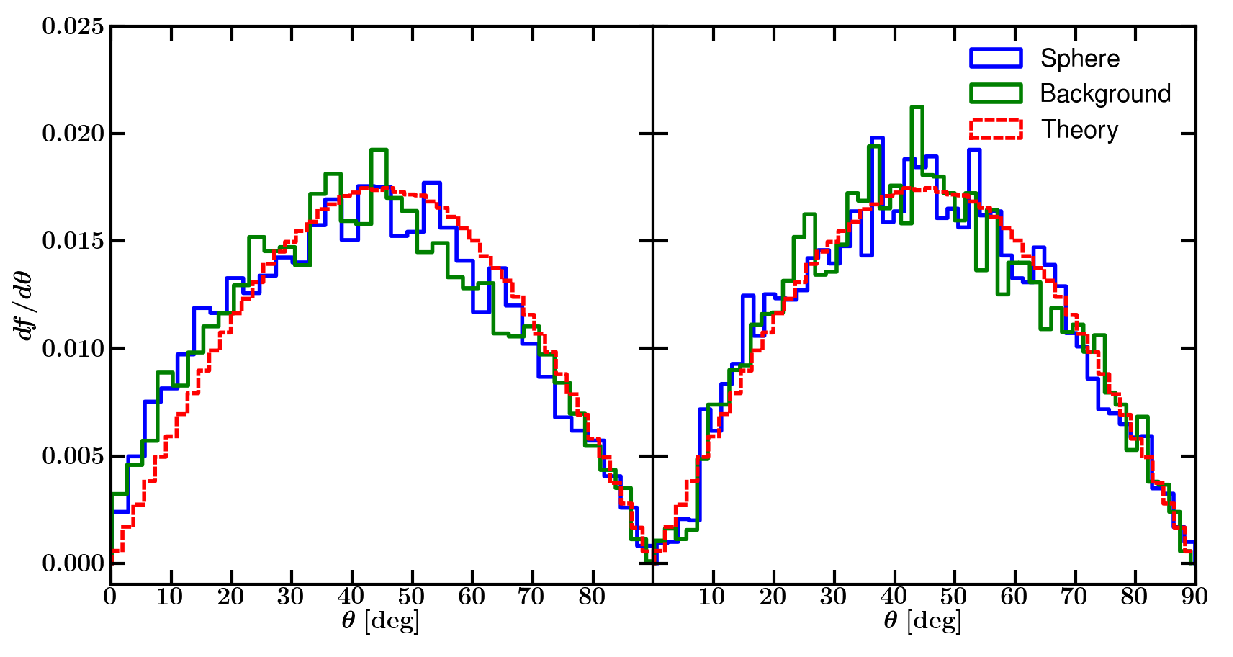}\\
\includegraphics[width=0.4\textwidth]{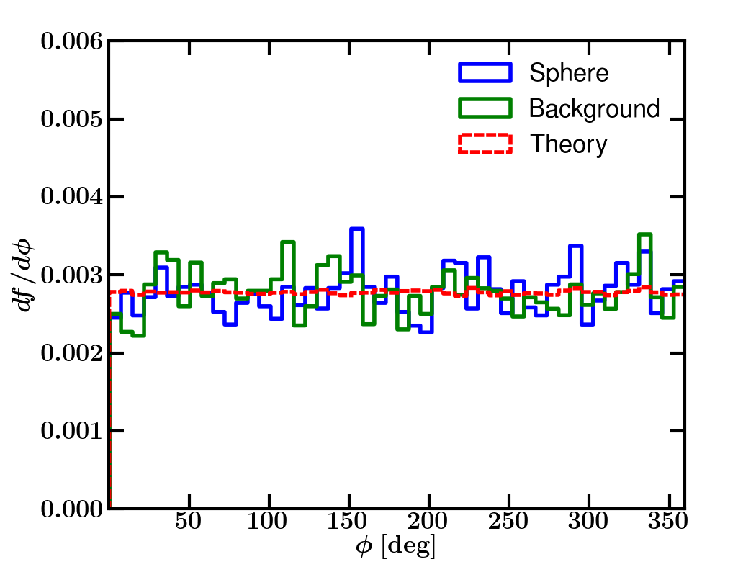}
\includegraphics[width=0.4\textwidth]{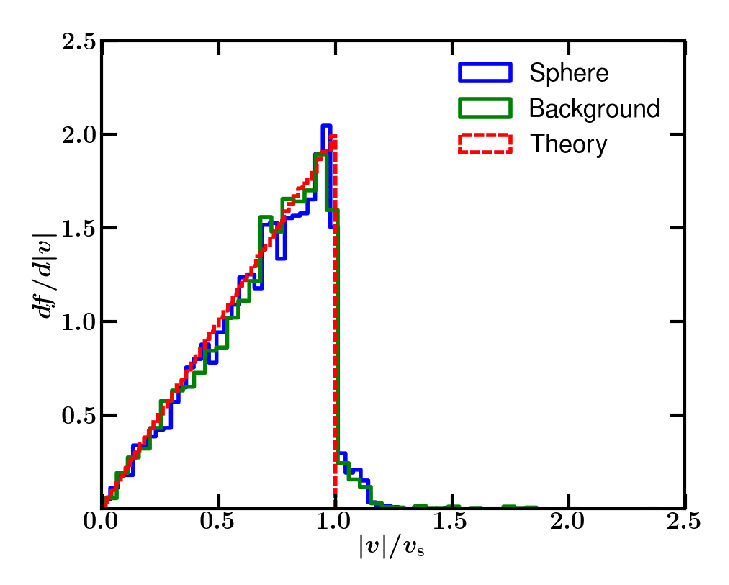}
\caption{This set of figures from \cite{Rocha13} shows the results from a test without gravity where a DM sphere is moving with uniform velocity with respect to a uniform background using the formalism described in Eq.~\ref{eq:rho_ij_rocha}. The top row shows the distribution in $\theta$ after scattering compared to the theory prediction for single elastic scatterings. The right panel shows better agreement when multiple scattering events are removed. The bottom row shows the distribution after scattering in speed and azimuthal angle. Note the excess in the speed distribution due to multiple scattering events.\label{fig:sidm_sim_tests}}
\end{figure}

A number of authors have performed tests of their simulation codes, both to test the results against analytical predictions, and to check for numerical convergence. One test for which the expected outcome can be calculated analytically is the so-called \emph{wind tunnel} test, in which particles are sent through a uniform field of stationary background particles. The rate at which particles scatter from this background can be calculated from equation~\eqref{eq:R}, and the distribution of speeds and directions of scattered particles can be calculated by converting the differential scattering cross section from the center-of-mass frame of each collision, to the rest-frame of the simulation (typically the rest-frame of the background particles). Refs. \cite{Rocha13} and \cite{Robertson17BC} both presented this test, with results that matched the analytical predictions (see e.g.~Fig.~\ref{fig:sidm_sim_tests}).

Another test with an analytical prediction is the rate of DM scattering in a DM halo with a known distribution function. A good example that has been used in the literature is a Hernquist profile \cite{1990ApJ...356..359H}, which has a similar density profile to the Navarro--Frenk--White (NFW) profile \cite{Navarro:1995iw} found in CDM simulations \cite{Navarro:1995iw}, except for a sharper fall off in density at large radius. This profile has an analytical distribution function, which allows for an analytical calculation of the expected scattering rate as a function of radius and also means that simulation initial conditions (which in the absence of DM self-interactions would have a stable density profile as a function of time) can be easily generated.  A test with the Hernquist profile was proposed and used by Vogelsberger et al. \cite{Vogelsberger:2012ku} to validate their choice of scattering kernel (Eq.~\ref{eq:rho_ij_vog}). \add{Note that it is not important whether Hernquist profiles form in SIDM (they do not); this test merely compares the analytic prediction for a given distribution function to the simulation results.}

\begin{figure}
\centering
\includegraphics[width=0.6\textwidth]{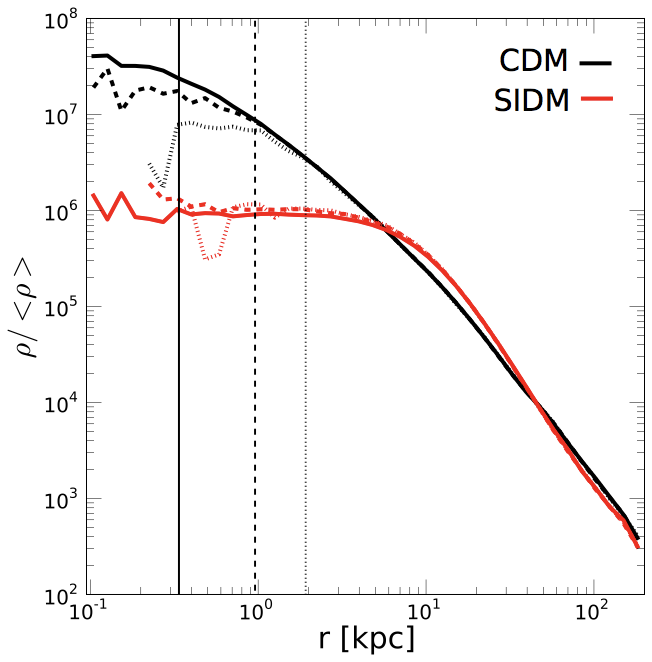}
\caption{Density profile of a MW-like DM halo (Aquarius Aq-A) at three resolution levels for both CDM (black) and isotropic SIDM with a cross section of 10 ${\rm cm}^2/{\rm g}$ (red). The resolution levels correspond to particle masses of $4.9 \times 10^4 \msun$ (solid), $3.9 \times 10^5 \msun$ (dashed), and $3.1 \times 10^6 \msun$ (dotted). The vertical lines mark the gravitational softening length (2.8 times the Plummer-equivalent softening length). The figure is adapted from Ref.~\cite{Vogelsberger:2012ku}.}
\label{fig:density_profile_convergence}
\end{figure}

Next, we discuss the question of numerical convergence, i.e. whether the profiles are converged with respect to a change in the mass resolution of the simulations.  Given the stochastic nature of SIDM scattering, it would be natural to think that a large number of particles is required in SIDM simulations, so that this stochastic interaction is adequately sampled. Given much of the interest around SIDM focuses on the effects it has on particle density profiles of halos, it is important to determine the minimum radius to which the density profiles from simulations can be trusted. \add{Interestingly, for a given particle mass in the simulation, the density profiles of SIDM halos are \red{better resolved} to smaller radii than their CDM counterparts, as can be seen in
Fig. \ref{fig:density_profile_convergence}.} The reason for this is that numerical effects in CDM simulations typically form cores, so when the physical model itself is one where a core is expected to form, these numerical effects are less significant. In fact, one of the drivers of non-convergence in the inner regions of CDM halos is the spurious two-body gravitational scattering between particles.\footnote{These interactions are unphysical, because the importance of these gravitational scattering events decreases with decreasing particle mass, such that they should be irrelevant for the Universe's DM fluid which is presumed to be made up of a very large number of particles with very low masses compared with the mass of simulation particles.} The cross section for these gravitational scattering events is a dramatic departure from the correct behaviour of CDM, but adds only a small perturbation to SIDM which has a high physical cross section for DM--DM scattering.

Finally, we briefly discuss the limitations of the Monte Carlo implementation of self-interactions that has been described above. As mentioned at the beginning of this section, this implementation is tailored to work in the intermediate regime between collisionless dynamics of CDM, and the fully collisional hydrodynamic limit. Therefore, for those models of self-interactions where the interaction rate can be much higher (typically in cases where the average momentum transfer is low), a different implementation is needed to correctly describe the evolution of the system. As a concrete example, consider a differential cross section such that DM particles experience a very large number of collisions, each with tiny scattering angles. The resulting effect is that DM particles lose momentum when moving through a sea of other DM particles; the momentum loss of the infalling particle is converted into an increase of energy of the surrounding particles, in the sense that the bulk motion of particles (i.e., the bulk kinetic energy) is turned into heat. In spirit, this is similar to the effect of dynamical friction, but can have a different velocity dependence depending on the details of the interaction cross section (e.g. \cite{Kahlhoefer:2013dca,Harvey:2015hha}). The implementation of such interactions in simulations depend on the range of the interactions. For long range interactions, which introduce an additional $1/r^2$ force on DM particles, the interactions can be implemented using the same framework used for computing the gravitational forces between simulation particles \cite{Kesden:2006vz}, in a completely deterministic manner. For short range interactions, different implementations are being explored. For example, \cite{Kummer:2019yrb} have attempted to use smoothed particle hydrodynamics methods to capture SIDM effects in models with frequent short range interactions, while \cite{2020arXiv201210277F} used an `effective drag force' approach. This area remains a field of active research, and is important to extend the reach of using simulations to explore a wider range of self-interaction models.

\section{Physical Effects in Galaxies and Clusters}\label{sec:pheno}

We begin this section with a discussion of the mechanism by which DM self-interactions alter the density profile of a DM halo: the transport of heat from hot to cold regions of the halo. In \S\ref{subsect:SIDM_density_profile} we show that this typically reduces the central density of DM halos, although things become more complicated when halos are baryon-dominated in their inner regions. This reduction in central density is only transitory, and in \S\ref{sec:core_collapse} we discuss the ultimate fate of SIDM halos, a process known as ``gravothermal collapse''. We then go on to discuss some consequences of these altered DM density profiles, including the lack of dynamical friction for objects orbiting near the centre of cored SIDM halos (\S\ref{sec:dynamicalfriction}), and the enhanced tidal stripping of SIDM subhalos due to their lower binding energies (\S\ref{sec:enhanced_stripping_core}).

We then look beyond the radial density profiles, and discuss how SIDM affects the shapes of halos (i.e. how they depart from being spherically symmetric) in \S\ref{sect:shapes}, with the primary effect being that halos become rounder with SIDM, although this is also complicated when baryons are considered. Then in \S\ref{sec:drag} we look at the effects that inter-halo DM self-interactions have in merging systems, focusing on the drag force induced by the exchange of momentum through self-interactions as well as the mass loss that happens when high-velocity collisions unbind particles from their halos.

Particle physics models that give rise to significant self-interactions can also affect the evolution of DM density fluctuations in the early universe. These changes to the matter power spectrum can affect the Cosmic Microwave Background (CMB) as well as the number density of low-mass halos, and we present current constraints in \S\ref{sect:LSS}. Finally in \S\ref{sec:unique} we discuss degeneracies that exist between DM self-interactions and other physical processes that could have similar effects. These include the baryonic physics relevant for galaxy formation, departures from cold and collisionless DM other than SIDM, and modifications to General Relativity.

\subsection{Core formation with SIDM}
\label{subsect:SIDM_density_profile}

Halos grow through hierarchical mergers in SIDM models (as in CDM models). The interactions between DM particles allows for energy transfer from one region of the halo to another. For moderate cross sections $\sigma/m_\text{DM} \lesssim 10 \rm cm^2/g$, the predictions of $\Lambda$CDM for the large-scale structure correlations and the growth of halos over time carries over unchanged to $\Lambda$SIDM models. For $\sigma/m_\text{DM} \gg 10 \rm cm^2/g$, the time scale for thermalization becomes shorter than the dynamical time scale for galactic halos and the predictions for the large-scale structure should be re-evaluated.

\begin{figure}
\centering
\includegraphics[width=0.8\textwidth]{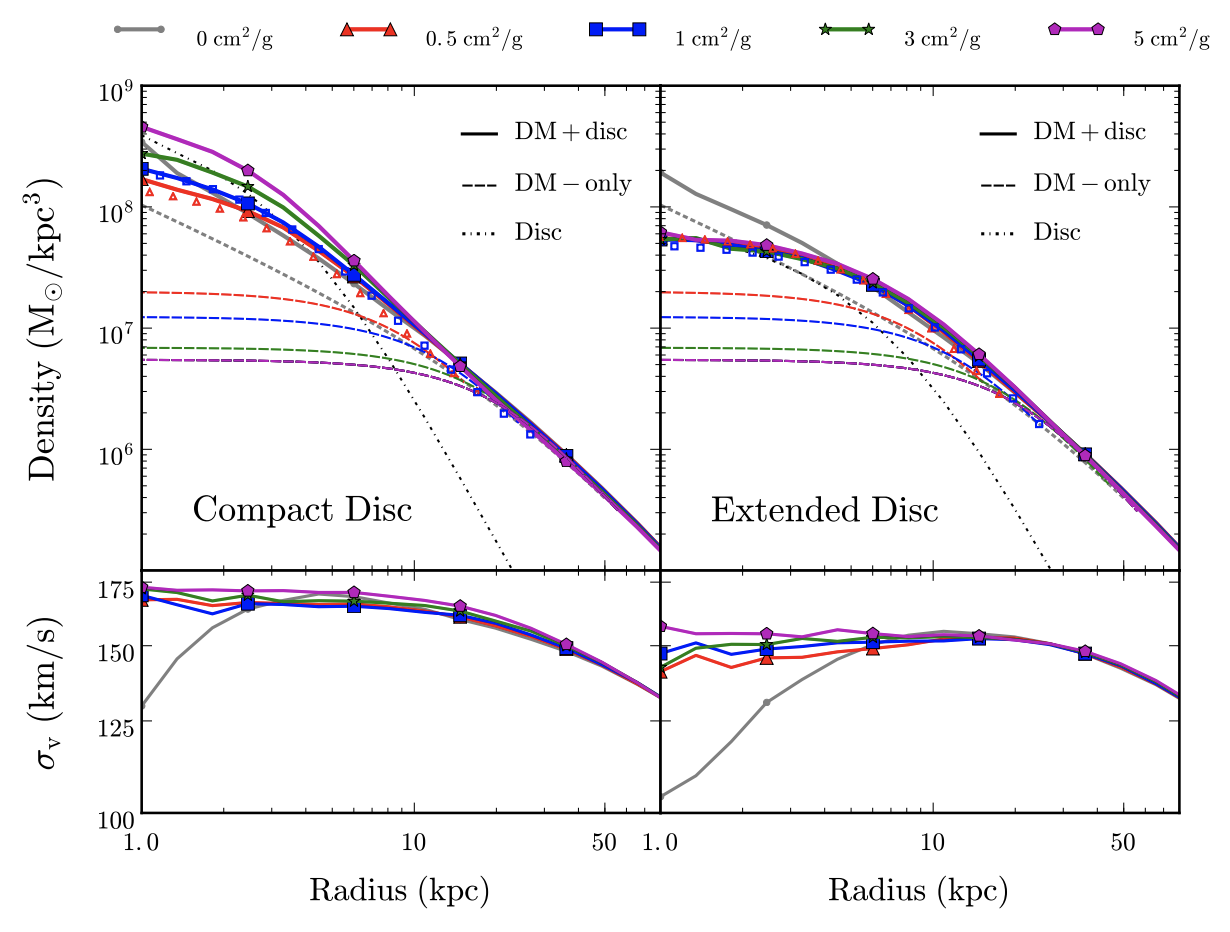}\\
\caption{Figure from \cite{Sameie:2018chj} showing the DM density (top) and velocity dispersions (bottom) in the central regions of halos in $N$-body simulation with SIDM and a central disk. The left panel shows the density and dispersion in the presence of a compact baryonic disk and the right panel corresponds to an extended disk. In the absence of a baryonic potential the main effect of introducing self-interactions is thermalization and the formation of a central core. A baryonic disk in the centre can shorten the period of core expansion and trigger core contraction. }
\label{fig:sameie_rho}
\end{figure}

As a halo grows, self-interactions transfer heat from the outer (hotter) parts to the inner (colder) parts of the halo (see the bottom panels of Fig.~\ref{fig:sameie_rho} for example velocity dispersion profiles of CDM and SIDM halos). Over time, the inner region of the halo becomes isothermal. This drive towards a constant temperature is a complicated process but the end result can be understood by focusing attention on the region of the halo where DM particles have had only one or fewer interactions over the age of the halo. Simulations show that in this region, the density profile of the halo is very similar to what would have resulted had the interaction strength been dialed to zero. Thus, the outer profile is well-described by the NFW profile, while SIDM in the inner regions is approximately isothermal. The radius at which the behaviour transitions is approximately given by the radius where the scattering rate multiplied by the age of the halo, $t_\text{age}$, is unity. This radius is known as $r_1$, which is defined through \cite{Rocha13, Kaplinghat:2015aga}:
\begin{eqnarray}\label{eq:sidm_density-rate_of_int}
\rho_s \tilde{\rho}(r_1/r_s) \langle v_{\rm rel}\rangle t_{\rm age} \sigma/m_\text{DM}= 1\nonumber \;,\\
\implies 2.5 \frac{V_{\rm max}^3}{G_{\rm N}R_{\rm max}^2} \tilde{\rho}(r_1/r_s) t_{\rm age} \sigma/m = 1
\end{eqnarray}
where $\langle v_{\rm rel}\rangle$ is the average relative velocity between DM particles. The quantity $\rho_s\langle v_{\rm rel}\rangle \propto V_{\rm max}^3/R_{\rm max}^2$ where $R_{\rm max}=2.16 r_s$ is the radius where the circular velocity reaches a maximum in an NFW profile.
and $\tilde{\rho}(x)=x^{-1}(1+x)^{-2}$ is the functional form of the NFW profile $\rho_{\rm NFW}(r)= \rho_s \tilde{\rho}(r/r_s)$. In the above equation, $V_{\rm max}$ is the maximum circular velocity within the halo (i.e. the maximum value of $\sqrt{G_{\rm{N}} M(<r) / r}$) and $R_{\rm max}$ is the radius at which this maximum occurs. We have implicitly assumed a spherical halo in making these calculations, but that is sufficient to understand the behavior of $r_1$. The concentration-mass relation in standard $\Lambda$CDM dictates that $R_{\rm max} \propto V_{\rm max}^{1.3-1.5}$ and hence \eqref{eq:sidm_density-rate_of_int} implies that $r_1$ is an approximately fixed fraction of $r_s$~\cite{Rocha13}. For a cross section of $1 {\rm cm}^2/{\rm g}$ and $t_{\rm age}=10 \,{\rm Gyr}$, $r_1 \simeq r_s$, such that the region inside $r_s$ is changed by self-interactions across a wide range of mass scales.

For cross sections much larger than $1 \rm cm^2/g$ the core size does not continue to grow. This is because the CDM halo is already isothermal around $r=r_s$. This implies that there is a lower limit to the central density in SIDM halos, which seems to be around 2-3 times $\rho_s$~\cite{Koda:2015gwa,Elbert:2014bma,Essig:2018pzq,nishikawa2019}. As $t_{\rm age} \times \sigma/m_\text{DM}$ is dialed up, all SIDM halos (evolving in isolation without baryons) will hit this floor in the density and then enter a phase where the core starts to shrink, with the core density starting to increase \cite{Balberg2002,Koda:2015gwa,Elbert:2014bma}. 
This process is analogous to the gravothermal core collapse process that is already well-understood from studies of globular clusters and that we discuss further in \S\ref{sec:core_collapse}.

While the arguments above indicate where interactions are important in the halo, they do not tell us about the density profile of the halo interior to $r_1$. Since the density profile is isothermal~\cite{Kaplinghat:2013xca, Kaplinghat:2015aga, Kamada, 2021MNRAS.501.4610R}, we can write
\begin{equation}\label{eq:isothermal}
\rho_{\rm SIDM} = \rho_0 \exp \left( -\frac{\Phi_{\rm tot}(r)-\Phi_{\rm tot}(0)}{v_{\rm{1D}}^2} \right)
\end{equation}
where $v_{\rm{1D}}$ is the one dimensional velocity dispersion of DM particles inside $r_1$ and can be related to $\langle v_{\rm rel}\rangle$. The gravitational potential $\Phi_{\rm tot}$ is due to all of the mass in the halo including the baryons. Thus, the density profile of an SIDM halo is tied to the gravitational potential of the baryons, unless the baryons in the halo are dynamically unimportant (as in low surface brightness galaxies; LSBs).
For cross sections $\sigma/m_\text{DM} \gtrsim 1 \rm cm^2/g$, the core size is close to $r_1$ for the case where the baryons are dynamically unimportant. As the stellar density increases at fixed halo mass (that is, the stellar distribution becomes more compact), the core size shrinks in response to the increased gravitational potential of the stars. In the limit where the stars dominate the central potential, the core size is set entirely by the half-light radius of the stars. Thus, at fixed halo mass, SIDM predicts that the core size is between (roughly) $r_{\rm half}$ of the stars and $r_1$ \cite{Kaplinghat:2013xca}. The central density $\rho_0$ increases as the core size shrinks due to the gravitational influence of the baryons.

These effects can be seen in Fig.~\ref{fig:sameie_rho}, which shows the density profiles and velocity dispersion profiles from $N$-body simulations of SIDM, in the presence of a baryonic potential. The halo in question has a decreasing central density with increasing cross section in the DM-only case. However, with the addition of the gravitational potential due to a compact disc,  the central densities increase above that expected with CDM, and increase with increasing cross section. The lower-panels show the velocity dispersion profiles ($\sigma_\text{v}$ in the figure is $v_{\rm{1D}}$ in our notation), which are isothermal in the inner regions as expected from the discussion above.

Returning our attention to the isothermal density profile inside of $r_1$, we note that it has two parameters that have yet to be determined: the central density, $\rho_0$, and temperature / velocity dispersion, $v_\text{1D}$. A very simple model where the isothermal mass profile is matched continuously and smoothly (both density and mass are continuous) at $r_1$ to the NFW profile has been shown to be an excellent fit to a wide range of SIDM simulations, both with and without baryons \cite{2021MNRAS.501.4610R}. One may consider many other matching procedures, for example matching at a radius where one has more or fewer interactions than 1. However, this simple procedure of matching at $r_1$ does a good job. Investigations into obtaining a better fitting function to the density profile would be useful, for example, matching at $N$ interactions where $N\simeq 1$ but is a mild function of halo mass or $V_{\rm max}$~\cite{Ren:2018jpt}.

\subsection{Gravothermal collapse of SIDM halos}\label{sec:core_collapse}

The transformation of the cusp at the halo center into a core due to elastic self-interactions is a transitory phase that leads to a quasi-equilibrium configuration once the core has achieved its maximum size, which is about the size of the radius where the velocity dispersion profile peaks. Prior to this stage, the transfer of energy due to elastic collisions occurs from the outside in, since the velocity dispersion profile has a positive gradient in the inner region (see lower panel of Fig.~\ref{fig:sameie_rho}).
There is a net flux of heat from the regions close to the maximum of the velocity dispersion to the halo centre (e.g. \cite{Colin2002}). Once the core reaches its maximum size, however, subsequent collisions cause a heat flux from the inside out since the velocity dispersion profile has a negative slope in the outer regions. This condition triggers the gravothermal collapse phase of the inner region of the SIDM halo.

Gravothermal collapse \cite{LyndenBell1968} is a well known process in globular clusters, where the inner regions have a negative specific heat that is smaller in magnitude than the positive specific heat of the outer regions when they are constrained to remain within a certain volume. In the case of an SIDM halo both the inner and outer regions have negative heat capacity, so that the flow of energy from the centre outwards heats the centre and cools the outskirts. It is this runaway process that is responsible for the collapse.
In the case of globular clusters, the collapse is prevented by the formation of binary stars. In the case of an SIDM halo, since interactions are purely elastic, the core contracts to form a cusp and ultimately collapses to form a black hole.
This phase can be followed using the gravothermal fluid approximation \cite{LyndenBell1980,Koda2011,Pollack:2014rja,Shapiro:2014oha,Shapiro:2018vju}, which combines the equations of hydrostatic equilibrium, the heat conduction law (Fourier's law) and the first law of thermodynamics. These equations contain three characteristic scales in the center of the halo---a time scale given by the relaxation time, and two length scales, the mean free path and the Jeans length. \red{There is no concrete derivation for the effective conductivity in the long mean free path (LMFP) regime, and a semi-empirical relation must be used to interpolate between the LMFP regime and the short mean free path (SMFP) regime where the fluid approximation can be applied.}
This relation to interpolate the conductivity between the LMFP and SMFP regimes introduces a constant that must be calibrated to SIDM {\it N}-body simulations \cite{Koda2011}.

Recently, investigations of the gravothermal fluid model have been extended to velocity-dependent cross sections \cite{Outmezguine:2022bhq,Yang:2022hkm,Yang:2022zkd}. In the LMFP regime, the gravothermal equations admit an approximate universality that allow, with an appropriate averaging, a velocity-dependent model to be mapped on a velocity-independent model~\cite{Outmezguine:2022bhq}. This seems to be reflected in the evolution of halo profiles in idealized N-body simulations~\cite{Yang:2022hkm}. Although further work with idealized and cosmological N-body simulations is required, there is promise here of a great simplification of the SIDM model space in as far as predictions for halo profiles of field galaxies \add{(i.e. isolated galaxies, not in group or cluster environments)} is concerned.

The gravothermal fluid approximation provides a time scale for the onset of the gravothermal collapse phase in terms of the relaxation time scale at the characteristic radius of the halo $r_s$ \cite{Balberg2002,Koda2011,Pollack:2014rja}:
$t_{\rm gc}/t_{\rm 0}\sim 400$, where
$t_{\rm 0}=1/(a\rho_0 v_{\rm 1D}\sigma/m_\text{DM})$
is the interaction time (i.e. $1/R$, where $R$ the scattering rate defined in equation~\ref{eq:R}) in the core with central density $\rho_0 = 2.4 \rho_s$, $v_{1D} = 0.65 V_{\rm max}$ and $a=4/\sqrt{\pi}$~\cite{Koda2011,Outmezguine:2022bhq}.
For the median densities and velocities of galaxy-sized halos, the collapse time exceeds a Hubble time as long as the cross section per unit mass is $\lesssim$ $\mathcal{O}(50 \, {\rm cm^2/g})$. \red{Note that there are no constraints that prohibit such large cross sections for velocities of order $100~\kms$ or smaller. However, the collapse time varies inversely with the cube of the concentration parameter~\cite{Essig:2018pzq,nishikawa2019}, which implies that a halo with concentration about 0.25 dex (roughly 2-$\sigma$) higher would collapse within a Hubble time even for a cross section of about $10 \, {\rm cm^2/g}$}.

\begin{figure}[t]
  \centering
  \includegraphics[width=0.49\columnwidth]{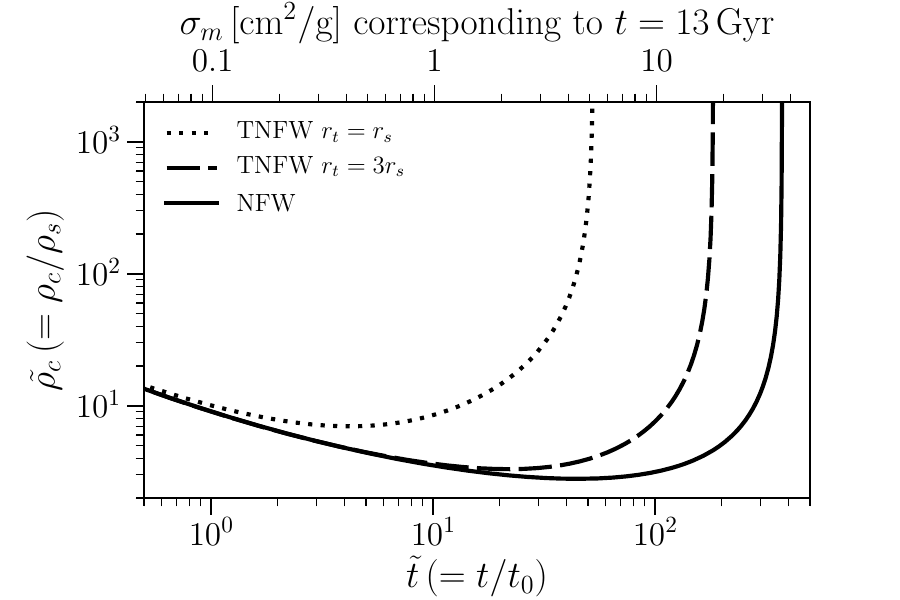} \includegraphics[width=0.48\textwidth]{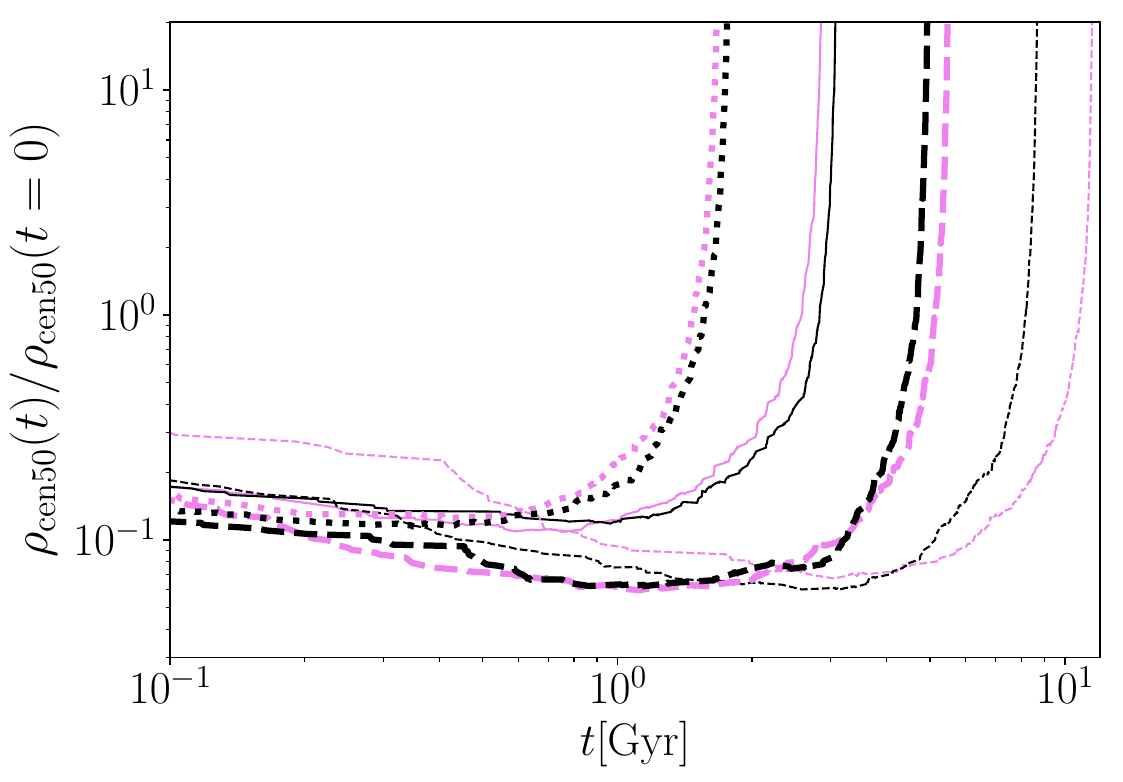}
  \caption{\emph{Left panel}: Gravothermal evolution for the central density of a halo with an initial NFW profile (solid) and with truncated NFW profiles, with truncation radii at $r_s$ (dotted) and $3r_s$ (dashed).
    The evolution of the central density is normalized to the NFW scale density $\rho_s$ and shown as a function of time, normalized to an interaction time scale $t_0 \propto 1/(\rho_s V_{\rm max} \sigma/m_\text{DM})$.
    The top axis shows the self-interaction cross section per mass needed for the associated dimensionless time on the bottom axis to correspond to 13 Gyr of evolution.
    From Ref.~\cite{nishikawa2019}.  \emph{Right panel:} Evolution of the central density of a $10^{10.5} M_\odot$ halo, relative to the density at the start of the simulation, as a function of concentration $c$ and isolation criterion, for a constant cross section of $\sigma = 6\,$cm$^2$/g. Magenta lines denote halos evolved in isolation.  Black lines show halos on a radial orbit within a group-scale halo, but with evaporation by the host turned off in order to highlight the effects of gravitational tides on the halo evolution. \red{Turning off evaporation is also the correct thing to do for SIDM models with significant velocity dependence where the cross section falls below a few $\rm cm^2/g$ at velocities of $200\ \rm km/s$}.  From right to left, the concentrations are 45, 60, 75 and 90. Figure by Z. C. Zeng, adapted from \cite{Zeng:2021ldo}.}
  \label{fig:gravothermal_with_truncation}
\end{figure}

An additional mechanism to speed up core collapse is tidal truncation of satellite galaxies. An investigation using the gravothermal equations suggests that the temporal evolution can be quickened sufficiently such that present-day subhalos could be in the core collapse phase (density increasing with time) even for moderate values of $\sigma/m_\text{DM} \sim 10~\mathrm{cm}^2/\mathrm{g}$~\cite{nishikawa2019,Zeng:2021ldo}. The effect of tidal truncation on halo collapse is demonstrated in Fig.~\ref{fig:gravothermal_with_truncation} for an initially NFW halo.
The disruption is modeled as an abrupt truncation, so that beyond the truncation radius $r_t$ (taken to be the NFW scale radius $r_s$), the density is $\rho(r) = \rho_\mathrm{NFW}(r_t) \times (r_t / r)^5$.
For an isolated halo, the core is gradually formed and persists through today. However, if the halo is initially truncated, the steepened outer slope creates a temperature gradient within the halo that allows for more rapid heat transfer and faster collapse.

Core collapse will contribute to the diversity of halo profile shapes, as demonstrated in $N$-body simulations~\cite{Sameie:2019zfo,Zavala:2019sjk,Kahlhoefer:2019oyt,Turner:2020vlf,Zeng:2021ldo, Correa:2022dey}. Moreover, the process of core collapse is faster for halos with higher concentrations, and the formation of compact, dense cores increases the probability of the halo surviving the tidal disruption event~\cite{Kahlhoefer:2019oyt}. {}While further studies and simulations are needed to test these predictions of SIDM against data, there is an intriguing anticorrelation between the central densities of bright MW dwarf spheroidal galaxies and their inferred orbital pericenter distances from Gaia~\cite{Kaplinghat:2019svz}\footnote{\url{https://sci.esa.int/web/gaia}}, which could be a window into the core collapse process. In simulations that do not allow large cross-sections, the sign of the correlation is reversed if the halos are still in the core-expansion phase for low cross-sections \cite{Ebisu:2021bjh}.

\red{We see from Fig.~\ref{fig:gravothermal_with_truncation} that the two satellite halos (black) with higher concentrations have longer collapse timescales than their isolated counterparts (magenta), but the trend is opposite for low-concentration halos. This is because higher concentrations speed up the core collapse generally, so that the isolated collapse timescales are close to the pericenter crossing time $\sim$1 Gyr and hence collapse starts before significant tidal disruption occurs. In the cases with high concentation, the collapse time is is similar to the time of the first pericenter (see also~\cite{Zheng}, fig. 6).}

Note that tidal heating and SIDM-driven evaporation may also delay or halt core collapse, potentially further enhancing the diversity of halo central densities \cite{Zeng:2021ldo}. On the other hand, the presence of a deep baryonic potential can accelerate core collapse as highlighted in Ref.~\cite{Elbert,Sameie:2018chj} and this can have a range of astrophysical consequences~\cite{Yang:2021kdf,Jiang:2022aqw}.

{There are several observational signatures of core collapse that have been discussed in literature in addition to the diversity of galaxy cricular velocity profiles. The steep inner densities for low mass subhalos that live in cluster environments can be probed through perturbations to the strong lensing arcs in galaxy clusters. \red{Ref.}~\cite{Nadler:2023nrd} recently showed that core collapse can explain the steep inner density profile of the dense substructure perturbing SDSSJ0946+1006. \red{Ref.}~\cite{Minor:2020hic} found that the concentration of the substructure was $>3\sigma$ higher than the median concentration in CDM for the inferred mass. \red{Ref.}~\cite{Yang:2022mxl} showed using cosmological zoom-in simulations of Milky Way like objects, that as large as $10\%$ of isolated halos around the Milky Way and nearly $20\%$ of low mass subhalos are expected to be in the core-collapse phase, they find that there is a higher probability of finding collapsed objects near massive halos for interaction cross-sections in currently allowed regions of parameter space. \red{Ref.}~\cite{Shah:2023qcw} demonstrated that even in large statistical samples of halos ranging from LMC to the cluster scale, a significant fraction of low mass subhalos are expected to be in the collapse phase either through tidal acceleration or even through general isolated collapse channels due to the enhanced inner densities near peaks.}  One of the relatively less-explored consequences is the formation of central black holes~\cite{Feng:2020kxv}. It is worth noting that black holes are expected as the end state in the core of a SIDM halo at late times with or without the presence of baryons or a central seed black hole~\cite{Balberg:2001qg,nishikawa2019,Feng:2021rst}.

\red{The core collapse phase is the longest phase in the evolution of a SIDM halo and subhalo~\cite{Koda:2015gwa,nishikawa2019,Outmezguine:2022bhq} relative to the total time for collapse. { The latter effect also leads to a bimodality of the slopes in the innermost density profiles of subhalos.\cite{Shah:2023qcw}} In models with large cross sections, it is natural to expect that many of the halos and subhalos will be in this phase of evolution.  A full investigation of the different aspects of core collapse promises to be a fruitful avenue for further research and an exciting area for observations. }

\subsection{Lack of dynamical friction in cored halos}\label{sec:dynamicalfriction}

An object, like a satellite galaxy, passing through a field of discrete particles, like stars or DM particles, experiences a drag force from the gravitational interaction between the object and field particles---dynamical friction \cite{chandrasekhar1943}.  In the context of galaxy formation theory, the main role dynamical friction plays is to cause satellite galaxy orbits to shrink, causing the satellite to merge with the host \cite{white1983,weinberg1989}.  Three different interpretations of dynamical friction lead to the same quantitative description of satellite orbital decay:  1. Slow field particles are accelerated by the fast object, leading to a negative acceleration of the fast object \cite{chandrasekhar1943,tremaine1984} 2. Slow field particles are deflected behind the fast object, creating a wake that pulls back on the fast object \cite{mulder1983}. 3. The fast object loses energy and angular momentum through a series of resonances with field particles \cite{tremaine1984,weinberg1986}.  For cuspy DM halos, the decay time is short (of order a Hubble time or less) for satellite-to-host mass ratios greater than 1:100 \cite{BoylanKolchin:2007ku}.

However, satellites orbiting in \emph{cored} DM halos do not experience dynamical friction.  Instead, satellites or central galaxies trapped in the core experience (nearly) undamped oscillatory motion about the center of the halo.  This unusual effect was first revealed to explain the existence of globular clusters in the MW dwarf spheroidal galaxy Fornax, which should have sunk and dissolved if Fornax inhabited a cuspy DM halo \cite{read2006,Goerdt:2008pw}. Ref. \cite{read2006} showed that this effect could be explained in the context of Interpretation \#3.  The gravitational potential of a cored halo is identical to a harmonic oscillator in the core; there is only one natural frequency for particle orbits.  Therefore, the satellite does not sweep through a series of particle resonances, like the cusped halo case; there is only one resonance.  {The resonance model of dynamical friction thus predicts no energy loss through multiple particle resonances.}  This effect can also be explained in terms of Interpretations \#1 \& \#2 of dynamical friction.  The authors of Ref. \cite{petts2015,petts2016} showed that there are few particles in the cores that are slower than a satellite on a circular orbit, owing to the isothermal nature of the particle velocity distribution.

\begin{figure}
\centering
\includegraphics[width=1\textwidth]{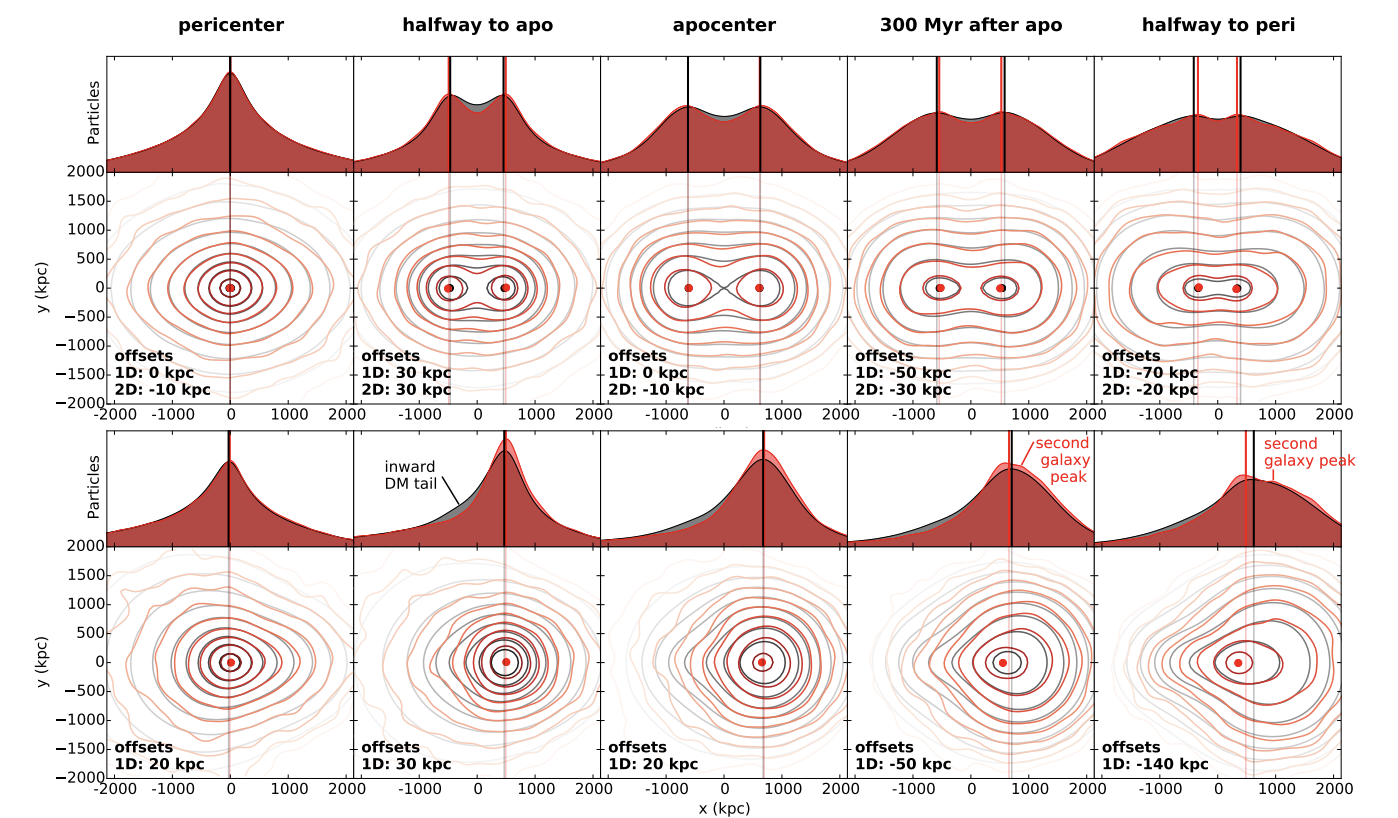}

\caption{Figure from \cite{Kim:2016ujt} showing the evolution of DM and galaxies {densities} in a merger of halos in SIDM. The top panel shows the distribution in both the halos while the bottom panel shows the distribution in one of the merging halos. The BCGs remain offset and do not merge due to reduced dynamical friction in the core. {In both panels the black contours correspond to dark matter evolution and the red contours correspond to galaxy evolution.}}
\label{fig:BCG-offset-kim}
\end{figure}

Ref. \cite{Kim:2016ujt} found that mergers between halos can trigger the oscillation of galaxies in the cores of SIDM halos.  The effect is especially noticeable on cluster scales because of the potentially large core sizes, and especially for the brightest cluster galaxies (BCGs). Figure \ref{fig:BCG-offset-kim} from \cite{Kim:2016ujt} demonstrates this effect, the BCGs of the individual clusters in a merging system do not sink to the center but instead oscillate about it. Thus, the sloshing of galaxies in halo cores is potentially a smoking gun of cored DM halos.  In \cite{Kim:2016ujt} this effect was primarily studied in the context of equal-mass mergers, and in dissipationless simulations, DM-only simulations. We discuss further work that require to determine if the BCG sloshing is a generic prediction of SIDM for less extreme mergers. {In principle such effects can also \red{be present} in CDM halos, that maybe cored due to baryonic feedback processes, we discuss further analysis and observational implications in Section \ref{sec:BCG}.}

\subsection{Enhanced gravitational tidal stripping of cored subhalos}
\label{sec:enhanced_stripping_core}

One of the original motivations for SIDM was the removal of subhalos through the process of evaporation \cite{Spergel:1999mh}, in order to solve what at the time was a pressing ``missing satellite problem" \cite{Klypin:1999uc,Moore:1999nt}.  Although neither the original problem---the missing satellite problem is no longer a problem \cite{Drlica-Wagner:2015ufc,Jethwa:2016gra,Kim:2017iwr,Nadler2021DM}, and in fact, there may be a ``too many satellites" problem in the MW \cite{Kim:2017iwr,Kim2021}---nor the physical mechanism---evaporation---are primary drivers for contemporary subhalo evolution considerations outside core-collapse scenarios \cite{Dooley:2016ajo,Zeng:2021ldo}, it is useful to consider evaporation as a launch point to a more general and observationally relevant discussion of subhalo evolution.

What happens when a smaller halo falls into a larger one?  The original Spergel \& Steinhard (2000) picture \cite{Spergel:1999mh} focused on evaporation.  Because the typical orbital speed of a subhalo is much larger than the escape velocity of a particle from the subhalo, scattering between subhalo and host-halo DM particles usually leads to neither remaining bound to the subhalo. This need not be true if scattering by large angles is strongly suppressed relative to small-angle scattering, but even in this case a large number of interactions can push subhalo particle above the subhalo escape velocity, leading to `cumulative evaporation' \cite{Kahlhoefer:2013dca,Kummer:2017bhr}.
Thus, the subhalo loses mass from these evaporative scatters, a process that is accelerated as the subhalo expands and becomes less bound.

While a theoretically attractive paradigm, simulations show that this effect changes the subhalo mass function negligibly unless the momentum transfer cross section is larger than $\sigma/m_\text{DM} > 10 \cmg$ for isotropic scattering \cite{Vogelsberger:2012ku,Dooley:2016ajo} (although see \cite{Nadler2020SIDM} for an opposing view).  Such cross sections are excluded to high significance on cluster scales in particular \cite{Randall:2007ph,Kaplinghat:2015aga,Kim:2016ujt}. To explain why such a large cross section is required, we note that a subhalo's orbital velocity will be similar to the velocity of individual host-halo DM particles, such that the rate at which subhalo DM particles scatter with host-halo
DM particles will be similar to the rate at which host-halo DM particles (at the position of the subhalo) scatter with one another. {Significant evaporation is therefore only expected for subhalos that pass through distances $r \lesssim r_1$ of the host (see Eq. \ref{eq:sidm_density-rate_of_int}), where we expect SIDM also to affect the DM density profile of the host halo. For example, in the MW we expect $r_1$ to be $20-30$ kpc for cross sections of order unity. Most subhalos have pericenter distances significantly larger than this and therefore the majority would not be affected significantly affected by evaporation. However, for those subhalos that are at distances smaller than a few times $r_1$, the evaporation effect could be significant.}

Although the subhalo mass function is essentially unaffected by self-interactions, the evolution of the satellite contained within the subhalos may be significantly altered on account of gravity alone.  For dark-matter-dominated systems like dwarf galaxies, self interactions lead to the formation of cores. Cored halos are less bound than cusped halos with the same mass.  Well outside the cored region of the halo, the density profile of the halo is nearly identical to a CDM halo of the same mass, as is the gravitational potential.  Mass loss from the halo is hence similar in the CDM and SIDM cases until the tidal radius lies near the core radius, at which point the divergence in tidal evolution accelerates \cite{Dooley:2016ajo}.  This occurs in the absence of scattering, and is a purely gravitational effect \cite{Penarrubia:2010jk}.

\begin{figure}
\centering
\includegraphics[width=1\textwidth]{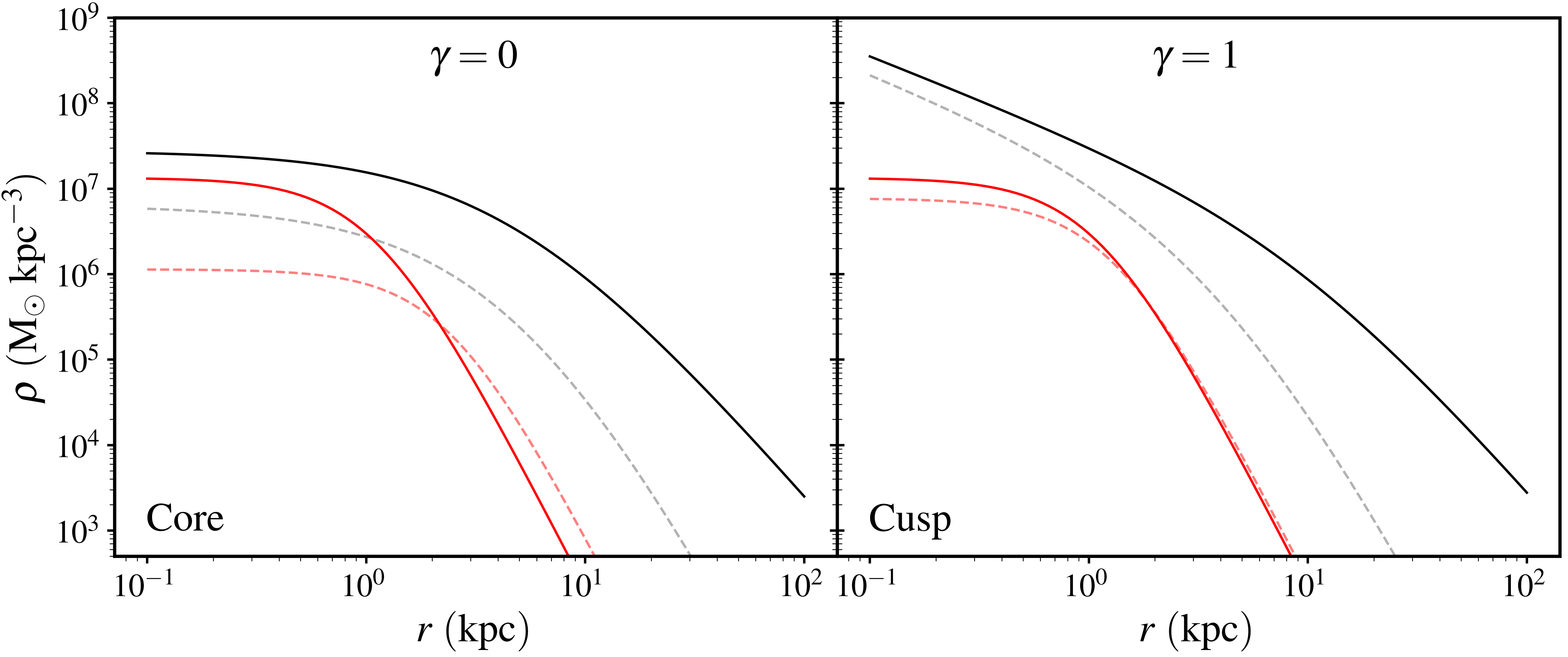}
\caption{Mass loss and stellar extent in cored (left) and cuspy (right) subhalos.  The halos originally have masses of $3.6\times 10^{10}$ M$_\odot$ (solid black line), with a stellar component of $7.5\times 10^7$ M$_\odot$ embedded within (red), and orbit a cluster-mass halo.  After 98\% of the halo mass has been stripped (dotted lines), the stellar component is larger in half-light radius if embedded in a cored halo (left) rather than a cusped halo (right).  Reproduced from Ref. \cite{Carleton2019}. }
\label{fig:Carleton_fig_1}
\end{figure}

What does this mean for the baryonic satellite at the center of the halo?  There are actually two effects relevant for subhalo evolution, on account of the baryons lying within the core region.  First, even if the tidal radius is well outside the core region, DM particles on eccentric orbits whose pericenters lie within the core region may be tidally stripped at apocenter.  This reduces the mass of the core region, and leads to an expansion of the satellite and dark core as the system equilibrates.  The effect is more severe for cored than cusped halos, leading to a greater expansion of the satellite half-light radius \cite{Dooley:2016ajo}. The expansion of the half-light radius was also invoked in the evaporation case (removal of core DM particles by scattering rather than tides) to expand elliptical galaxies in cluster environments, which would lead to a (so far unobserved) dependence of the fundamental plane of galaxies on environment \cite{Gnedin:2000ea}. Second, once the tidal radius does approach the core radius, tidal stripping of both DM and stars proceeds more quickly than in the cusped case \cite{Penarrubia:2010jk,Dooley:2016ajo}.  These effects are illustrated in Fig.~\ref{fig:Carleton_fig_1}, which shows the evolution of the DM halo (black) and stellar component (red) of a dwarf galaxy simulated in a cluster environment, as a function of the slope of the inner dark-matter halo profile.  After 98\% of the halo mass has been stripped, the remaining stellar mass is much more extended if it is embedded in the cored halo than if it were in a cusped halo.

In both cases, tidal effects will be more severe in the presence of a massive galaxy at the center of the host galaxy beyond the DM-only case considered by Ref. \cite{Dooley:2016ajo}.  The presence of a disk dramatically alters the host gravitational potential within a few scale lengths of the disk, leading to a substantial drop in the tidal radius of the satellite \cite{Penarrubia:2010jk}.  Because baryonic feedback may also lead to core formation in and enhanced tidal evolution of dwarf satellites \cite{Brooks:2012ah}, distinguishing between SIDM and baryonic core formation requires a careful assessment of core size as a function of subhalo mass for each effect.

An additional implication is that the subhalo radial distribution in the host may be much less centrally concentrated than in DM-only simulations, on account of the extra tidal field from the host galaxy \cite{DOnghia:2009xhq,2017MNRAS.467.4383S,Garrison-Kimmel:2017zes}.  Satellite disruption will be enhanced preferentially for small-pericenter orbits.  Although this population is relatively small compared to the overall satellite population, it is important for the interpretation of the missing satellites problem \cite{Kim:2017iwr} and prospects for substructure lensing probes of DM physics \cite{Nierenberg:2017vlg}.

Finally, we note that the global properties of stellar halos---the moderately phase-mixed debris from disrupted satellites---are independent of SIDM model \cite{Dooley:2016ajo}.  Most of the mass of the stellar halo comes from the few most massive satellites to fall into the host because of the steepness of the stellar mass--halo mass relation and the halo mass function \cite{Deason:2016wld}.  These satellites quickly sink to the center of the host by dynamical friction.  The mass loss is driven by the rapidly shrinking tidal radius, accompanying the rapidly decaying orbit.  The orbital decay is independent of SIDM model, and hence, the stellar halos have bulk properties identical to the CDM case.  See Ref. \cite{Dooley:2016ajo} for an extended discussion of the conditions under which SIDM might affect stellar halo observables.

In short, the relative evolution of SIDM and CDM satellites is driven by gravity rather than non-gravitational scattering.  Self-interactions matter in the sense that they set the core size of halos, and hence the tidal radius.  Once the core is set, though, the evolution can be modeled as if halos are made up of collisionless particles, unless the evaporation rate is very high \cite{Zeng:2021ldo}. {This is especially true in the core-expansion phase; if the interaction cross-sections are such that core-collapse timescales are short, the interplay between gravity and self-interactions becomes more complicated as discussed in \ref{sec:core_collapse}.}

\subsection{Halo shapes}
\label{sect:shapes}

Our discussion above has focused on the spherically-averaged density profile of SIDM halos. The shape of the SIDM halo is another key discriminant between SIDM and CDM halos. This is because the self-interactions that thermalize the inner part of the halo also isotropize the DM particle orbits and lead to axis ratios closer to unity (meaning more spherical) in the inner regions. These two predictions are in contrast to the predictions of CDM, where the velocity dispersion anisotropy can be substantial and the shapes of halos are distinctly triaxial in the center.

\begin{figure}
\centering
\includegraphics[trim=0 262 0 0,clip,width=0.98\textwidth]{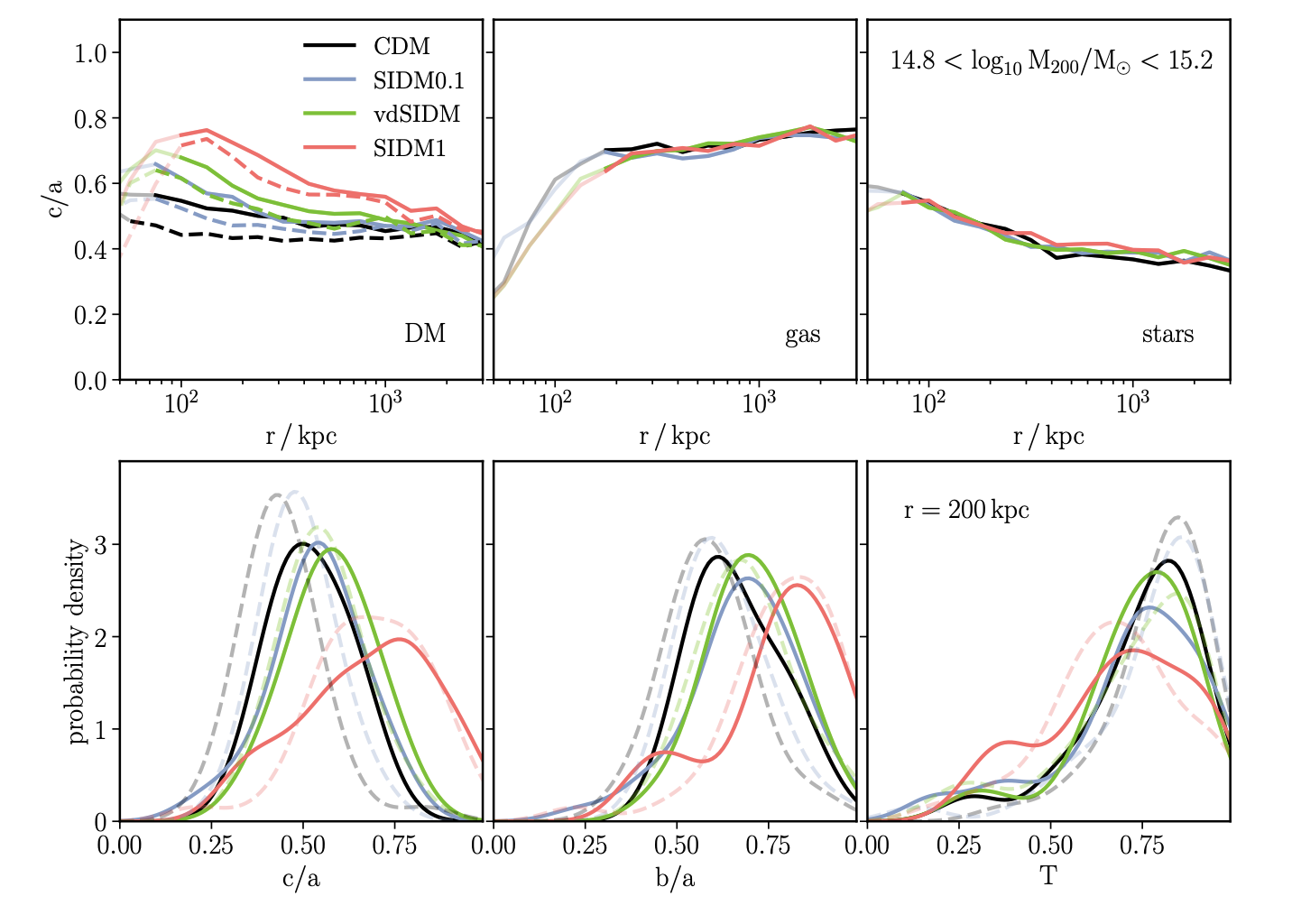}\\
\caption{This figure from \cite{Robertson:2018anx} shows the median minor-to-major axis ratios of different components of simulated galaxy clusters in  the BAHAMAS-SIDM cosmological simulations. SIDM1 and SIDM0.1 correspond to $1$ and $0.1 \, \rm{cm^2 \, g^{-1}}$, respectively, while vdSIDM is an SIDM model with a velocity-dependent cross section. The panels show the shapes of the DM (left), gas (middle) and stars (right), with dashed lines corresponding to the DM-only results. While SIDM leads to significantly more spherical DM distributions than CDM, this is not obviously reflected in the shapes of the stars or gas.}
\label{fig:shape}
\end{figure}

An example of the effects of SIDM on the shapes of halos for galaxy clusters is shown in Fig.~\ref{fig:shape}. SIDM makes halos rounder, especially towards the centre. Baryons also affect halo shapes, but the difference between CDM and $1 \, \rm{cm^2 \, g^{-1}}$ is significantly larger than the change when going from DM-only to including baryons. While SIDM makes the DM distribution rounder, it does not (at least for the clusters simulated in \cite{Robertson:2018anx}) significantly affect the shapes of either the gaseous or stellar distributions.

It is important to note that for a given cross section and halo age, the degree of impact of self interactions on the structural properties of halos is not the same in each case. Simulations have shown that of order one scattering per particle is enough to significantly affect the density profiles (see above).
However, the central (typically cored) regions do not get fully spherical (e.g. \cite{Peter:2012jh,Brinckmann2018}) and they retain a degree of orbital anisotropy \cite{Brinckmann2018}. An interesting fact in this regard is that SIDM halos with a dominant disk in the center will have a distinctly non-spherical DM halo~\cite{Kaplinghat:2013xca} because the halo shape will be set by the potential of the stellar disk.

\subsection{Drag force and mass loss due to self-interactions}
\label{sec:drag}

In the presence of DM self-interactions, DM halos moving through a background DM density experience both evaporation, as discussed in \S\ref{sec:enhanced_stripping_core}, and deceleration~\cite{Gnedin:2000ea,Markevitch:2003at,Ackerman:mha}:
\begin{equation}
 \frac{\dot{M}}{M} = - R_\mathrm{e} \, , \qquad \frac{\dot{v}}{v} = - R_\mathrm{d} \; ,
\end{equation}
where $R_\mathrm{e}$ and $R_\mathrm{d}$ denote the evaporation and deceleration rates, respectively. In terms of the total scattering rate $R$ defined in eq.~(\ref{eq:R}), these rates can be written as $R_\mathrm{e,d} = \chi_\mathrm{e,d} \, R$, where $\chi_\mathrm{e}$ and $\chi_\mathrm{d}$ denote the fraction of scattering events leading to evaporation and the average fraction of momentum lost with each scattering event, respectively. One then obtains the simple expressions
\begin{align}
 \frac{\dot{M}}{M} = - \chi_\mathrm{e} \frac{\sigma}{m_\text{DM}} \rho \, v \, , \qquad
 F_\text{drag} = - \chi_\mathrm{d} \, \sigma \, \rho \, v^2 \; ,
\end{align}
where in the second expression we have defined $F_\text{drag} \equiv m_\text{DM} \, \dot{v}$ in order to recover the well-known form of a drag force.

\textcolor{black}{The evaporation and deceleration fractions $\chi_\mathrm{e}$ and $\chi_\mathrm{d}$ depend on the ratio of the escape velocity to the velocity of the halo relative to the background density and on the differential cross section $d\sigma/d\theta$~\cite{Kummer:2017bhr} {(see eq.~(\ref{eq:longrange}))}.
Roughly speaking, the evaporation fraction is reduced for a tightly bound DM halo (high escape speed), while the deceleration rate is enhanced (and vice versa)~\cite{Kahlhoefer:2013dca}.}
Conversely, the evaporation fraction is enhanced for larger relative velocities, while the deceleration fraction is suppressed~\cite{Kim:2016ujt}. The relative importance of drag and evaporation depends sensitively on the angular and velocity dependence of the self-interaction cross section. For example, frequent interactions with low momentum transfer can lead to a a sizeable drag force on infalling objects while preventing complete evaporation~\cite{Kummer:2017bhr}. \add{Typically, as discussed in \S\ref{sec:particle} the angular dependence and low momentum transfer regime becomes relevant when $\xi>1$ in eq.~(\ref{eq:longrange}), i.e. $v/c>m_{\rm med}/m_{\rm DM}$. Note that this condition also puts a lower bound parameter space of the normalization constant,
$\sigma_0=\frac{\hbar^2 c^2}{v^{4}} \left(\frac{\alpha}{m_{\rm DM}}\right)^2$, in eq.~(\ref{eq:longrange}) for a given velocity scale at a given dark matter particle mass $m_{\rm DM}$ and coupling constant $\alpha_{\rm DM}$. For example for a Milky Way like halo with $v_{\rm rel}\sim 200 \kms$ drag effects will appear at $\sigma_0/m> 0.006 \cmg$ for a $100$ GeV dark matter particle (with $\alpha_{\rm DM}$ of order $v_{\rm rel}/c$, where eq. ~(\ref{eq:longrange}) is valid).}

Furthermore, it was pointed out in Ref.~\cite{Kahlhoefer:2013dca} that the deceleration of a DM halo due to DM self-interactions generically leads to a ``heating'' of the halo, as the directed motion of the entire halo is converted into the random motion of individual DM particles. This heating will cause some of the DM particles to obtain a kinetic energy exceeding their binding energy and escape from the DM halo, thus contributing to the evaporation rate. Writing the rate of this \emph{cumulative evaporation} as
\begin{equation}
 \frac{\dot{E}}{E} \approx \frac{\mathrm{d} v^2 / \mathrm{d}t}{v^2} = - R_\mathrm{c} \; ,
\end{equation}
one typically finds $R_\mathrm{c} \approx R_\mathrm{d}$~\cite{Kahlhoefer:2013dca}. In other words, it is generally impossible to have deceleration without at least some amount of evaporation.

While evaporation does not lead to a complete disruption of the subhalo, the main consequence of evaporation due to self-interactions is a decrease of the mass to light ($M/L$) ratios of halos in interacting systems~\cite{Clowe:2003tk}. This phenomenon is relevant both in major and minor mergers. In minor mergers, which are more common, the subhalo mass-to-light ratio is expected to be smaller than the one of typical clusters after one pericenter passage. While DM particles are scattered away into the ambient medium, the stars in the galaxies do not experience scattering, reducing the $M/L$ ratio. Ref.~\cite{Kim:2016ujt} points out however that evaporation of DM particles also can also lead to a loss of stars due to the reduced gravitational binding energy. At least for equal-mass mergers, this can potentially compensate the decrease in $M/L$, making it more difficult to obtain robust predictions. One related consequence of evaporation of subhalo particles is the thickening or puffing up of galaxies that they harbor, due to the diminished gravitational binding energy from the subhalo. However, it may be difficult to disentangle the thickenings of disks and puffing up of ellipticals due to SIDM from that due to mass loss from tidal forces in the cluster itself.

Finally, evaporation can also lead to a deformation of the shape of the DM halo. The reason is that DM particles escaping from a DM halo typically do so in the direction opposite to its velocity (relative to the background DM density). Thus, SIDM may affect the formation of DM trails, which are much more asymmetric than the ones induced by tidal forces~\cite{Kahlhoefer:2013dca}. This effect can be searched for by measuring the skewness of DM distributions~\cite{Harvey:2016bqd,Taylor:2017ipx}.

The drag force experienced by SIDM halos has a number of observational consequences. The main one is an offset between the centroids of stars and DM in merging or interacting systems, most obvious in the case of merging clusters where the DM would end up between the dissipational gas and effectively interactionless galaxies~\cite{Williams:2011pm}. Merging clusters are rare, however, and a much more common situation is the infall of galaxies into more massive halos, which is the process responsible for growing larger structures in hierarchical structure formation.

The challenge with satellite galaxies is that the much smaller lensing signal precludes statistically significant comparison of the stellar and lensing mass centroids. A possible way forward is by statistically combining a large number of minor and major mergers~\cite{Harvey:2015hha,Wittman:2017gxn}, which can potentially lead to an improved sensitivity (see section~\ref{sec:mergers}). Even relatively small star--halo offsets may however be detectable indirectly by means of their effects on the structure of stellar disks, and in particular the generation of a characteristic U-shaped warping. We discuss this signal in more detail in Sec.~\ref{sec:warps}.

In isotropic interactions the effect of drag, however, is subdominant. Simulations of mergers of massive clusters \cite{Kim:2016ujt} have shown that offsets greater than $50\,\mathrm{kpc}$ are unlikely in merging systems where DM interacts isotropically, these offsets are moreover short lived and disappear in less than a dynamical time after pericenter passage. For observed offsets as large as $100\,\mathrm{kpc}$, cross sections much higher than $1\cmg$ are needed which will cause the infalling subhalo to evaporate in the first place. Interestingly they highlight an alternative and more robust signature, whereby DM self-interactions alter the evolution of cluster merger, triggering a core sloshing of galaxies. As clusters merge and the system relaxes, the central BCG or the centroid of the galaxy distribution is offset from the DM distribution and oscillates around the barycenter, a phenomenon that does not arise in the collisionless DM scenario. The oscillation of the BCG, that is mainly produced due to reduced dynamical friction from the cored merger remnant (see section~\ref{sec:dynamicalfriction}), can be as large as a few $100\,\mathrm{kpc}$ for cross sections of $\sm\le 1 \cmg$.

While the main focus of observational tests in merging systems remains on the offsets between stars and DM and on $M/L$ ratios, it is important to note that DM self-interactions can affect the orbits of subhalos within their hosts, thereby altering the distribution of galaxies in halos. For example subhalos on radial orbits that have small pericenters, pass through denser regions of the host, undergoing more interactions than subhalos that are on tangential orbits. Depending on the nature of the cross section, this may cause radially-orbiting objects to lose more energy due to drag forces than tangential ones. This leads to objects on radial orbits being trapped near the centre changing the velocity distribution of subhalos within their hosts in comparison to CDM (\cite{Banerjee:2019bjp}). The splashback radius, which is the boundary of the multistreaming region of a halo, is the apocenter of recently accreted objects \cite{Diemer:2014xya,Adhikari:2014lna}.  The slope of the density profile of a halo rapidly falls off in a narrow localized region around this radius. Drag forces on subhalos due to self interactions lead to a loss of momentum that may cause the splashback radius to shift to smaller cluster-centric distances. The splashback radius is observed as a minimum in the slope of the number density profile of galaxies \cite{More:2016vgs,Baxter:2017csy,Chang:2017hjt}, the location of this feature can change depending on the nature of the cross section of interactions and the relative importance of drag forces.

\subsection{Large-scale structure of SIDM models}
\label{sect:LSS}

For canonical SIDM models where the \emph{only} relevant dark sector interaction following DM chemical decoupling is DM self-scattering, the success of the standard $\Lambda$CDM paradigm on large cosmological scales is retained. For these models, mass and momentum conservation ensures that the evolution of large-scale density perturbations that are still in the linear regime are unaffected by the presence of DM self-interactions \cite{Cyr-Racine:2015ihg}. The linear SIDM power spectrum is thus identical to its CDM counterpart on large cosmological scales for the canonical models.

In realistic particle physics implementations of SIDM models, however, new DM interaction beyond self-scattering are often present, which could potentially modify the evolution of DM density fluctuations. This occurs for instance in models where DM self-interactions are mediated by a relativistic or massless particle. For these, a large scattering rate between DM and the thermal bath of relativistic mediator particles at early times can efficiently erase DM density fluctuations in a process analogous to Silk damping \cite{Silk:1967kq}. A similar phenomenon occurs in theories where the force mediator is massive but is itself coupled to relativistic particles at early times.\footnote{From a model-building perspective, such a coupling might be needed to avoid having the mediator overclose the Universe (see e.g. Ref.~\cite{Huo:2017vef}).}
In both cases, the key ingredient responsible for the modified evolution of DM density fluctuations is the presence of relativistic particles coupling (directly or indirectly) to DM. Examples of SIDM models where such coupling occurs include:

\begin{itemize}
\item[] DM interacting with a massless dark photon that mixes with the SM photon \cite{HOLDOM1986196,HOLDOM198665,Dolgov:2013una,McDermott:2010pa}.
\item[] DM interacting with a massless dark photon that does not mix with the SM photon \cite{Feng:2009mn,Ackerman:mha,Ko:2016uft,Agrawal:2016quu,Ko:2017uyb}.
\item[] Atomic DM \cite{Goldberg:1986nk,Foot:2002iy,Kaplan:2009de,Kaplan:2011yj,CyrRacine:2012fz,Cline:2012is,Fan:2013tia,Fan:2013yva,Cline:2013pca,Cline:2013zca,Choquette:2015mca,Boddy:2016bbu,Agrawal:2017rvu}.
\item[] DM interacting with neutrinos \cite{Boehm:2000gq,Boehm:2001hm,Boehm:2004th,Mangano:2006mp,Serra:2009uu,Wilkinson:2014ksa,Schewtschenko:2014fca,Escudero:2015yka}.
\item[] DM interacting with photons \cite{Boehm:2000gq,Boehm:2001hm,Boehm:2004th,Sigurdson:2003vy,Sigurdson:2004zp,Wilkinson:2013kia,Kamada:2013sh,Boehm:2014vja,Schewtschenko:2014fca,Kamada:2016qjo}.
\item[] DM interacting with the baryon-photon bath \cite{Chen:2002yh,Sigurdson:2004zp,Dvorkin:2013cea,Ali-Haimoud:2015pwa,Gluscevic:2017ywp,Boddy:2018kfv,Xu:2018efh,Slatyer:2018aqg,Boddy:2018wzy,Nadler:2019zrb,DES:2020fxi,Maamari:2020aqz,Ali-Haimoud:2021lka,Nguyen:2021cnb,Rogers:2021byl,Boddy:2022tyt}
\item[] DM interacting with a non-abelian gauge boson \cite{Boddy:2014yra,Boddy:2014qxa,Buen-Abad:2015ova,Lesgourgues:2015wza,Ko:2016fcd,Krall:2017xcw,Pan:2018zha}.
\item[] DM interacting with sterile neutrinos or other types of ``dark'' radiation \cite{Aarssen:2012fx,Bringmann:2013vra,Chu:2014lja,2014PhLB..739...62K,Chacko:2016kgg,Huo:2017vef}.
\end{itemize}

The evolution of DM fluctuations in the presence of significant interactions with a relativistic species at early times mimics that of the SM baryons before the epoch of recombination \cite{Peebles:1968ja,1968ZhETF..55..278Z,Seager:1999km}. In the latter case, the photons and baryons form a tightly-coupled fluid with a large sound speed, allowing for the propagation of acoustic waves (known as the baryon acoustic oscillation, BAO) to large cosmological distances. However, the finite photon mean free path $\lambda_\gamma$ within the plasma weakens the pressure support of the acoustic oscillations, leading to significant damping of their amplitude on scales smaller than $\lambda_\gamma$. The net result of this is the presence of damped baryonic oscillations in the matter power spectrum at late times, which have been observed  in galaxy redshift surveys \cite{Eisenstein:2005su, Cole:2005sx}.

\begin{figure}[t]
\includegraphics[width=0.497\columnwidth]{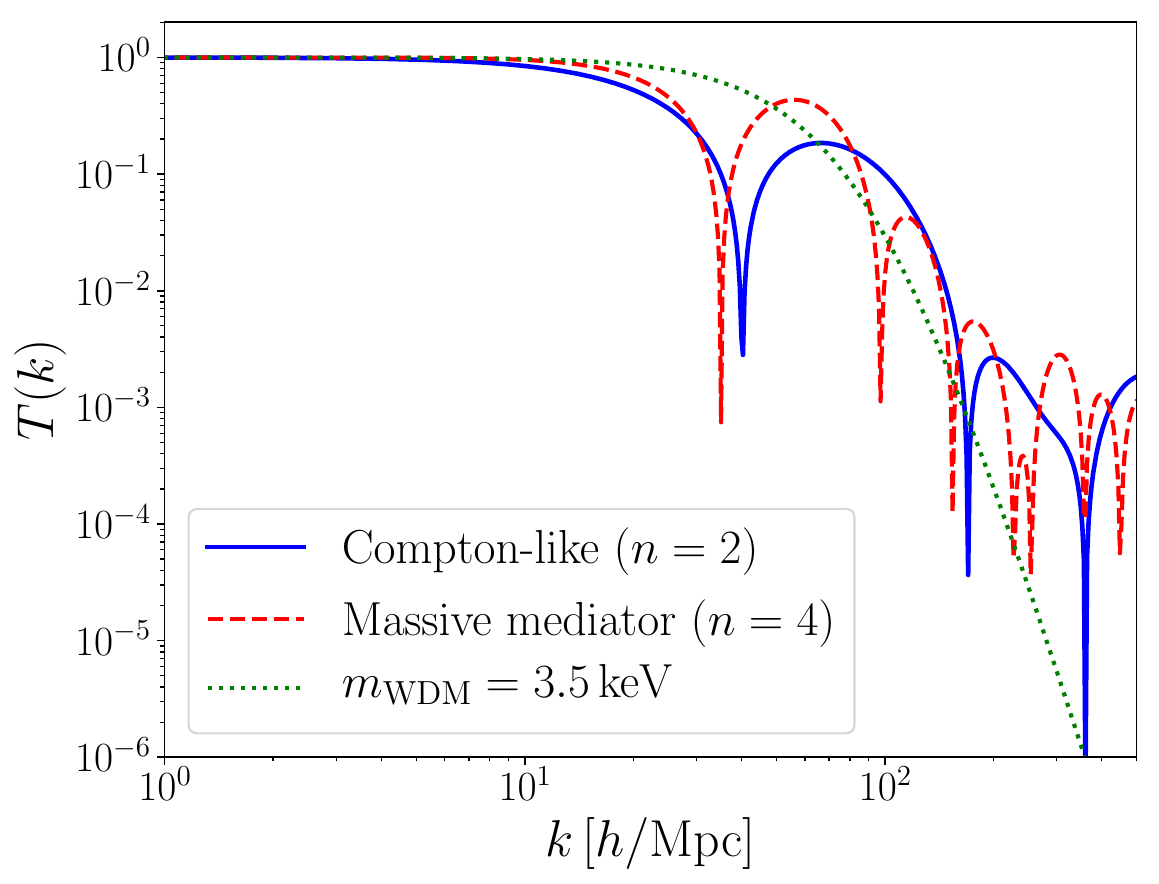}
\includegraphics[width=0.497\columnwidth]{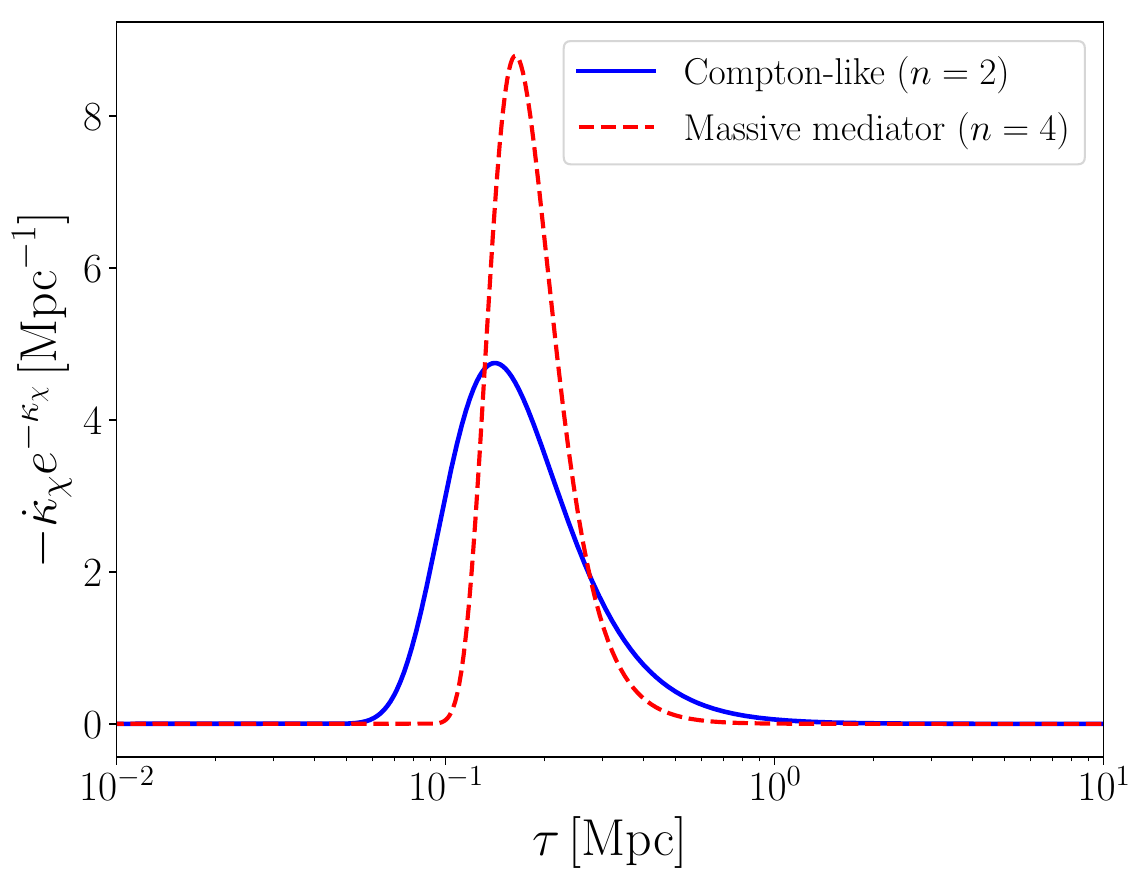}
\caption{\emph{Left panel}: Interacting DM transfer function ($T(k) \equiv (P_{\rm SIDM}(k)/P_{\rm CDM}(k))^{1/2}$) for two different types of interaction between DM and a relativistic species. Each model is parametrized with an index $n$, which corresponds to the momentum dependence of the squared matrix element describing the scattering process, $|\mathcal{M}|^2\propto (p/m_{\rm DM})^{n-2}$, where $p$ is the momentum of the incoming relativistic particle. Here, the values of the coupling strength are chosen such that all models shown have the same DM drag epoch. As shown in the legend, the $n=2$ model corresponds to a Compton-like interaction between DM and the dark radiation, while the $n=4$ case corresponds to DM interacting with dark radiation via a massive (vector or scalar) mediator. For comparison, we also display the transfer function for a warm DM model with thermal mass $m_{\rm WDM} = 3.5$ keV. \emph{Right panel}: DM drag visibility function plotted against conformal time $\tau$ for the same models as the left panel. The DM drag visibility function is essentially the probability distribution function for the time at which a DM particle last scatters off the relativistic species. Figure adapted from Ref.~\cite{Cyr-Racine:2015ihg}. \label{fig:Transfer_for_diff_n}}
\end{figure}

A similar situation occurs for DM interacting with a relativistic species at early times, except that the amplitude of the ``dark'' acoustic oscillation (DAO) in the matter power spectrum can be significantly larger than the BAO due to the higher abundance of DM as compared to baryons. For the same reason, the damping caused by the diffusion of the light species out of DM overdensities can have a dramatic impact on structure formation, essentially erasing all density perturbations on scales below the relativistic species' mean free path. Of course, the exact scale at which this occurs as well as the shape of the damping envelope of the matter power spectrum depend on the type and strength of the coupling between DM and the relativistic species (e.g., see Refs.~\cite{Feng:2009mn,CyrRacine:2012fz}). More precisely, the shapes of the oscillations and damping envelope strongly depend on the width of the DM drag visibility function (defined in analogy with the baryon drag visibility function \cite{Hu:1995en}), which itself reflects the dependence of the scattering amplitude on the momentum of the incoming relativistic species. This is illustrated in Fig.~\ref{fig:Transfer_for_diff_n} where we observe that both the oscillation wavelength and the width of the damping envelope decrease as the index $n$ describing the momentum dependence of the squared matrix element is increased (here, $|\mathcal{M}|^2\propto (p/m_{\rm DM})^{n-2}$, where $p$ is momentum of the incoming relativistic species that is scattering with DM).

In SIDM models where DM remains coupled to the relativistic species until close to the epoch of hydrogen recombination, the suppressed DM density fluctuations affect the CMB temperature and polarization power spectra in characteristic ways \cite{Cyr-Racine:2013fsa}. Generally speaking, DM must kinetically decouple from the relativistic species before the epoch of matter-radiation equality to avoid tensions with CMB measurements, unless only a small DM fraction couples to the radiation \cite{Cyr-Racine:2013fsa,Archidiacono:2017slj,Pan:2018zha}. More stringent constraints on the presence of DAO and damping in the matter power spectrum can be put using Local Group satellite galaxy counts and the Lyman-$\alpha$ forest \cite{Buckley:2014hja,Boehm:2014vja,2016MNRAS.460.1399V,Schewtschenko:2014fca,Schewtschenko:2015rno,Huo:2017vef}, and the reionization history of the Universe \cite{2018MNRAS.tmp..794L,Das:2017nub}. These bounds are primarily driven by the suppression in the DM power spectrum and the correlated suppressed abundance of low-mass halos in the late Universe, which lead to a deficit of satellite galaxies around the MW and could also delay the formation of the first stars.

DM halos with masses below the scale at which the SIDM halo mass function starts to deviate from its CDM counterpart, due to the early coupling between DM and the relativistic species, experience a delay in their formation. Since the typical inner density of DM halos is proportional to the mean cosmological DM density at the epoch at which they form, these halos tend to be less dense then their CDM counterparts. Therefore, SIDM models with significant damping in their matter power spectrum on scales corresponding to dwarf galaxies yield very diffuse DM halos at those mass scales that are difficult to reconcile with observations \cite{2016MNRAS.460.1399V}. An important corollary is that the constraints on the SIDM cross section and on the allowed amount of suppression in the matter power spectrum are not independent from each other~\cite{Aarssen:2012fx,Boddy:2016bbu,Bringmann:2016ilk,2016MNRAS.460.1399V}. Ideally, for DM models where both self-interaction and damping are present, these two quantities should be constrained simultaneously.

\subsection{Are these physical effects unique?}
\label{sec:unique}

\vspace{3mm}

\noindent To interpret the observations of any of the phenomena described in the previous sections as evidence for self-interactions of DM, it is important to consider whether they could arise by other means, for example baryonic physics, modified gravity or alternative theories of DM. In this subsection, we summarise the main alternative types of physics that could be observationally degenerate with certain SIDM phenomenology, along with the best methods to distinguish them from SIDM.

\subsubsection{Degeneracies with baryonic physics}

As we have seen, a key prediction of SIDM is the formation of DM cores in the central regions of halos, in contrast to cuspy profiles seen in pure CDM simulations. However, there are several reasons to expect the CDM prediction of cusps to be altered by baryonic physics in the process of galaxy formation. The simplest consequence of the dissipative formation of a cold baryonic disk at halo centers is that DM is drawn inward gravitationally in a process known as ``adiabatic contraction''~\cite{Blumenthal, Gnedin_1, Gnedin_2}. This process further steepens the central density profile. However, stars, supernovae, and energetic processes originating in the galactic center may subsequently inject energy and momentum into the halo by means of feedback processes, causing it to expand (e.g. \cite{Maccio, PG}). Due to the highly non-linear nature of this feedback effect (and unlike adiabatic contraction), its magnitude cannot be calculated analytically, necessitating the use of hydrodynamical simulations which can track the baryonic processes. However, even high-resolution simulations cannot cover the full range of scales that are important for the processes leading to feedback---from the collapse of giant molecular clouds at $\mathcal{O}(\text{pc})$ to the baryon cycle of the circumgalactic medium at $\mathcal{O}(\text{Mpc})$. The smallest relevant scales are therefore below the grid size of all but the highest resolution simulations, introducing a dependence on uncertain, empirically-calibrated ``sub-grid models''. Different simulations use different sub-grid models, such as the prescriptions for star formation, the coupling of stellar radiation and winds to DM, supernova rates, cooling, and gas re-accretion, and therefore end up with different predictions for observables.

While some simulations predict adiabatic contraction to remain the dominant effect on the halo density profile (e.g.~\cite{Schaller:2014uwa,Illustris}), others produce cores. Of notable interest are high-resolution galaxy simulations such as FIRE \cite{Fire-1, Fire-2}, NIHAO \cite{Nihao} and MaGICC \cite{Brook,Stinson}, which have been designed to investigate the effects of multiple interconnected feedback mechanisms in a cosmological setting. FIRE is a suite of zoom simulations of galaxies ranging from ultra-faint to MW mass, with sub-pc and $\sim 100 M_\odot$ spatial and mass resolution respectively.
Significantly, it is found in these simulations that both radiative and supernova feedback are necessary (and sufficient) for reproducing the low galaxy formation efficiency expressed in the relations between stellar and halo mass, and between stellar mass and specific star formation rate. A similar conclusion is reached in the NIHAO suite of 100 galaxies using a comparable feedback scheme, as well as in the MaGICC simulation which incorporates an early source of thermal stellar feedback.

\begin{figure}[t]
\centering
\includegraphics[width=0.6\columnwidth]{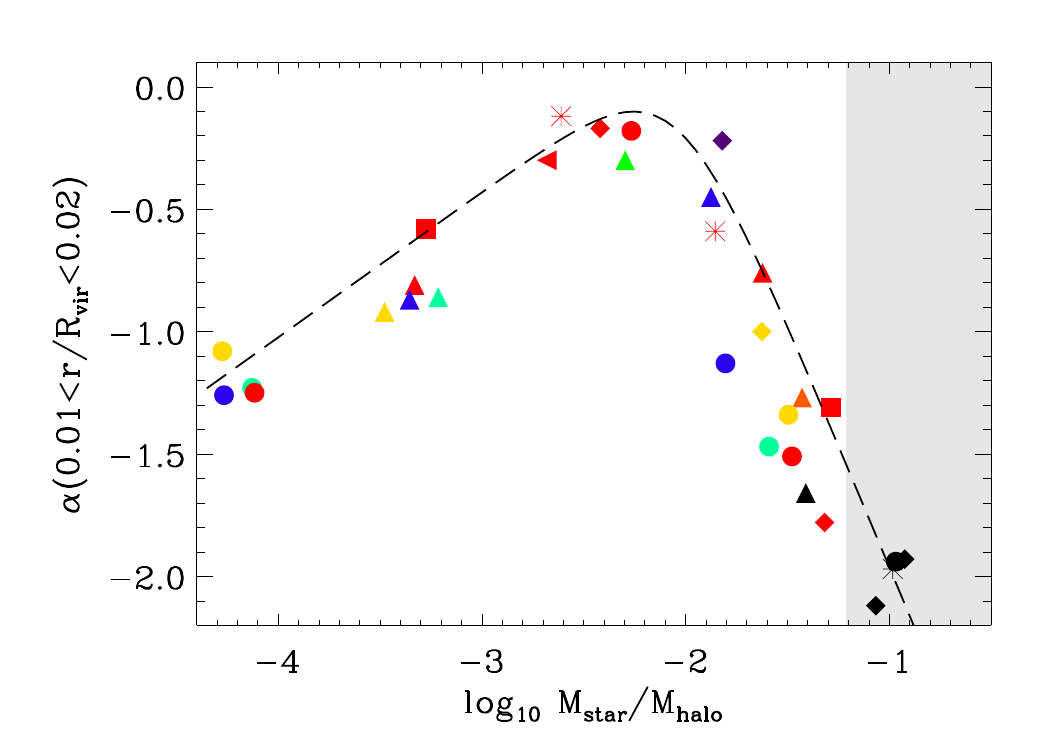}
\caption{Slope of the halo density profile between 1 and 2\% of $R_\text{vir}$ vs $M_*/M_\text{halo}$ for 31 galaxies from the MAGICC simulation. Reproduced from Ref.~\cite{DC2}.} \label{fig:dicintio}
\end{figure}

Besides regulating star formation, these feedback schemes impart energy to the DM, causing it to move out of the central regions inhabited by the galaxy. The result is a flattening of the initial NFW cusp into a core. In particular, a clear dependence is found between the central slope of the DM profile and $M_*/M_\text{halo}$, which determines the amount of energy injected into the halo relative to the depth of its potential well (e.g.~\cite{DC2}). Fig. \ref{fig:dicintio} shows such an example from the MaGICC simulation. Core formation through feedback is efficient at $10^{-3} \lesssim \log(M_*/M_\text{halo}) \lesssim 10^{-2}$, corresponding to $10^6 \lesssim M_*/M_\odot \lesssim 10^{10}$, but ineffective on either side (see Fig.~\ref{fig:dicintio}), and these results were used to create parameterized halo profiles that incorporate baryonic effects (e.g.~\cite{DC1}).
\red{A detailed Bayesian comparison of the fits with the parameterized halo profile shows that these profiles do as well as the analytic SIDM profiles at fitting the rotation curves in the SPARC database~\cite{Zentner:2022xux}. It would be interesting to see how closely the density profile parameters fit to the observations, track the same parameters for the galaxies in the MaGICC simulations used to created the parameterized halo profile.}

At low halo masses, the dominant source of feedback energy is from supernovae. However, both star formation and supernova rates fall towards lower halo mass as galaxy formation efficiency declines, so that below a certain mass ($M_\text{vir} \sim 10^8 - 10^9 M_\odot$), it is unlikely that sufficient supernovae energy is released over the entire formation history to flatten the cusp~\cite{SN_energetics}. Even if core formation were possible for the dwarf spheroidal galaxies currently at the sensitivity limit of surveys, it would be very unlikely for lower-$M_*$ ultra-faint galaxies that will be detected by next-generation surveys such as the Legacy Survey of Space and Time (LSST; \cite{Ivezic_LSST, LSST})\footnote{\url{http://www.lsst.org}} on the Vera C. Rubin Observatory. These would be predicted unambiguously to have cusps even when accounting for the effect of baryons (but see \cite{Orkney:2021wmt}), so that observations of cores would provide strong evidence for new DM physics.

\red{Hydrodynamic simulations in $\Lambda$CDM have traditionally had difficulty in accounting for the diversity in the rotation curves of galaxies with a range of surface brightness at fixed halo mass: the strong feedback necessary to push DM out of the central regions also moves the stars, which are similarly collisionless, out. This makes it difficult to form high surface brightness galaxies with steeply rising rotation curves ~\cite{Santos-Santos, Oman:2015xda, Kaplinghat:2019dhn} (but see \cite{Santos-Santos:2019vrw,roper2022} and Sec.~\ref{sec:RCs}). This is not a known issue for SIDM, however, because the stars are not directly affected by the processes that set the DM profile. In the SIDM model, thermalization leads to more compact DM cores in more luminous galaxies, which naturally explains the flat outer rotation curves of high surface brightness galaxies as well as the variation in halo profiles with galaxy size~\cite{Ren:2018jpt, Kamada, Sameie:2019zfo}. The SIDM halo profile remains as steep as the $\Lambda$CDM one in galaxies where the gravitational potential is dominated by stars and similar to $V_{\rm max}^2$ \cite{Kaplinghat:2013xca,Elbert,Sameie:2018chj}. In cases where the stellar disk is not dynamically important, the SIDM predictions for the core size are tied closely to the NFW scale radius and depend sensitively on the concentration parameter~\cite{Kamada}. }

\red{Since high and low concentration halos behave differently for SIDM (the former are subject to gravothermal collapse and the latter to core formation), a moderately large interaction cross section increases the range of inner DM densities in satellite galaxies, bringing them more in accord with the ultra-faint dwarf spheroidals around the MW \cite{Kahlhoefer_diversity}. SIDM is also expected to affect the orbital trajectories of the satellites within their host halos \cite{Robles:2019mfq}, which may provide an additional means of differentiating the model from one with CDM and baryons. In summary, mass profiles of ultra-faint dwarfs combined with detailed kinematics of low and high surface brightness galaxies
show promise in being able to
distinguish SIDM effects from those due to strong feedback in CDM halos. In addition, age and metallicity gradients may also prove useful as discriminators~\cite{Burger}.}

It is worth noting that it is possible for the effects of stellar feedback on star formation and halo density profiles to be different in SIDM than CDM \cite{Elbert, 2019MNRAS.484.4563D}, since energy exchange causes significant fluctuations in the inner potentials of halos at early times. However, hydrodynamical SIDM simulations indicate that the consequences of SIDM are more robust to the inclusion of baryonic physics than those of CDM: in both $M_*=10^8 M_\odot$ \cite{Vogelsberger:2014pda, Fry:2015rta} and $10^7 < M_*/M_\odot < 10^8$ \cite{sidm_on_fire} galaxies, inner DM density slopes are little affected by baryons.

\subsubsection{Degeneracies with novel dark matter or gravitational physics}

Besides baryonic effects, the redistribution of dynamical mass in SIDM could be mimicked by other types of DM interaction, or by modified gravity. By giving the DM particle a thermal energy, warm DM models reduce the clustering on small scales as well as lower the amplitude of the mass function at the dwarf scale by inhibiting the collapse of small halos \cite{Bode_WDM}.
Interactions between DM and baryons could also significantly alter halo density profiles~\cite{Khoury_2,Khoury_1}. The suppression of small-scale power these models imply is likely the best means to break their degeneracies with SIDM, which in the canonical case would not alter the halo mass function (though see \S\ref{sect:LSS}). Small-scale power may be effectively probed using the Lyman-$\alpha$ forest,
which has been used to constrain the temperature of DM \cite{lya1, lya2}.

The model perhaps the most similar to SIDM in terms of astrophysical phenomenology is ``fuzzy DM'', in which the DM forms soliton structures (akin to a Bose-Einstein condensate) in the central regions of halos. This is also caused by axion dark matter and scalar field models. The solitonic cores of fuzzy DM have an approximately uniform density profile, ameliorating the cusp/core problem, and, like warm DM, these models reduce the amount of substructure.
\cite{Bernal, Hu_FuzzyDM}. \add{The size of the core is proportional to the Compton wavelength $1/(m v)$ where $m$ is the mass of the DM particle and $v$ its characteristic velocity. Since $v$ increases with halo mass, for fixed $m$ the core size scales inversely with halo mass, related to a relation seen in simulations between core and halo mass~\cite{fuzzy_sim_1,fuzzy_sim_2,fuzzy_sim_3}. This seems to be inconsistent with spiral galaxy rotation curves~\cite{Bar_fuzzy,Robles_fuzzy}. This trend is opposite to that of SIDM (assuming insignificant velocity dependence of the cross-section), providing a convenient means of distinguishing the models observationally. The fuzzy DM particle mass required to form cores of the correct density is also at odds with the mass required to have enough substructure in many fuzzy DM models \cite{hayashi2021}, and, since it removes low-mass substructure, it is constrained by the Lyman-$\alpha$ forest~\cite{lya3}. Finally, fuzzy DM generates unique signatures such as interference patterns which heat stellar orbits and alter stellar streams and lensing \cite{interference_1, interference_2, interference_3}, and presents more complex dynamical friction behaviour than CDM \cite{df1, df2}.}

Because the inference of the halo density profile is purely dynamical, an alternative to altering the clustering properties of DM is to modify the law of gravity. A paradigmatic example is Modified Newtonian Dynamics (MOND)~\cite{Milgrom_1, Milgrom_2, Milgrom_3}, which alters either Newton's second law or the gravitational inverse square law to fit galaxy kinematics from the baryon mass alone. The model postulates a critical acceleration scale, $a_0 \simeq 1.2 \times 10^{-10} \: \text{m/s}^2$: above this the dynamics is Newtonian, while much below it $F \propto a^2$, leading to asymptotically flat rotation curves. $a_0$ corresponds to a critical baryon surface density $\Sigma_0 \equiv a_0/G_\text{N} \simeq 1 \: \text{kg/m}^2$ above which the ``phantom'' DM mass (reconstructed by a Newtonian analysis) falls off.
\textcolor{black}{MOND has been used to fit rotation curves and velocity dispersion profiles of galaxies over a wide range of masses, and a number of relativistic parent theories have been proposed (e.g. \cite{Skordis}; see~\cite{Famaey, Banik} for reviews). In terms of rotation curves, the relation between the total radial acceleration and the radial acceleration due to baryons is not a single curve in SIDM models but the scatter around the average is small \cite{Ren:2018jpt}. Thus, SIDM looks like MOND in an averaged sense and it is not clear if any further insight could be teased out of rotation curves. These theories could be distinguished if they predict deviations on large scales, including the CMB, matter power spectrum and properties of clusters.}

Another class of modified gravity theories
that can impact mass profiles of galaxies
are the scalar--tensor theories (e.g.~\cite{Clifton}), which supplements the metric tensor with one or more scalar fields. These fields induce a new interaction (``fifth force''), which boosts rotation and dispersion velocities. In the well-studied $f(R)$ model~\cite{fR_1,fR_2} (which is equivalent to a scalar-tensor theory \cite{Burrage_Sakstein_chameleon}), the scalar field effectively increases the \add{gravitational interaction strength $G$ above Newton's constant $G_\text{N}$}---and hence the squared velocity for given enclosed mass---by a factor of $1/3$ on scales below the Compton wavelength of the scalar field. Most viable theories including $f(R)$ implement a ``screening mechanism'' to hide the scalar field in high density regions such as the MW where observational constraints are strong (e.g.~\cite{Chameleon,Vainshtein,Kinetic, Joyce}.
This still allows a fifth force in dwarf galaxies of much lower density, however, and may make an underlying cuspy matter distribution appear core-like~\cite{Lombriser_core, Naik:2019moz} or otherwise alter the inferred dynamical density profile. Other modified gravity models such as symmetrons may have important effects on galaxy dynamics too~\cite{Burrage_symmetron,Burrage_MDAR}.

Finally, we note that modified gravity may mimic or confuse other SIDM signals too. In Sec.~\ref{sec:warps} we describe how SIDM can lead to warps, asymmetries and thickening of galaxy disks: this may also occur in the presence of a partially-screened fifth force. In particular, under chameleon, symmetron or dilaton screening the DM (and gas) in a low-mass galaxy may be unscreened, while the stars are sufficiently massive to screen themselves~\cite{Hui}. This induces an offset between the stellar and halo centres of mass in the direction of the external fifth-force field, generating a potential gradient that warps the disk~\cite{Jain_Vanderplas, Desmond_warp, Desmond_fR}. This can also cause asymmetries in rotation curves~\cite{Vikram}. However, the warp is in roughly the opposite direction to that predicted by SIDM and also has a different dependence on gravitational environment, enabling the two types of physics to be disentangled. These effects are described in detail in the modified gravity arm of the Novel Probes Project \cite{Baker:2019gxo}.

\section{Observations of SIDM phenomenology}\label{sec:observations}

An impressive diversity of observations have been used to test for SIDM. These include observations of satellite and dwarf galaxies, spirals and massive ellipticals and clusters using techniques ranging from optical spectroscopy and imaging (including strongly lensed images) to X-ray observations.

This section covers observations that have provided useful constraints and are expected to deliver improvements in the coming years. \S\ref{sec:strong_lensing} is focused on strong lensing, starting with the `classic' elliptical galaxy lenses and galaxy clusters, and moving on to the subtler effects due to subhalos within the lens galaxy and along the line of sight. \S\ref{sec:stellar_streams} discusses stellar streams of the MW. \S\ref{sec:Xray_WL} discusses the shape of galaxies' and clusters' mass profiles as probed by X-ray and weak lensing measurements. Cluster mergers, including the Bullet Cluster, are discussed in \S\ref{sec:mergers}, while \S\ref{sec:BCG} describes the kinematics and locations of the brightest galaxies in clusters (BCGs).
Dwarf galaxies in the Local Group are discussed in \S\ref{sec:dwarfs} and the rotation curves of spirals in \S\ref{sec:RCs}. Ultra-diffuse galaxies and warped disk galaxies are discussed in \S\ref{sec:UDGs} and \S\ref{sec:warps} respectively.

\subsection{Strong gravitational lensing in clusters, groups and large ellipticals}
\label{sec:strong_lensing}

In a strong gravitational lens, a background source is multiply imaged and magnified due to the distorting effects of a foreground deflector such as a massive galaxy or cluster of galaxies. The positions and magnifications of the images of the background source depend on the first and second derivatives of the projected Newtonian gravitational potential, which in turn is related to the mass distribution of the deflector projected along the line of sight. Here we discuss two ways in which strong gravitational lensing can potentially test SIDM. First we discuss strong gravitational lensing constraints on the \emph{total} mass distribution of massive elliptical galaxies as well as galaxy groups and clusters. In the second subsection we discuss how strong gravitational lensing is sensitive to the mass distribution of DM \emph{substructure}.

\subsubsection{The total deflector mass distribution \label{sec:lens-main-halo}}

\emph{Galaxy-scale lenses:} Several dedicated surveys have focused particularly on using strong gravitational lensing to study early-type galaxies, including the Sloan Lensing Advanced Camera for Surveys (SLACS, \cite{Bolton++06}) survey, the Strong Lensing Legacy Survey (SL2S, \cite{Gavazzi++12}), and the Boss Emission Line Survey (BELLS, \cite{Brownstein++12}). In the simplest approach, the slope of the mass distribution of a gravitational lens can be measured using two quantities: the Einstein radius, within which strong lensing precisely constrains the projected enclosed mass, and the stellar velocity dispersion, which provides a three dimensional mass measure within typically half of the effective radius of the deflector \cite{Treu++02,Treu++04}. Alternatively, with deep data and detailed modeling, the radial distortion of lensed images provides a direct measurement of the local slope of the mass distribution  \cite{Warren++03,Suyu++06,Vegetti++09,Schuldt:190102896S}. A robust result arising from these studies is that the \emph{total} density profile, including DM and baryons, is well approximated by a power-law with $\rho_{\rm tot}\propto r^{\gamma}$ with $\gamma\sim -2$, and small intrinsic scatter of $\simeq 0.16$ \cite{Koopmans:2006iu,Sonnenfeld:2013ApJ,Oldham++18}.

Separating the baryonic and DM density profiles is challenging because of the significant contribution of stars to the mass budget in the regime probed by strong lensing. Uncertainties in the stellar mass-to-light ratio, arising especially from the stellar initial mass function (IMF) in massive elliptical galaxies, preclude a direct subtraction of the stellar mass. The combination of the Einstein radius and velocity dispersion constrains the normalization of the stellar and DM density profiles \emph{if} the latter's slope is assumed \cite{Treu++10,Sonnenfeld++15}. Measuring both the normalization and slope requires additional information, which can come from several sources: (1) the relationship between stellar and halo mass \cite{Auger++10}, (2) ensemble constraints derived from samples of strong lenses assumed to have homologous structures \cite{Grillo++12,Oguri++14,Sonnenfeld++15}, (3) the local density slope as measured from the detailed structure of the multiple images \cite{Oldham++18}, (4) in rare cases, multiple Einstein rings \cite{Sonnenfeld++12}, and (5) spatially resolved stellar velocity dispersions and spectroscopic diagnostics of the IMF \cite{Spiniello+11,Barnabe++13}. These studies have reported inner DM density slopes ranging from $\beta_{\rm DM} \approx 0.5-2$, where $\rho_{\rm DM} \propto r^{-\beta_{\rm DM}}$, which emphasizes the difficulty of the measurements and the possibility of large galaxy-to-galaxy variations. Further progress in understanding the DM profiles in these galaxies will require more precise constraints on their stellar mass distributions, which may come from improvements in detailed stellar population models \cite{Conroy++12} or microlensing of quasars \cite{Pooley++12,Schechter++14}.

\begin{figure}[t]
\centering
\includegraphics[width=0.6\columnwidth]{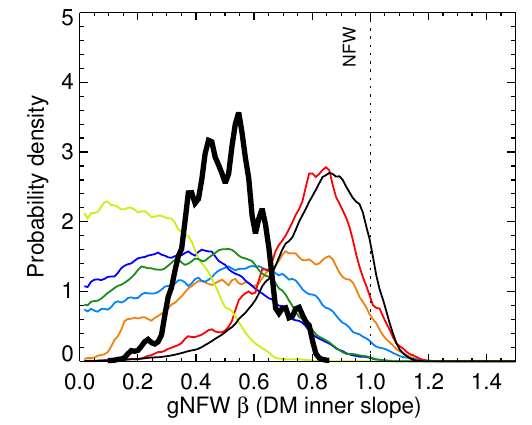}
\caption{Marginalized posterior probability densities for the DM inner slope from a sample of seven massive ($M_{200}=0.4-2\times 10^{15}$ M$_\odot$) relaxed galaxy clusters at $z=0.19-0.31$ (from Ref.~\cite{Newman++13b}). The thick black line shows the joint constraint from all clusters, while the thin lines show the individual posteriors from each cluster. The vertical dotted line shows the standard CDM expectation.}. \label{fig:cluster_dm_profile}
\end{figure}

\emph{Group- and cluster-scale lenses:} The aforementioned samples of strong lenses are primarily galaxy-scale lenses residing in $\sim10^{13} M_{\odot}$ halos \cite{Gavazzi++07}. The DM profile is more readily constrained in more massive halos due to their lower baryon fractions and richer observational constraints. Massive galaxy clusters in $\sim10^{15} M_{\odot}$ halos often present several multiple image systems with Einstein radii $\sim 50$~kpc. Spatially resolved stellar velocity dispersions within the BCG  probe the mass distribution within $\sim 20$~kpc of the cluster center, while weak lensing and X-ray emission probe the mass distribution on $\sim$Mpc scales \cite{Sand:2003bp,Sand:2007hm,Newman++11,Newman++13a,Newman++13b}. Although there is significant covariance in an individual cluster between the inner DM profile and the stellar $M_*/L$ within the BCG, this can be resolved by combining constraints from an ensemble of clusters under the assumption that their BCGs share a common IMF that is not known \emph{a priori}. An analysis of 7 clusters in this way \cite{Newman++13b} showed that the DM density profile is shallower than NFW within a typical radius of $\sim 30$~kpc and has a mean inner slope of $\beta_{\rm DM} = 0.50 \pm 17$ (c.f.~$\beta_{\rm DM} = 1$ for NFW, see Fig.~\ref{fig:cluster_dm_profile}). At smaller mass scales, new samples of group-scale lenses have been identified in panoramic imaging surveys \cite{More++12,Stark++13}, and these can be used to investigate the DM distribution in intermediate-mass halos of $\sim 10^{14} M_{\odot}$. These group-scale systems appear to be consistent with NFW-like profiles \cite{Newman++15}, suggesting that shallow DM profiles may be confined to massive clusters. Ref.~\cite{Oldham++18} also found tentative evidence that the inner DM density slope may not be universal, but instead appears to vary with the halo mass or environment.

A systematic variation in the DM density slope with halo mass might reflect trends in the formation histories of the galaxies they host. BCGs are primarily assembled through dissipationless mergers that might evacuate DM from the inner halo via dynamical processes \cite{ElZant++04,Laporte++15}, whereas in galaxy or group-scale halos, dissipational processes are probably more important, and these might instead compress the inner halo \cite{Blumenthal++86,Lackner++10}. The net effect of these baryonic processes is not well understood, especially in high-mass halos. Within the context of SIDM, the gravitational coupling of DM to the baryon distribution could produce a trend of the DM density slope with halo mass \cite{Kaplinghat:2013xca,Kaplinghat:2015aga,Kamada,Ren:2018jpt}; this would arise from the underlying variation of the central baryon fraction, not from baryon-DM interactions during galaxy formation.

\red{The measurements from cluster lenses and BCG stellar velocity dispersion measurements make it possible to constrain the density profiles of galaxy clusters down to scales of $\mathcal{O}(10\ \rm kpc)$. The resulting constraints on the SIDM cross section are very strong because a cross section of $\sim$1 $\rm cm^2/g$ can thermalize a cluster halo out to scales larger than 100 kpc, which would be inconsistent with existing constraints~\cite{Kaplinghat:2015aga,Elbert}. The fact that the cross section is so large means that the central dispersion $v_{\rm 1D}\simeq V_{\rm max}/\sqrt{3}$, which is much larger than the stellar potential well thereby reducing the effect of baryons on the SIDM density profile (see Eq.~\ref{eq:isothermal}). }

\red{These arguments were exploited to put stringent constraints on SIDM models~\cite{Kaplinghat:2015aga} at the level of about $0.2 \ \rm cm^2/g$. This method has been refined and extended by recent work to produce more robust constraints~\cite{Andrade:2020lqq,Sagunski:2020spe}. Eight clusters with generally relaxed profiles and a large set of lensed images were analyzed in Ref.~\cite{Andrade:2020lqq}. The images in these systems were spread over a large range of radii and allowed the strong lensing data by itself to place constraints on the density profile. Seven out of the eight clusters had median cross section values below $0.15 \ \rm cm^2/g$ while one of the clusters, MACS2129, was an outlier in many respects and preferred a larger cross section of $0.28 \ \rm cm^2/g$. Interestingly, MACS2129 was the only system for which one of the images couldn't be reproduced, and it preferred an anomalously high mass-to-light ratio compared to the other clusters, which is relevant because smaller mass-to-light ratios correlate with smaller core (higher DM densities). It is likely that MACS2129 requires more lensed images or stellar velocity dispersion data to provide a better model for the cluster's DM density profile. Combining all eight clusters, the upper limit on the cross section is $0.13 \ \rm cm^2/g$ at the 95\% C.L. with the mean $\langle v_{\rm rel}\rangle$ of about $1500\ \rm km/s$~\cite{Andrade:2020lqq}. It is important to note that systematic errors at the level of $0.1\ \rm cm^2/g$ are possible~\cite{Andrade:2020lqq} and more work along these lines is warranted given the stringent constraints that are now possible. }

\red{Ref.~\cite{Sagunski:2020spe} analyzed the seven clusters from Ref.~\cite{Newman++15} the span $1000-2000\ \rm km/s$ in relative velocities. They also include constraints from the stellar velocity dispersion data. Their upper limit from the clusters is $0.35 \ \rm cm^2/g$ at 95\% confidence. They take adiabatic contraction of the halo into account (which Ref.~\cite{Andrade:2020lqq} do not), which weakens the upper limit by $0.07 \ \rm cm^2/g$. The average $\langle v_{\rm rel}\rangle$ for the sample of seven clusters is $1900 \ \rm km/s$. Group scale lenses with average $\langle v_{\rm rel}$ of $1150 \ \rm km/s$ were also analyzed by the same method and a much less stringent bound of $1.1 \rm cm^2/g$ was obtained with the best-fit cross section of $0.5\pm 0.2 \ \rm cm^2/g$. We note that the constraints in Ref.~\cite{Sagunski:2020spe} are weaker than in Ref.~\cite{Andrade:2020lqq}. This could be due to the fact that Ref.~\cite{Andrade:2020lqq} created new lens models that did as well as, or outperformed, existing models for those clusters, and the full projected density profile was used to constrain the SIDM halo profile. More work on this topic is clearly warranted given the stringent constraints that are possible.}

\subsubsection{Constraints from resolved subhalos and line-of-sight halos \label{sec:sl-resolved}}

Because strong lensing observables depend on the mass distribution of the deflector, perturbations to the smooth mass distribution from low-mass DM halos in and out of the lens can alter observed image positions and fluxes \cite{Mao:1998aa,Metcalf:ad,Chiba:aa,10.1111/j.1365-2966.2010.17795.x,Sengul:2020yya}. Of particular interest for constraining SIDM are the perturbations caused by low mass ($M_{\rm vir}<10^{9} M_\odot$) subhalos and line-of-sight halos which are believed to be largely dominated by DM and thus less affected by baryons than higher mass structure (see Sec.~\ref{sec:unique}). A handful of low-mass perturbers have been identified along strongly lensed arcs thus far \cite{Vegetti_2010_2,Vegetti_2012,2014MNRAS.442.2017V,Hezaveh:2016ltk,Ritondale:2018cvp} through a technique called gravitational imaging \cite{Koopmans:aa,Vegetti++09}, and one system has been re-analyzed and shown to be consistent with a line-of-sight halo~\cite{Sengul:2021lxe}.

The probability of observing significant perturbations to a smooth gravitational lens depends in part on the mass distribution of DM subhalos (i.e. their mass function) as well as their internal mass distributions. References \cite{Nierenberg:2017vlg, Nierenberg++14} showed that an NFW perturber for example has a larger region of influence relative to an SIS perturber ($\rho \propto r^{-2}$) for fixed aperture mass. This can be understood by the fact that a shallower mass profile has to have a larger total virial mass to achieve the same aperture mass as a steeper mass profile. This higher total mass  effectively extends the region in which the perturber can influence lensed images, thus making it possible to jointly infer typical perturber mass distributions and the total subhalo mass function. Reference \cite{Vegetti++14} demonstrated that if a perturber falls on a resolved arc, high enough signal to noise data can distinguish between the effects of an SIS and a generalized NFW mass profile.

The sensitivity of strong gravitational lensing to the perturber mass profile is promising for the use of this method as a test of SIDM, which can predict halos that have either steeper or shallower inner density profiles than NFW halos at these low masses, depending if they are in the core expansion or core collapse phase of their evolution (see Sec.~\ref{sec:core_collapse}). Since the core collapse of SIDM subhalos is intrinsically linked to the amount of tidal disruption they experience \cite{nishikawa2019,Kahlhoefer:2019oyt,Zeng:2021ldo}, it will primarily affect subhalos of the main lens galaxy that have small pericenters. It is thus possible that in SIDM models with a sufficiently large cross section, the relative lensing contribution from subhalos compared to that from the line-of-sight is modified due to the former potentially having denser density profiles (resulting in them being better lenses) while the latter have (mostly) low-density cored profiles, which are more difficult to detect at high significance through strong lensing. Moreover, SIDM subhalos that are in their core expansion phase can experience severe tidal disruption \cite{Dooley:2016ajo} once their tidal radii becomes of the order of their core radii (see  Sec.~\ref{sec:enhanced_stripping_core}), resulting in a change to the subhalo mass function that might be detectable through strong lensing.

\red{Recent work has shown that for dense enough subhalos both the mass and {\it effective} slope of the mass profile of the subhalos~\cite{Minor:2020hic} or its concentration~\cite{Minor:2020bmp} could be measured. The effective slope for the perturber would be measured in the region where we expect them to have the largest observable effect \cite{Sengul:2022edu}. This is analogous to the situation for the perturber mass~\cite{Minor:2016jou,Despali:2017ksx}. The possibility of making measurements in the slope--mass plane opens up a new discovery mode for dark sector dark matter since CDM and SIDM models could look very different in the slope--mass plane; more work is needed to flesh out the predictions. Recent work in Ref.~\cite{Minor:2020hic} indicates that the dark substructure detected in SDSSJ0946+1006~\cite{Vegetti:2009cz} has a density profile that is steeper than expected for dark subhalos in CDM~\cite{Minor:2020hic}. In this context, there is a the exciting possibility of an excess of compact perturbers seen in galaxy-galaxy strong lensing observed in galaxy clusters~\cite{Meneghetti:2020yif} that don't seem to be present in hydrodynamical simulations based on CDM~\cite{Meneghetti:2022apr} but which could be consistent with SIDM models~\cite{Yang:2021kdf}. { Analysis based on N-body simulations of SIDM \cite{Nadler:2023nrd} shows that core-core collapse of subhalos in Group mass objects can explain the presence of highly concentrated perturbers in these environments, consistent with the findings of \cite{Minor:2020hic}.} The connections of these lines of evidence for compact perturbers to dark sector physics is still in its infancy. In large surveys (e.g., Euclid \cite{EUCLID}\footnote{\url{http://sci.esa.int/euclid/}} and LSST), the information about individually detected perturbers in the slope-mass plane could be combined with constraints on the subhalo mass function~\cite{Gilman:2021sdr} to increase the discriminatory power. Together, the prospects for constraining dark matter physics with strong lensing is very promising. }

\subsubsection{Constraints from the collective effect of unresolved subhalos and line-of-sight halos \label{sec:sl-unresolved}}

\begin{figure}[t]
\centering
\includegraphics[width=0.7\columnwidth]{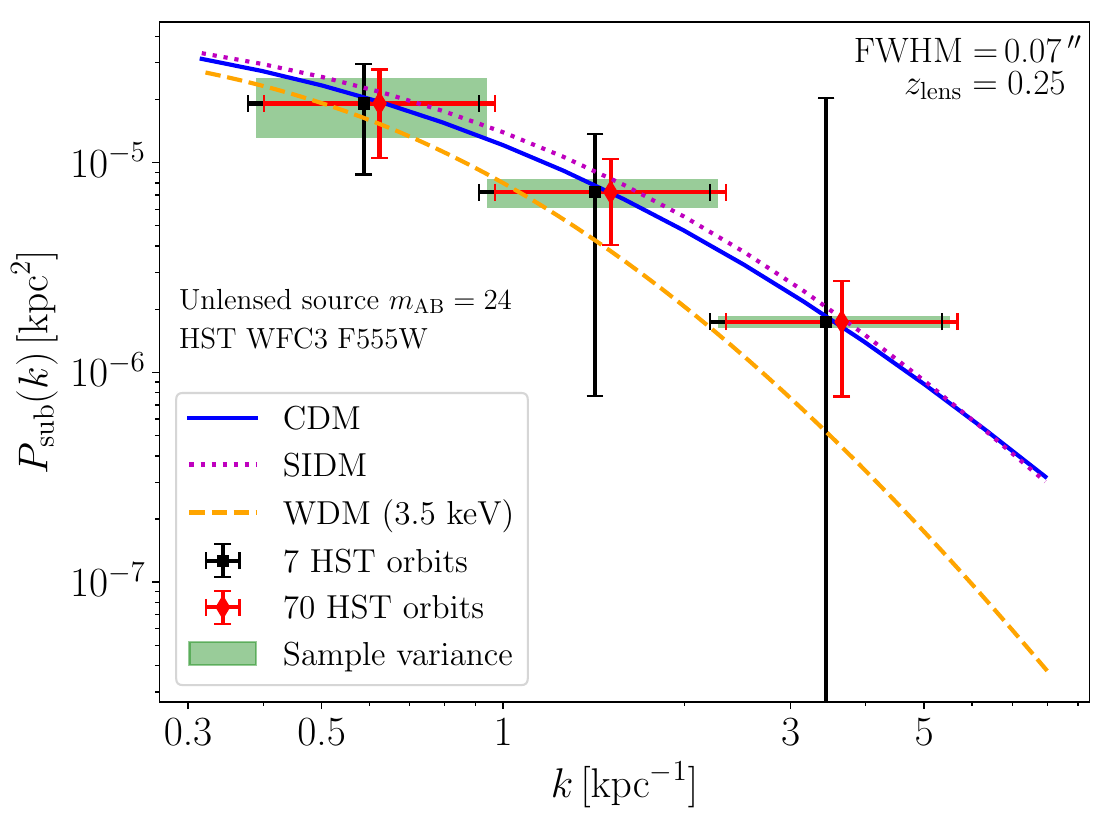}
\caption{Fisher forecast for the substructure convergence power spectrum in three logarithmic wavenumber bins. We consider here observations with the wide-field camera 3 (WFC3) aboard the Hubble Space Telescope (HST) using the F555W filter, resulting in a point-spread function FWHM of $0.07$ arcsec. The source is placed at $z_{\rm src}=0.6$ with an unlensed magnitude $m_{\rm AB}=24$. The error bars show the $1$-$\sigma$ regions, while the green rectangles display the sample variance contribution within each bin. The black points display the forecasted error bars assuming an image exposure corresponding to 7 HST orbits, while the red error bars show the forecast for a 70-orbit total exposure (in both cases we conservatively assume that only half of each orbit is available for observation). For clarity, the wavenumber bin center for each observational scenario shown have been offset by $6\%$, with the green rectangle showing the true wavenumber bin used in the analysis. The blue solid line shows the fiducial substructure power spectrum model used in the forecast, which corresponds to a CDM population of subhalos modeled with truncated NFW profiles. The dotted magenta line shows the power spectrum for SIDM, assuming a subhalo core size equals to $70\%$ of the scale radius. For comparison, the orange dashed line shows the substructure power spectrum for a thermal relic warm DM with mass of 3.5 keV. Adapted from Ref.~\cite{Cyr-Racine:2018htu}. {Note that the models are predicted assuming subhalos are in a core-expansion phase.}} \label{fig:pksub}
\end{figure}

Beyond the individual detection of massive DM halos through gravitational imaging, it is also possible to constrain SIDM by looking at the collective impact on lensed observables of the entire subhalo and line-of-sight halo population. Indeed, most DM halos contained in the small cylinder stretching from the source to the observer are either too light or lie too far in projection from the lensed images to be directly detection with high confidence. However, their collective effect might be observable at a statistically significant level, especially when data from multiple gravitational lenses are combined.

Such an approach was originally taken in Ref.~\cite{Dalal:2002aa} using flux-ratio anomalies in quasar lenses, and recently refined using an updated sample of lenses in Refs.~\cite{Hsueh:2019ynk,Gilman:2019vca,Gilman:2019nap}. Of particular relevance to SIDM, Ref.~\cite{Gilman:2019bdm} showed that a sample of eleven quasar lenses can provide constraints on the mass-concentration relation of subhalos and line-of-sight halos. \red{With 50 lenses, it would be possible to put constraints on the self-interaction cross section and its velocity dependence~\cite{Gilman:2021sdr}. {In an extension of the work developed in Ref.~\cite{Gilman:2019bdm}, Ref.~\cite{PhysRevD.107.103008} propose a formalism to infer constraints from quasar lenses in model parameter spaces where core-collapse is allowed, while the data does not allow for a large fraction ($f>80\%$) of all low mass halos to be collapsed (a scenario likely in dissipative dark matter models) it also implies that an SIDM interpretation is viable only in the presence of significant core-collapse.}}
Beyond quasars, extended lensed images could also be used to constrain the properties of the low-mass DM halo population (see e.g.~Refs.~\cite{Birrer2017,Daylan:2017kfh}). In this case, it is useful to describe the DM halo population in terms of its convergence power spectrum, as originally proposed in Ref.~\cite{Hezaveh_2014} (see also \cite{Chatterjee2018}).

To develop intuition about whether a convergence power spectrum measurements could be used to distinguish between CDM and SIDM, Ref.~\cite{Rivero:2017mao} developed a general formalism to compute from first principles the power spectrum of the convergence field for different populations of subhalos. It was shown here that the power spectrum can be mainly described by three quantities: a low-k amplitude, that depends on the subhalo abundance and on specific statistical moments of the subhalo mass function; on a turnover (intermediate) scale, that probes the truncation radius of the largest subhalos in the system; and on the small-scale slope, that probes the subhalo inner density profile. This latter property appears key to distinguish SIDM from CDM since a population of cored subhalos shows a small-scale ($\gtrsim 100$ kpc$^{-1}$) power spectrum slope that is much steeper than that of a population of truncated NFW subhalos (going as $1/k^8$ as opposed to the $1/k^4$). These findings were confirmed numerically in Refs.~\cite{Rivero:2018bcd,Brennan:2018jhq}, where it was also noted that one could in principle detect the presence of both self-interaction and of a cutoff in the subhalo mass function by measuring the amplitude and slope of the convergence power spectrum on scales $k\sim$ 1-10 kpc$^{-1}$. These conclusions did not, however, take into account the possible impact of subhalo core collapse. Ref.~\cite{Cyr-Racine:2018htu} performed a detailed forecast for the sensitivity of current and future observations to the convergence power spectrum (see Fig.~\ref{fig:pksub}). There, it was found that even deep space-based images at optical wavelengths are likely not sufficient to distinguish between CDM and SIDM \add{(in the absence of core-collapse)}, although the possible presence of a mass function cutoff could be detectable. Higher resolution interferometric data are likely to be required to distinguish SIDM from CDM. So far, only an upper bound \cite{Bayer:2018vhy} on the convergence power spectrum has been measured. {In the presence of core-collapse, where the inner slopes of collapsed halos can significantly steeper, even compared to CDM, the work of \cite{Dhanasingham:2022nox,Dhanasingham:2023thg} indicates that higher-order statistics of the convergence map can be used to disting among different scenarios of core-collapsing SIDM, core-forming SIDM, and warm dark matter models; a conclusion which is amplified by the ability to separate line-of-sight from lens-plane structure in the signal.}

\subsection{Stellar streams in the Milky Way}
\label{sec:stellar_streams}

Stellar streams, particularly those produced by tidally disrupted stars from globular clusters have very small velocity dispersions. These cold streams can be gravitationally perturbed by substructure in the Milky Way halos making them excellent probes of low mass subhalos \cite{Johnston:2001wh,Carlberg:2009ae,Yoon:2010iy,Erkal:2015kqa,Bovy:2016irg, Bonaca:2018fek, Banik:2018pjp,ibata2020}. Subhalos are expected to alter orbits of the closest stream stars creating characteristic gaps that can be observed to infer the time of encounter and the mass of the perturber. Observable gaps can probe subhalos in the mass range $10^5-10^6 M_\odot$, pushing the limit significantly below the galaxy formation threshold, allowing us to probe dark substructure. \red{The long thin stellar stream GD-1 shows evidence for such a gap and a correlated spur of nearby stars, which could be consistent with an interaction with a perturber~\cite{Bonaca:2020psc}. This perturber has to be fairly compact with approximately $10^6-10^8$ in mass for a Hernquist profile with a scale radius of  $30\ \rm pc$ or smaller~\cite{Bonaca:2018fek}.}

\red{SIDM predictions can change the radial distribution, the numbers and the internal densities of the subhalos, as discussed in the previous section on strong lensing (Sec.~\ref{sec:strong_lensing}). This will result in quantitatively different predictions for the kinematic and density perturbations that are imprinted on the streams thereby providing a venue to constrain SIDM.  In particular, core collapse can lead to dense inner cores that could survive the extreme tidal effects near the disk. Densities of order $10 \rm M_\odot/pc^3$ and higher, similar to those required in Ref.~\cite{Bonaca:2018fek}, are possible within the inner part of a core-collapsed SIDM subhalo~\cite{nishikawa2019} but a more detailed discussion of the implications is lacking. Like with the strong lensing case, this preliminary analysis highlights a real possibility for discovering SIDM if more thin streams are discovered. More recent work in Ref.~\cite{Zhang:2024fib} shows that a collapse subhalo could explain the high density of the perturber of the GD-1 stellar stream.}

The most recent data from DES analysing 5000 deg$^2$ of sky from their year 3 data have found 11 new streams within 50 kpc  of our location in the MW, these streams are observed down to surface brightness of 23 $\rm mag/arcsec$ \cite{DES:2018imd}. While currently DES is the state-of-the-art, with the most sensitive wide-area view, with the advent of the Vera C. Rubin Observatory we are expected to see of the order of a hundred streams out to the virial radius of the MW halo \cite{DES:2018imd}. Moreover, spectroscopic programs like the S$^5$ survey and instruments like the Maunakea Spectroscopic Explorer telescope will allow us to follow up on some of the streams with spectroscopic data to observe the disruptions of the streams in velocity space \cite{Li:2019nud,li_s5_2019,li_s5_2022}.

\subsection{X-ray and weak lensing observations of clusters, groups and large ellipticals}
\label{sec:Xray_WL}

The overall distribution of DM in halos can be probed in a variety of ways. While rotation curves of galaxies can be measured where spectroscopic data is available, the gas distribution in X-ray and gravitational lensing have been used to probe the overall distribution and density profile of the DM out to the viral radius for larger systems. Current and future generation surveys like the Dark Energy Survey (DES; \cite{Abbott_DES})\footnote{\url{http://www.darkenergysurvey.org}}, Hyper Suprime Camera (HSC; \cite{Aihara_HSC})\footnote{\url{http://hsc.mtk.nao.ac.jp/ssp/}} survey and LSST will provide precision measurements of the DM halo density profiles through weak lensing of galaxies in the back ground of large, statistical samples of cluster and group mass halos.

\begin{figure}[t]
\centering
\includegraphics[width=0.7\columnwidth]{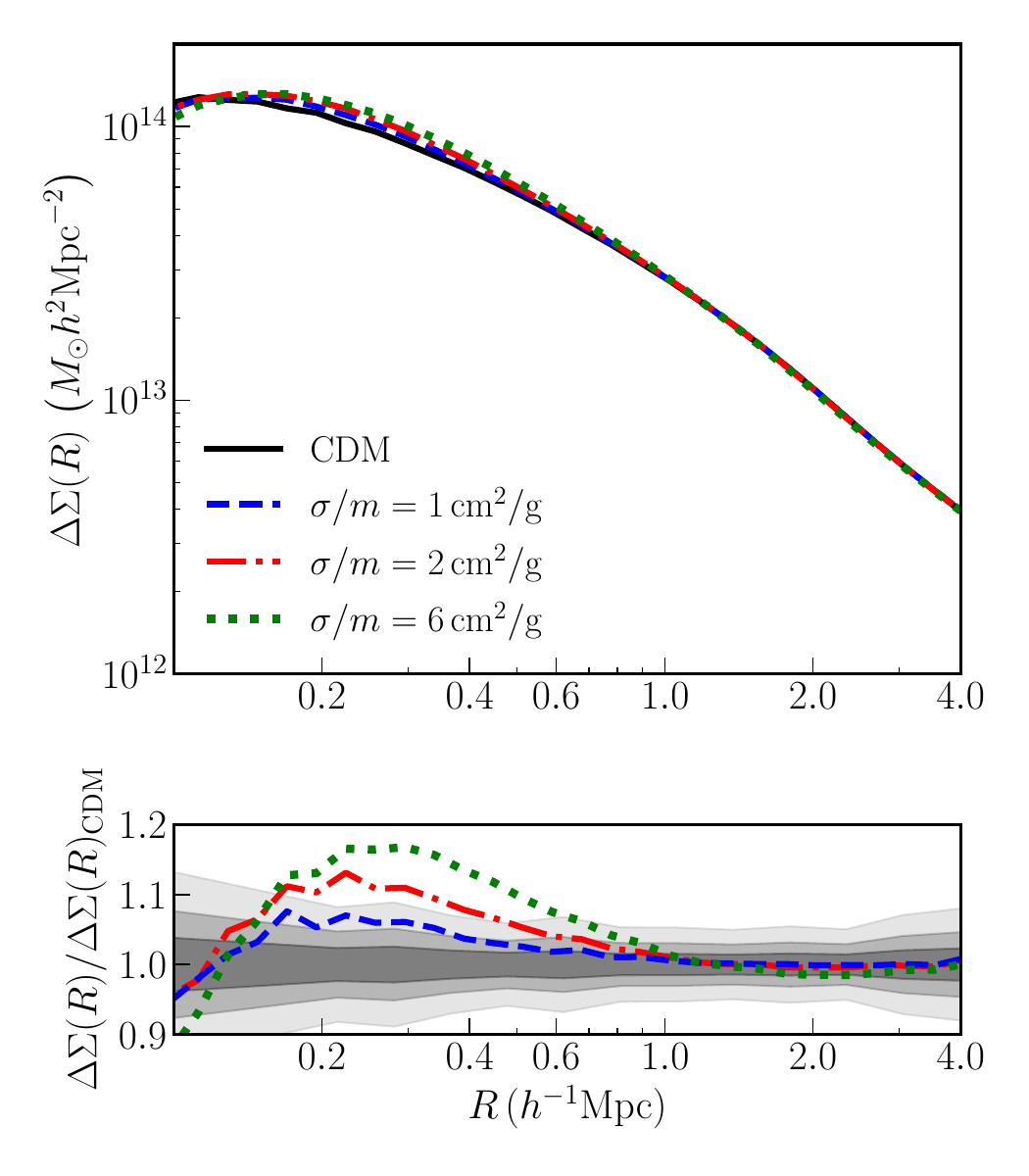}
\caption{The stacked lensing signal, $\Delta\Sigma$, of cluster mass DM halos from simulations. The scales plotted are just just beyond the scale radius of the DM halo. The density profiles steepen in this region. The lightest grey band is the current measurement error from DES Y1 scaled to Y3 and LSST sky coverage as they get darker respectively. (This plot has been recreated based on Ref. \cite{Banerjee:2019bjp} with additional estimates for LSST). } \label{fig:cluster_lensing}
\end{figure}

While the formation of a core is the most dramatic effect on the density profile of DM halos in the presence of self-interactions, mass conservation implies that the density at distances larger than the core becomes steeper than the density profile in the CDM scenario. While the latter is a relatively smaller effect, the scale at which this feature is expected, especially for clusters, is less affected by baryonic physics than the core. Further, the projected density profiles of clusters on these scales can be measured accurately through weak lensing measurements. Fig. \ref{fig:cluster_lensing} from \cite{Banerjee:2019bjp} shows the stacked projected density profile, $\Delta \Sigma$ of cluster/group mass objects ($M>10^{14} h^{-1} M_\odot$, $v\sim 1000$ km s$^{-1}$) from $N$-body simulations of SIDM with different cross sections. The red curve in the figure corresponds to the current best constraints on the cross section from the Bullet Cluster \cite{Robertson17BC} of $2$ cm$^2/$g (see Sec.~\ref{sec:mergers}). The bottom panel shows the ratio of the expected weak lensing signal for various SIDM cross sections and CDM. The lightest grey bands show the measurement error bars for a similar sample of objects from the DES Year 1 data. The progressively darker bands show the projected error bars from DES Year 3 and LSST, taking into account their deeper coverage. This observable, therefore, can potentially provide much stronger constraints on the SIDM cross section at cluster and group scales, especially once data from LSST becomes available.

Apart from the overall density profile, as mentioned previously DM self-interactions can alter the shape of the DM halo, making the regions of higher density more spherically symmetric. Constraints on scattering cross sections from shapes of halos have been obtained using the X-ray measurements of the shapes of the hot halo gas of NGC720. The shape and the twisting of X-ray isophotes carry information about the three dimensional shape of the potential. Measurements from NGC720 constrain the halo ellipticity to be $\sim 0.35\pm 0.03$, consistent with elliptical halos from CDM, this values is consistent with $\sigma/m_\text{DM} < 1$ cm$^2$/g \cite{Peter:2012jh}. An analysis of 12 clusters using X-ray emission and the Sunyaev-Zeldovich effect fitted Einasto profiles and calibrated with hydrodynamical SIDM simulations to achieve the bound $\sigma/m_\text{DM} < 0.19$ cm$^2$/g at a collision velocity $\sim$1000 km/s \cite{Ettori_Einasto}.

While the shape of DM halos was previously thought to be a promising probe of the nature of DM, subsequent studies have found that introducing baryonic physics can significantly complicate the inferences of cross sections from halo shapes \cite{Robertson:2018anx, 2022MNRAS.516.4543D}. \cite{2022MNRAS.516.4543D} find that the distribution of shapes in SIDM and CDM in the inner parts of clusters become much more similar in the presence of baryons. {The Milky Way is measured to be quite spherical with axis ratios, $c/a\sim 1$ (\cite{2016ApJ...833...31B}) at intermediate length scales of about $20$ kpc, in favor of SIDM models. However \cite{Sameie:2018chj} et al. show using the NIHAO hydrodynamical simulations \cite{2016MNRAS.462..663B} that CDM halos in the presence of baryons can also be produce more spherical objects on similar length scales making the observations consistent with CDM.} Additionally \cite{Sameie:2018chj} also show that the asymmetric potentials from central disks can make the shape of the DM halos aspherical near the center on short timescales of a few Gyr depending on the size of the disk. Note also that on cluster mass scales, while the DM can remain spherical near the center for cross sections of $\sigma/m_\text{DM}<1$ cm$^2/$g, as shown in Fig. \ref{fig:shape} from \cite{Robertson:2018anx}, the stars and the gas within the halo remain aspherical for a longer time and their shapes are much slower to respond to the change in the DM potential.

Weak lensing measurements of shapes of DM halos, stacked on the galaxy light profiles or using three point functions can measure the shape of dark halo directly as a function of radius. \cite{vanUitert:2016guv, Clampitt:2015wea, Shin:2017rch} measure shapes of Luminous Red Galaxies and cluster mass halos using weak lensing and distribution of satellite galaxies and find that the ellipticities are consistent with CDM, \cite{Shin:2017rch} for example place strong constraints on the axis ratios of RedMaPPer clusters in SDSS at $0.573\pm 0.039$, implying that the halo is overall quite aspherical.

\subsection{Major and minor mergers in groups and clusters (and associated observables)}
\label{sec:mergers}

Just like terrestrial particle accelerators allow us to study the interactions of fundamental particles, merging galaxy clusters can be thought of as ``cosmic accelerators'' probing the properties of DM. Indeed, the Bullet Cluster is arguably among the most sensitive and certainly the most famous probe of DM self-interactions~\cite{Clowe:2003tk}. What makes this system so unique is that the gas in the smaller of the two merging clusters exhibits a shock wave, which enables us to conclude with certainty that the system is being observed shortly after core passage. Combining this information with a lensing reconstruction of the system's DM profile~\cite{Bradac:2006er} enables us to infer the velocity and impact parameter of the collision~\cite{Springel:2007tu,Dawson:2012fx,Lage:2014yxa}. Typical estimates yield a velocity of around $v_\mathrm{c} \sim 4000\kms$ at core passsage and a projected (i.e.\ integrated) DM density of $\Sigma \sim 0.3\gcm$ along the path of the subcluster.

\begin{figure}[t]
\centering
\includegraphics[width=0.8\columnwidth]{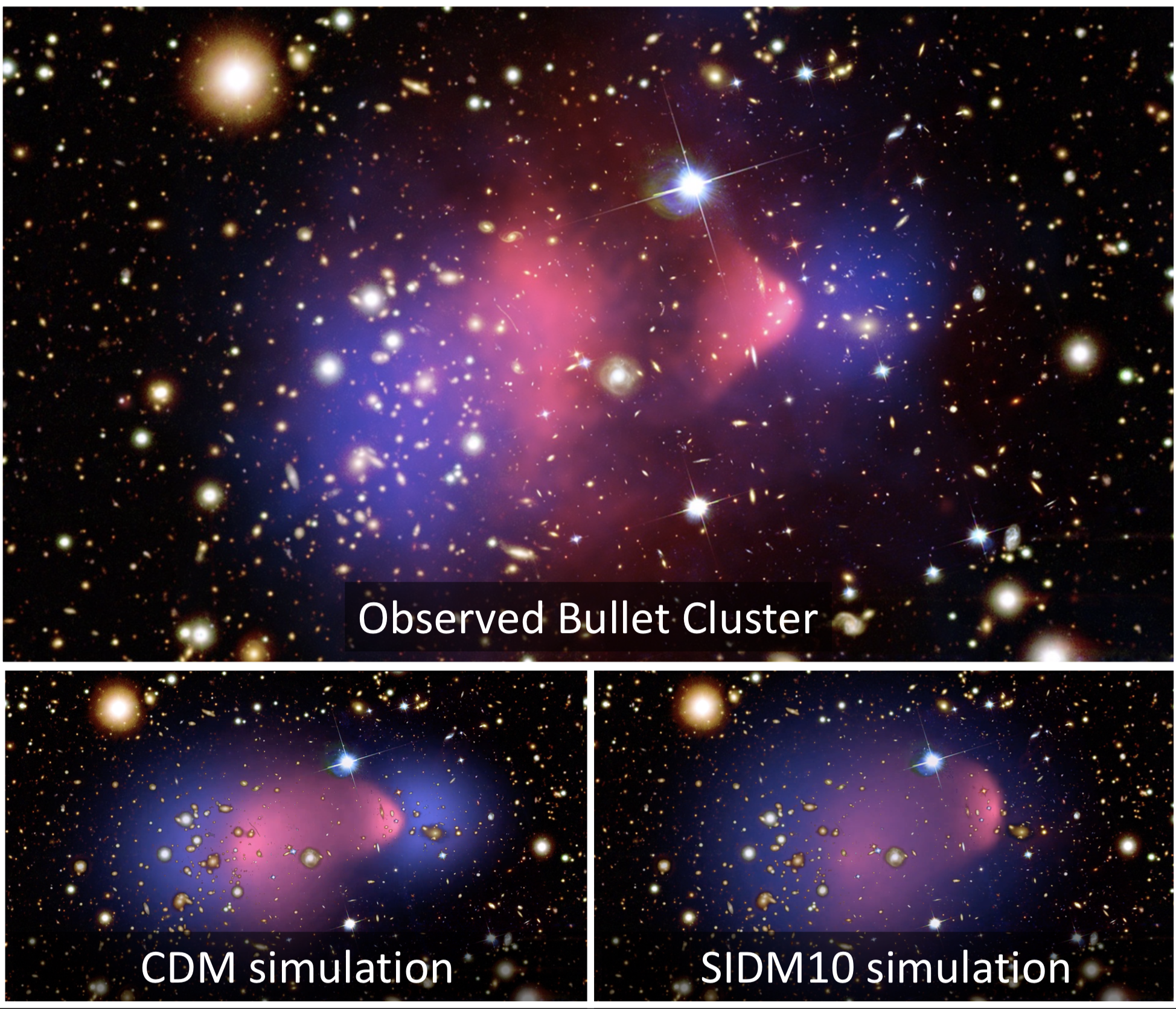}
\caption{Press release image of the Bullet Cluster (top) compared with simulations of a Bullet Cluster-like system \cite{Robertson:2016qef} with CDM (bottom-left) and SIDM with $\sigma/m_\text{DM} = 10 \, \rm{cm^2 \, g^{-1}}$ (bottom-right). Blue shows a map of the DM (inferred from gravitational lensing in the observed case), while pink shows the X-ray luminosity. The observed optical image has been overlaid onto the simulation images. The simulation images were made from the simulation snapshots where the separation between the two peaks in the galaxy distribution matched that in the observed case. Press-release image credit: X-ray: NASA/CXC/CfA/\cite{2006ESASP.604..723M}; Optical: NASA/STScI; Magellan/U.Arizona/\cite{2006ApJ...648L.109C}; Lensing Map: NASA/STScI; ESO WFI; Magellan/U.Arizona/\cite{2006ApJ...648L.109C}. Videos of the simulations evolving are available at \url{https://www.youtube.com/watch?v=rLx_TXhTXbs} and \url{https://www.youtube.com/watch?v=stvkDJVHPuQ}.} \label{fig:bullet_cluster}
\end{figure}

While the frequent claim that the Bullet Cluster demonstrates DM to be collisionless is clearly exaggerated,  
it is true that it places strong bounds on the DM self-interaction cross section. Given the projected DM density of the main cluster halo along the path of the subcluster, the main halo is expected to become optically thick for DM self-interactions with $\sigma/m_\text{DM} \gtrsim 3 \, \rm{cm^2 \, g^{-1}}$. This means that cross sections larger than this will lead to almost complete destruction of the subcluster's DM halo, as can be seen in Fig.~\ref{fig:bullet_cluster} for the case of a simulated Bullet Cluster with $\sigma/m_\text{DM} = 10 \, \rm{cm^2 \, g^{-1}}$.

The most stringent SIDM constraint from the Bullet Cluster comes from the measurement of the mass-to-light ratio ($M/L$) within $150 \kpc$ of the subcluster centre, which is found to be consistent both with that of the main cluster and with the typical range of $M/L$ ratios for clusters~\cite{Clowe:2003tk}. Ref.~\cite{Markevitch:2003at} argues that this agreement places an upper bound on the integrated mass loss by the subcluster and that it cannot have lost more than 20--30 percent of its mass within the $150 \kpc$ aperture. The resulting bound on isotropic, hard-sphere scattering is approximately $\sigma/m_\text{DM} \lesssim 0.7 \cmg$. It should be noted that this result is subject to some caveats, most notably that there is significant intrinsic scatter in $M/L$ of isolated clusters. Ref.~\cite{2007A&A...464..451P} find a 55\% scatter in $M/L$ within the virial radius of clusters, which would presumably increase if only looking within a small aperture. The consistency in $M/L$ between the main cluster and subcluster therefore does not exclude the subcluster having lost a significant fraction of its pre-collision DM mass, if it started with high $M/L$.

An independent constraint arises from the fact that the peak of the subcluster's projected density inferred from weak lensing measurements coincides with the galaxy centroid from visible light to within $1\sigma$. This observation rules out the presence of a substantial drag force, which would lead to the deceleration of the DM halo and hence a separation between DM and galaxies~\cite{Markevitch:2003at}. The observational results is that the DM is lagging behind the galaxies by $25 \pm 29 \kpc$ \cite{2006ApJ...648L.109C,Randall:2007ph}. The first numerical simulations of the system with SIDM translated this observation into the constraint $\sigma/m_\text{DM} < 1.25 \, \cmg$~\cite{Randall:2007ph}. However, more recent work has found this constraint to be over-stated, with a cross section of this magnitude producing offsets of only $10-15 \kpc$ \cite{Kahlhoefer:2013dca,Robertson17BC}. This difference is driven by the methods used to find the ``positions'' of the subcluster's DM and galaxy components, with the latter papers using methods more closely related to what was done observationally.

Despite the discovery of numerous merging galaxy clusters (and groups \cite{bullet_group}) in recent years, the Bullet Cluster remains the single most sensitive system, owing mostly to the superior projected density of the main cluster. Progress can potentially be made by the statistical combination of a large number of major and minor mergers, with the first attempt at this producing an upper limit on the SIDM cross section of $\sigma/m_\text{DM} < 0.47 \, \cmg$~\cite{Harvey:2015hha}. However, both the analytical model used to relate a DM--galaxy offset (strictly, the ratio of the DM--galaxy offset to the gas--galaxy offset) to a cross section, as well as the quality of the lensing maps used to infer DM positions, have been called into question~\cite{Wittman:2017gxn}. It remains to be seen whether further discoveries of merging groups and clusters can improve on the bounds obtained from the Bullet Cluster.

One system that has received particular interest in recent years is the Abell 3827 cluster. The advantage of this system is that strong gravitational lensing allows for a very precise reconstruction of the DM distribution. It has thus been possible to identify four galaxies within the core of the main cluster that all retain their own DM halos~\cite{Williams:2011pm,Mohammed:2014iya}. Intriguingly, early studies found a substantial separation between at least one of these DM halos and the associated stars~\cite{Massey:2015dkw}, which has been interpreted as positive evidence for SIDM~\cite{Kahlhoefer:2015vua}. Unfortunately, more detailed observations have not confirmed this separation and seem to imply that Abell 3827 is fully consistent with the predictions for collisionless DM~\cite{Massey:2017cwf}, in which the peaks in stellar and DM density should be coincident \cite{2015MNRAS.453L..58S}.

Another controversial system is the `Train Wreck Cluster' Abell 520, which has been claimed to exhibit a dark core, i.e.\ a region of \emph{enhanced} $M/L$ ratio close to the collision point~\cite{Mahdavi:2007yp,Jee:2012sr} (see however Ref.~\cite{Clowe:2012am} for a contradicting claim). While the most plausible explanation for such a dark core may be a projection effect due to the presence of filaments~\cite{Girardi:2008jp}, an interpretation in terms of SIDM is also conceivable. It is, however, unclear how elastic scattering of DM particles could lead to the formation of such a dark core, since the scattered DM particles would typically have too much kinetic energy to become gravitationally bound to each other. While it may be possible in principle to accommodate such an observation in more complex SIDM models (as discussed in Section~\ref{sec:extensions}), it remains challenging to bring the required self-interaction cross section into agreement with the bounds from the Bullet Cluster.

Finally, we mention the Musket Ball Cluster, which owes its name to the fact that it is both older and slower than the Bullet Cluster~\cite{Dawson:2011kf}. This system was initially considered to be particularly sensitive to DM self-interactions, due to the fact that it is much further progressed than other merging clusters. More detailed studies have however revealed, that it is typically not advantageous to observe a system long after core passage, because the DM halos quickly relax to their equilibrium position, so that the separation between DM and stars decreases rather than increases with time~\cite{Kahlhoefer:2013dca,Kim:2016ujt,Robertson:2016qef}.

To summarize, systems of merging clusters with the following properties would make them ideal candidates for constraining DM self-interactions:

\begin{enumerate}
 \item The integrated density of DM along the axis of merger for the system should be as large as possible, so that self-interactions have a larger effect.
 \item The system should allow for a precise reconstruction of the DM profile via (strong) gravitational lensing so that any offset between the light profile and the center-of-mass of the DM can be accurately measured.
 \item The system should be observed as shortly after core passage as possible, to prevent the DM particles from relaxing back to their equilibirum positions.
 \item The collision should be observed face on to minimize projection effects.

\end{enumerate}
Even if no single system is discovered that surpasses the Bullet Cluster in terms of sensitivity to SIDM, progress can still be made through the statistical combination of many colliding galaxy clusters. Expanding the catalogue of such systems is therefore a highly promising avenue in the search for SIDM.

\subsection{Incidence of multiple BCGs, their stellar kinematics and spatial separations}
\label{sec:BCG}

In Sec.~\ref{sec:dynamicalfriction}, we discussed the counterintuitive result that galaxies do not virialize if bound to the core of a cored DM halo.  In simulations of equal-mass mergers, Ref. \cite{Kim:2016ujt} found that such galaxies oscillate about the center of the halo with an amplitude equivalent to the core radius.  The easiest galaxies to definitively associate with cores are the biggest, the BCGs.  If the models of Ref. \cite{Kim:2016ujt} are correct and hold for more typical mergers where the masses of the two merging systems can be significantly different, then BCGs should be offset in both position and velocity from the center of a relaxed cluster. These studies from DM-only simulations were followed up in \cite{Harvey:2018uwf} to show that the effect of the BCG ``wobble" in a cored profile is present even in full hydrodynamic simulations that include star--formation and feedback. {In principle cores formed by baryonic feedback can also lead to BCG wobbles during merger. To study the combined effect of self-interactions and baryonic feedback, in a detailed study of offsets in simulations that include both SIDM and hydrodynamics, Ref. \cite{Harvey:2018uwf} show that the distribution of BCG offsets in equal mass mergers is biased high in the simulations that have both SIDM and baryons compared to those that have CDM only and baryons. The mean of the offset distribution increases as a function of the SIDM cross--section.}

\begin{figure}
\centering

\includegraphics[width=0.7\textwidth]{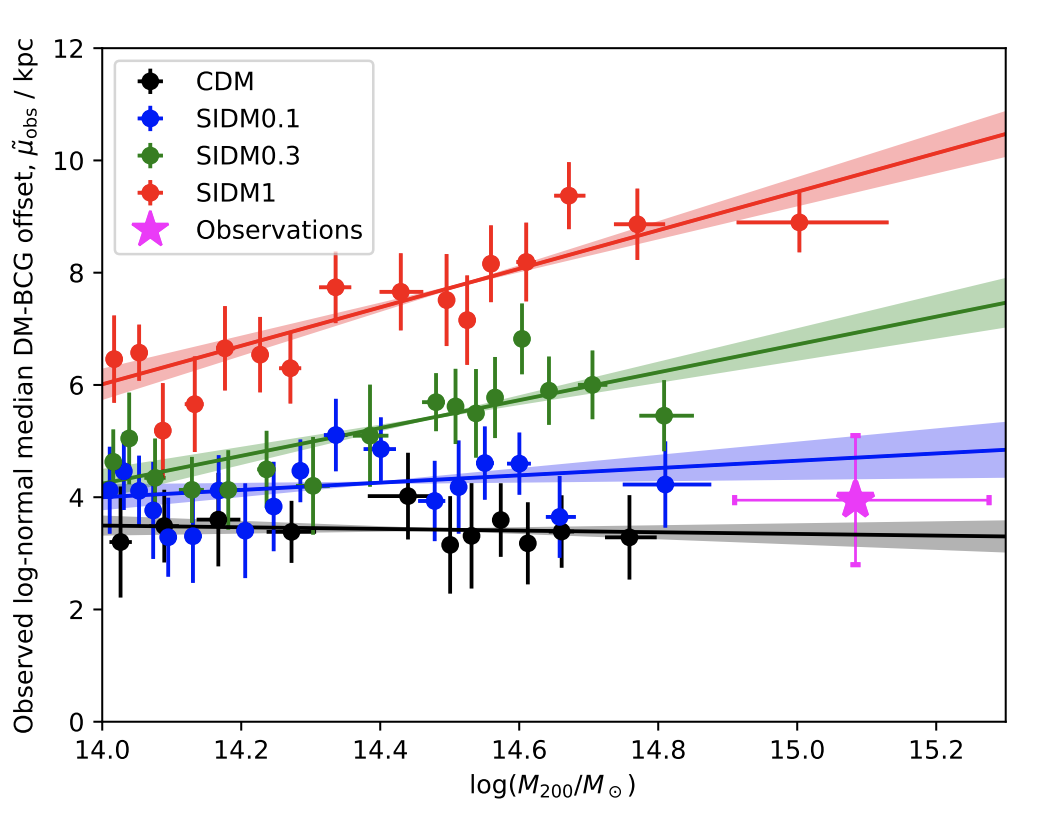}
\caption{Figure from \cite{Harvey:2018uwf} showing the evolution of the expected log-normal median DM-BCG offset in observations. The magenta symbol is the median BCG offset from the Local supercluster substructure survey measured in \cite{Harvey:2017afv}.}
\label{fig:BCG-offset-obs}
\end{figure}

It is in this context that some key observational results about galaxy cluster centers become even more interesting.  Because measurement of the density profile of clusters from weak lensing depends critically in accurate determination of cluster centers, the weak lensing community is highly invested in identifying the best observational tracer of halo centers.  From numerous studies, the main finding is that the position of the central galaxy (usually, but not always, the BCG \cite{Skibba:2010ez,lauer2014,hikage2017}) is by far the least noisy and most accurate tracer of the cluster center \cite{George:2012xd,vonderLinden:2012kh,lauer2014}.  Ref. \cite{lauer2014} finds that the median offset between the BCG and the X-ray center of massive clusters is about 10 kpc.  If true, and if the equal-mass merger prediction of Ref. \cite{Kim:2016ujt} holds more generally, the implied constraint on the SIDM hard-sphere cross section is of order $0.1~$cm$^2$/g.  Ref. \cite{Harvey:2017afv} claim a detection of BCG offsets, which can be verified with the large and well-studied cluster samples from current wide-field optical surveys. The measurement from \cite{Harvey:2017afv} is presented in Fig. \ref{fig:BCG-offset-obs}, along with the simulation results for the offset from \cite{Harvey:2018uwf}. \red{The current constraints from the measurement of BCG wobbles in the cores of galaxy clusters is $\sigma/m_\text{DM} < 0.39 \ \rm cm^2/g$ at the 95\% C.L.~\cite{Harvey:2018uwf} and data from future surveys like Euclid are expected to further improve these tight bounds. }

There are several important lines of investigation that need to be pursued to turn this promising probe of SIDM into a robust observable. First, more simulations are required to determine how generic the galaxy sloshing in cores is as a function of cluster assembly history (e.g., Ref. \cite{Robertson:2017mgj}).
Second, it is important that the position and velocity offsets  be evaluated in the context of both CDM and SIDM theory. This requires realistic mock observations to be created from  simulations. Third, from the observational side, the position and velocity offsets for BCGs should be measured more precisely, and as a function of halo mass.  In the case of position offsets, enormous datasets are already in hand from various imaging surveys, on account of the great interest in clusters in the dark energy/wide-field survey community.  Fourth, these offset measurements should be performed not just for a single BCG, but for the handful of BCGs, to trace the merger history of the cluster.  In the extreme case of an equal-mass merger with each component initially seeded with its own BCG, we expect two sloshing BCGs in the merger remnant sitting on opposite sides of the cluster center.

\subsection{Dwarf galaxies in the Local Group and beyond}
\label{sec:dwarfs}

Dwarf galaxies have long been targets of dark matter studies, especially the dwarf Spheroidal (dSph) satellites of the MW (see \cite{Bullock:2017xww,Sales:2022ich} for reviews).
While dwarf galaxies exist in all environments, observational considerations have led to a focus on \emph{satellite} dwarf galaxies.  For external systems, satellite systems yield a high return-on-investment for the deep imaging required to find faint dwarf galaxies.  Several surveys are now able to detect satellites in MW-like galaxies in the Local Volume and this number will grow in the coming years, especially in the era of deep wide-field imaging with next-generation ground- and space-based facilities \cite{danieli2017,Geha:2017iqy,smercina2017,crnojevic2019,Drlica-Wagner:2019xan,Carlin:2020qxv,habas2020,Mao:2020rga,davis2021,Garling:2021yru,mutlu-pakdil2021,carlsten2022full,sand2022,zaritsky2022}.  However the focus on dwarf
satellite galaxies observed around the MW and Andromeda \cite{Mateo1998,McConnachie2012}, and especially the MW, is driven by the wealth of data that are available for these systems that are difficult-to-impossible to obtain for more distant systems.  In this section, we discuss SIDM constraints with the MW satellite system specifically, because that is where most of the work has been performed, before broadening out to consider the potential of extragalactic dwarf satellite systems.

\begin{figure*}
\centering
\hspace{-1cm}
\includegraphics[height=6.0cm]{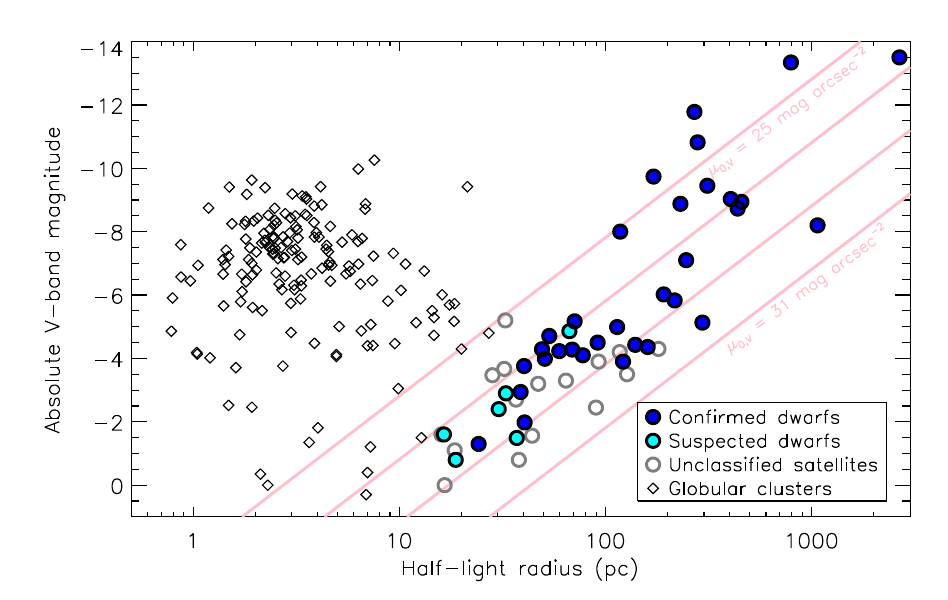}
\hspace{.1cm}
\includegraphics[height=5.8cm]{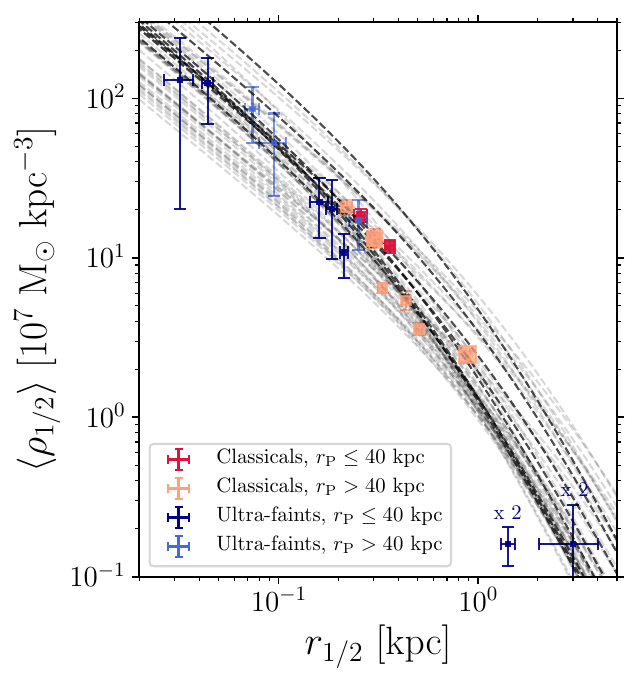}
 \caption{Plot on the left from Ref.~\cite{Simon2019} shows the MW dSphs in the V-band magnitude vs. half-light radius plane showing how deeper surveys have led to more dwarfs being uncovered and how the dSphs properties compare against MW globular clusters. Plot on the right from Ref.~\cite{Kaplinghat:2019svz} shows the inferred total mass density (almost all DM) within the half-light radius and the strong anti-correlation between the mass density and half-light radius and the large diversity in the properties of these dSphs. In the background are subhalo mass profiles from a MW halo ({\em Kauket} from ELVIS)~\cite{Garrison-Kimmel:2013eoa}.}
\label{fig:dsphs}
\end{figure*}

There are about 60 known satellite galaxies of the MW \cite{Simon2019}.  The dSph satellites (i.e., all MW satellites but the LMC and SMC) span an enormous range in luminosity, stellar mass, and size, from ultra-faint dwarfs such as Segue 1 with stellar mass of ${\cal O}(10^2) M_\odot$ and half-light radius of $30 \rm~pc$ to Fornax with about $10^7 M_\odot$ of stellar mass within about a kpc. The spread of satellite luminosities (expressed as V-band absolute magnitudes) and size (expressed in terms of stellar half-light radii) is shown in the left panel of Fig.~\ref{fig:dsphs}.  Many of the satellites have been targeted for follow-up spectroscopic observations, which lead to measurements of satellite internal stellar kinematics, systemic line-of-sight velocities, and chemical abundances from the individual stellar spectra.  Combined with proper motion data from the Gaia satellite, it is possible to infer 3D satellite orbits in the MW, including the pericenter distances that are important for assessing the tidal states of the satellites.  Notably, 3D satellite orbits, and spectroscopic star-by-star line-of-sight and chemical abundance measurements are currently only possible to obtain for Local Group galaxies with current current facilities.

From a theorist's perspective, these observables map onto properties of dark matter halos that may be affected by SIDM physics.  1.  The satellite luminosity function can be mapped to a subhalo mass function if the stellar-mass--halo-mass is well understood (and for which, in practice, there is significant uncertainty on these small scales, see, e.g., \cite{Munshi:2017czb}).  2.  The stellar kinematics can be tied to the central density of dark matter halos and their inner density profiles.  The dynamical masses of these dSphs are significantly in excess of their stellar masses, making them some of the most dark-matter-dominated systems observed. A simple way to estimate the dynamical masses of the dSphs is to use the results of Refs. \cite{Walker:2009zp, Wolf:2009tu}, who argue that the mass within the half-light radius can be related to the observed stellar dispersion for a wide range of halo profiles and anisotropies (see also \cite{Errani2018}).
In the right panel of Fig.~\ref{fig:dsphs}, densities within the half-light radius estimated in this way are shown revealing a striking anti-correlation of the density with  half-light radius, hinting at a density profile \cite{Kaplinghat:2019svz,Hayashi2020}.  Several attempts have been made to measure the density profiles of individual satellites' dark matter halos, but this is challenge on account of the paucity of stars in many galaxies, the degeneracy between orbital anisotropies and enclosed mass, the lack of internal proper motion measurements, and assumptions regarding equilibrium \cite{Kleyna:2003zt,Battaglia:2008jz, Amorisco:2011hb, Walker:2011zu,Breddels:2013dga,Strigari2017,Read2019,Read:2018pft,2021MNRAS.501..978R,Hayashi2020,Nguyen:2022ldb,Andrade2024}.  What is clear, though, is that the slopes of the density profiles is diverse, even if the uncertainties on individual measurements is large in many cases.  Some dwarfs (e.g., Draco) indicate they are in cuspy halos, while others (e.g., Fornax) indicate that their central density profiles are core.  As described in Introduction, the too-big-too-fail problem may be a manifestation of either an issue with subhalo mass functions or with the central densities of halos \cite{TBTF, BoylanKolchin:2011dk}, although we will treat it as a central density problem below.  

We discuss the physics of how SIDM affects the theory-space observables before addressing the current state of constraints and future directions.  The presence of elastic self-interactions changes the evolution of a satellite within the host halo relative to CDM in many different ways as listed below and discussed in more detail in Refs.~\cite{Kummer:2017bhr,Jiang:2022aqw}.
\begin{itemize}
    \item Dark ram-pressure stripping: the high velocity interactions of the subhalo particles with the host halo particles can lead to mass being unbound from the subhalo, accelerating mass loss (discussed in more detail in Sec.~\ref{sec:drag}).
    \item Deceleration of the subhalo due to self-interactions and enhanced dynamical friction because of orbital decay (although dynamical friction is slowed once the satellite decays into the core region of the host; Sec.~\ref{sec:dynamicalfriction} \& Sec.~\ref{sec:drag}).
    \item Thermal evolution, including core expansion (Sec.~\ref{subsect:SIDM_density_profile}) and collapse (Sec.~\ref{sec:core_collapse}), of the core impacted by the mass removal by tides (Sec.~\ref{sec:enhanced_stripping_core}).
    \item Response of the DM and stellar density profiles to the removal of mass by tidal stripping and ram-pressure stripping (Sec.~\ref{sec:enhanced_stripping_core}).
\end{itemize}
Interestingly, the different aspects of the orbital and thermal evolution of a subhalo are sensitive to the velocity dependence of the self-interaction cross section
\cite{Nadler2020SIDM,Jiang:2021foz,Zeng:2021ldo} and the presence of the disk~\cite{Correa2021,Silverman:2022bhs}. The thermalization of the subhalo is controlled by the cross section at relative velocities of about $4 V_{\rm max,\ subhalo}/\sqrt{\pi}$, while the ram-pressure stripping is controlled by the cross section at relative velocities of order $V_{\rm max,\ main\ halo}$\cite{Dooley:2016ajo, Penarrubia:2010jk, Nadler2020SIDM}. The gravothermal evolution of the SIDM subhalos is complicated by the fact that the subhalos' interactions drive the thermal evolution in the presence of mass loss and structural changes due to the tidal forces of the host halo\cite{nishikawa2019, Zeng:2021ldo}. This can lead to a wide diversity of subhalo central densities and density profiles \cite{Zeng:2023fnj}.  Moreover it also matters if the subhalos have fallen in as part of a larger structure like the LMC \cite{Nadler:2021rpo}.

In practice, the biggest effect of these physics is on the density profiles of the subhalos.  While SIDM can potentially alter the subhalo mass function, we argued in Sec.~\ref{sec:dynamicalfriction} that this is a minor effect unless the cross section is higher than allowed by observational constraints \cite{Turner:2020vlf}.  The orbital and radial distribution of satellites can be affected relative to CDM \cite{Robles:2019mfq}, but the full effect has yet to be systematically explored.  Thus, we focus on measurements of dwarf halo density profiles.

Observationally, the strongest constraints on SIDM from the MW satellite system arise from the central dark matter densities and halo profiles of the MW satellites.  The central densities cannot be explained by a constant cross section SIDM model---any cross section results in a central density that matches a typical classical dwarf galaxy (one with relatively recent star formation), assuming a reasonable stellar-to-halo mass relation, leads to central densities that are much too low for the low-mass ultrafaint dwarf galaxies \cite{Errani2018,Zavala:2019sjk,Kim:2021zzw,Silverman:2022bhs}.  By modeling observed internal kinematics and light profiles of individual satellites through the lens of SIDM, one may map out the velocity dependence of SIDM.  Using an analysis for the core-formation regime, Refs.~\cite{Hayashi:2020syu,Ebisu:2021bjh} find that a small handful of ultrafaint dwarfs prefer a cross section of much less than $10^{-1}$ cm$^2$ g$^{-1}$, even if some more massive dwarfs can be accommodated with cross sections an order of magnitude larger.  When core collapse is considered, Ref. \cite{Correa2021} finds that classical dwarf galaxies' central densities can be modeled with a velocity-dependent cross section of order $100$ cm$^2$ g$^{-1}$. Other work also suggests that classical dwarfs may show signs of core collapse \cite{Valli2018,Read:2018pft,Zavala:2019sjk,Kaplinghat:2019svz}.   Simulations show that the central densities of satellites may be very diverse if such large cross sections are allowed \cite{Turner:2020vlf,Jiang:2021foz,Zavala:2019sjk,Yang:2022mxl,Zeng:2023fnj}.

\begin{figure*}
\centering
\hspace{-1cm}
\includegraphics[width=6.8cm, height=6.5cm]{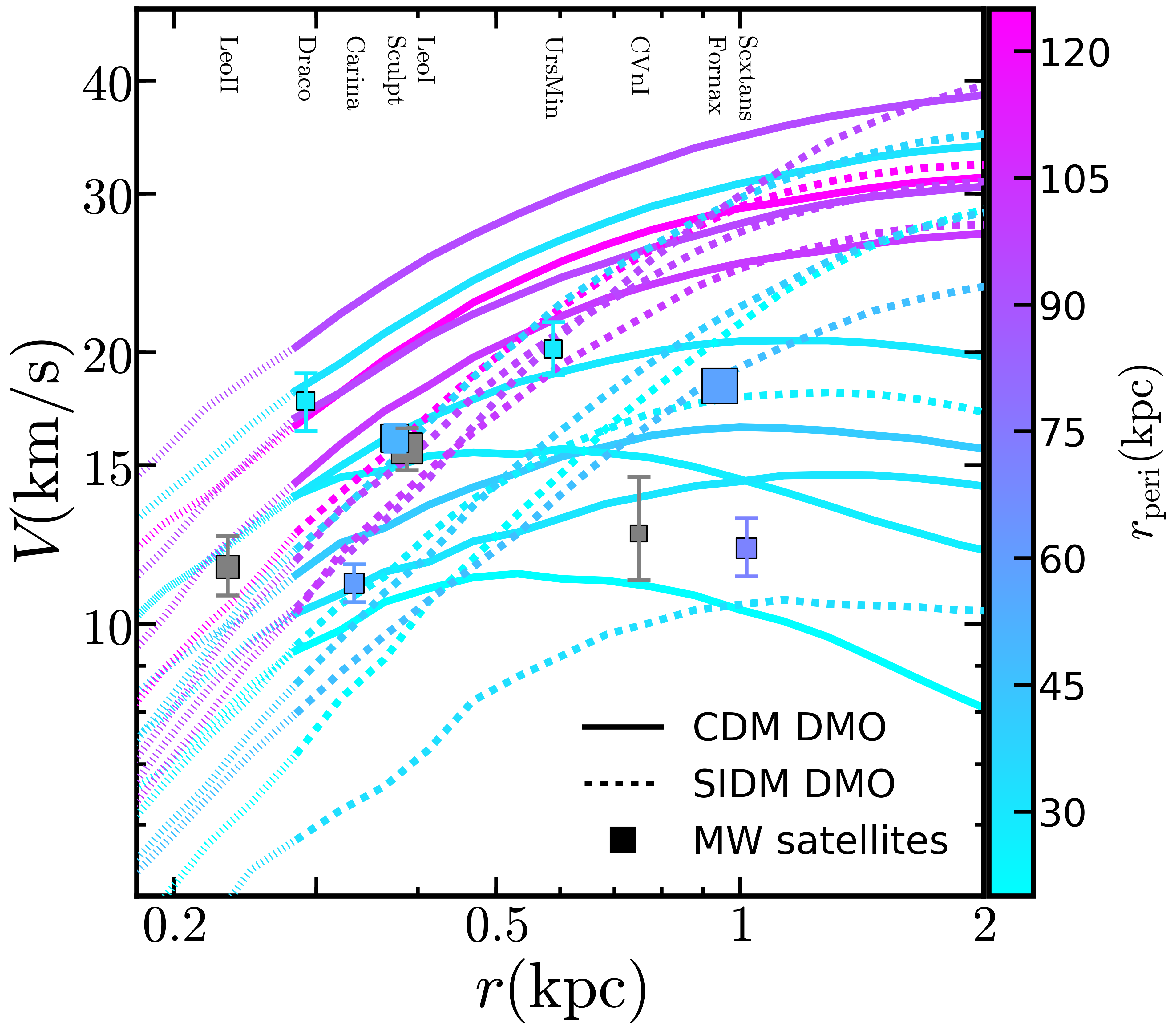}
\hspace{.2cm}
\includegraphics[width=6.8cm, height=6.5cm]{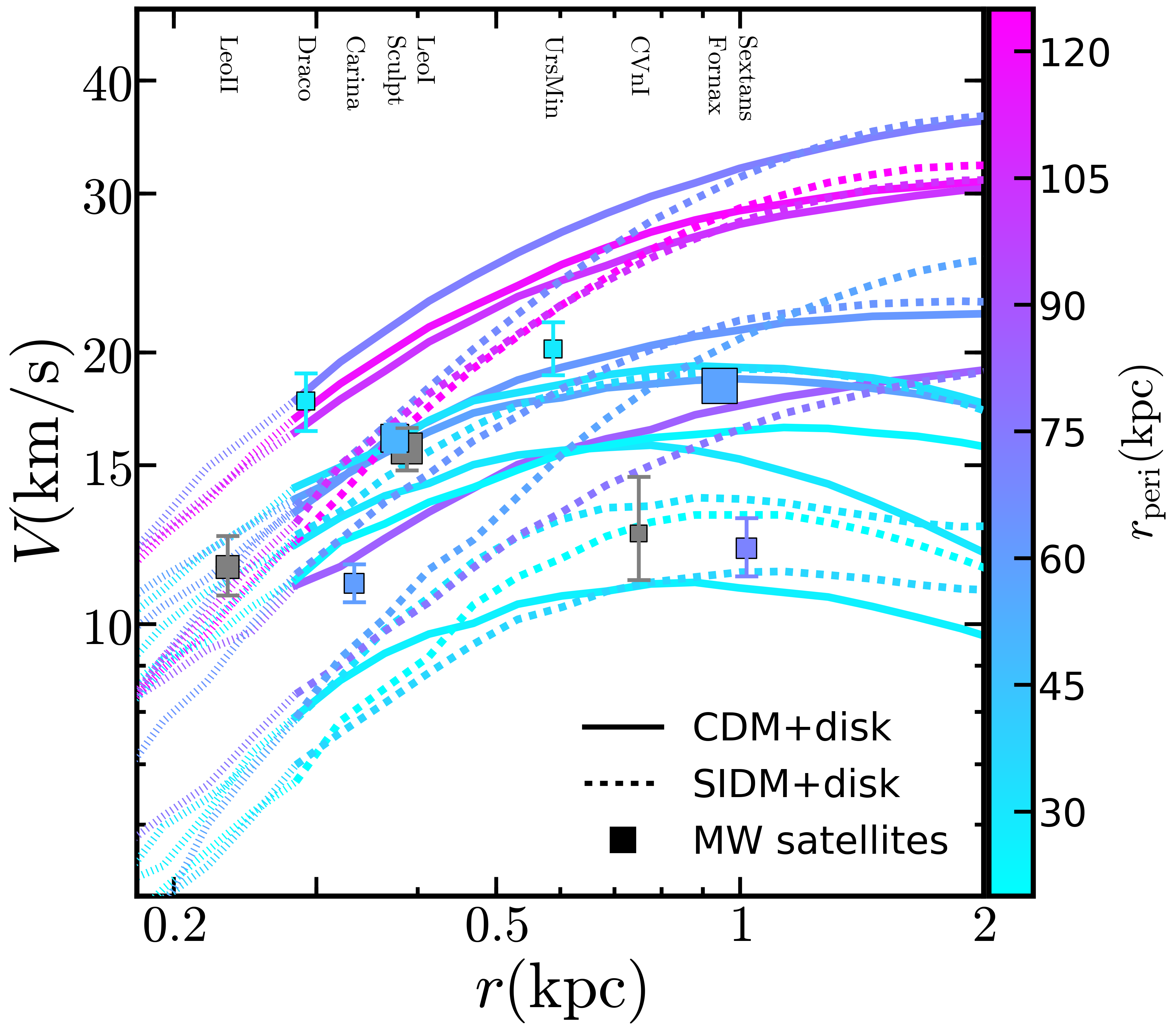}
 \caption{
 Circular velocity profiles for the ten subhalos with the largest peak velocity across time in the simulations with the disk and bulge potentials (right) and the DM-only simulations (left). Note that these are not necessarily the subhalos with the largest peak velocity today. The SIDM cross section over mass is set to $1 \rm cm^2/g$, \red{which implies that there is no possibility of the subhalos entering the core collapse phase}. The data for the classical MW satellites brighter than $2 \times 10^5  \, \mathrm{L}_\odot$ \cite{Wolf:2009tu} are denoted with filled squares with point sizes proportional to the logarithm of their stellar masses. The points and curves are color coded by the inferred pericenter distances; for the MW satellites the pericenter distances calculated in \cite{Fritz2018} using Gaia DR2 have been used. Gray symbols are satellites whose inferred pericenter distances have errors of the same order as their mean value, and are hence uninformative. The thin dashed (faded) lines indicate unresolved regions in the simulations.
The CDM halo has three subhalos as dense as Draco, while the $\sigma/m_\text{DM} =1 \, \rm cm^2/g$ halo does not have any. Note that for all models, with and without the disk, there is a trend that subhalos with smaller pericenters (cyan lines) have lower densities. The MW satellites do not seem to adhere to this trend. Figure adapted from Ref. \cite{Robles:2019mfq}.}
\label{fig:tbtf}
\end{figure*}

At the moment, most MW satellite SIDM constraints come from modeling individual satellites.  In practice, this means finding best-fit the halo mass, concentration, and SIDM cross sections for each individual galaxy, taking into account the orbital history of the object (see Refs. \cite{Correa2021,Sameie2020} for examples).  Future progress on SIDM constraints requires modeling the whole population of known MW satellites --- including the dense ultrafaint dwarf galaxies --- self-consistently within the framework of specific SIDM models.  This means simulating realizations of the MW system, with baryons.  As an example of the power of this approach, Fig.~\ref{fig:tbtf} shows circular velocity curves in simulations vs. classical dwarf data, considering models with with $\sigma/m_\text{DM} = 0~\rm{and}~1 \rm cm^2/g$, with and without a MW disk, and with pericenter distances labeled. The SIDM model with a disk fits the observed data better in this TBTF plot.  However, this plot does not include ultrafaint dwarf galaxies, which, if included, would dramatically alter our conclusions about how well $\sigma/m_\text{DM} = 1 \rm cm^2/g$ fits the satellite population.  The best constraints will only come when we are able to produce validated realizations of the whole MW satellite population, with a resolution to match the observable properties of the system.

SIDM constraints with MW satellites can additionally be strengthened through a variety of other observational and theory pathways.  On the observational side, new observational facilities will broaden the number of available satellites for measurements of central densities and grow the types of data sets that can be used to measure the slopes of density profiles.  Up to hundreds of additional satellites may be discovered with the Rubin Observatory \cite{hargis2014,Kim:2017iwr,Drlica-Wagner:2019xan,manwadkar2022}. Because those MW satellites are expected to be faint and distant, spectroscopic instruments need to be paired with large-aperture ($> 10$ m) telescopes in order to measure the internal kinematics of these new dwarfs \cite{Li:2019nud,simon2019BAAS,Bechtol:2022koa}.  These spectroscopic facilities will additionally enable more precise measurements of the central densities and slopes of known dwarfs, because they will enable us to go much farther down the luminosity function of stars for line-of-sight measurements \cite{Li:2019nud,simon2019BAAS}.  Many authors have shown that data sets of $\mathcal{O}(10^3-10^4)$ are needed to distinguish cores from cusps at all, but that measurements of the relative proper motions of stars within the galaxies are required to definitively break the degeneracy between the stellar velocity anisotropies and dark matter density profiles \cite{Strigari:2007vn,theia2017,Chang:2020rem,Guerra:2021ppq,read2021}.  Thus, while central density measurements should be achievable for all but the faintest future MW satellite discoveries, it is expected that measurements of the density profiles of the classical dwarf galaxies, and the brightest ultrafaint dwarf galaxies, should improve dramatically on timescales of about a decade.

\begin{figure*}
\centering
\includegraphics[width=14cm, height=12cm]{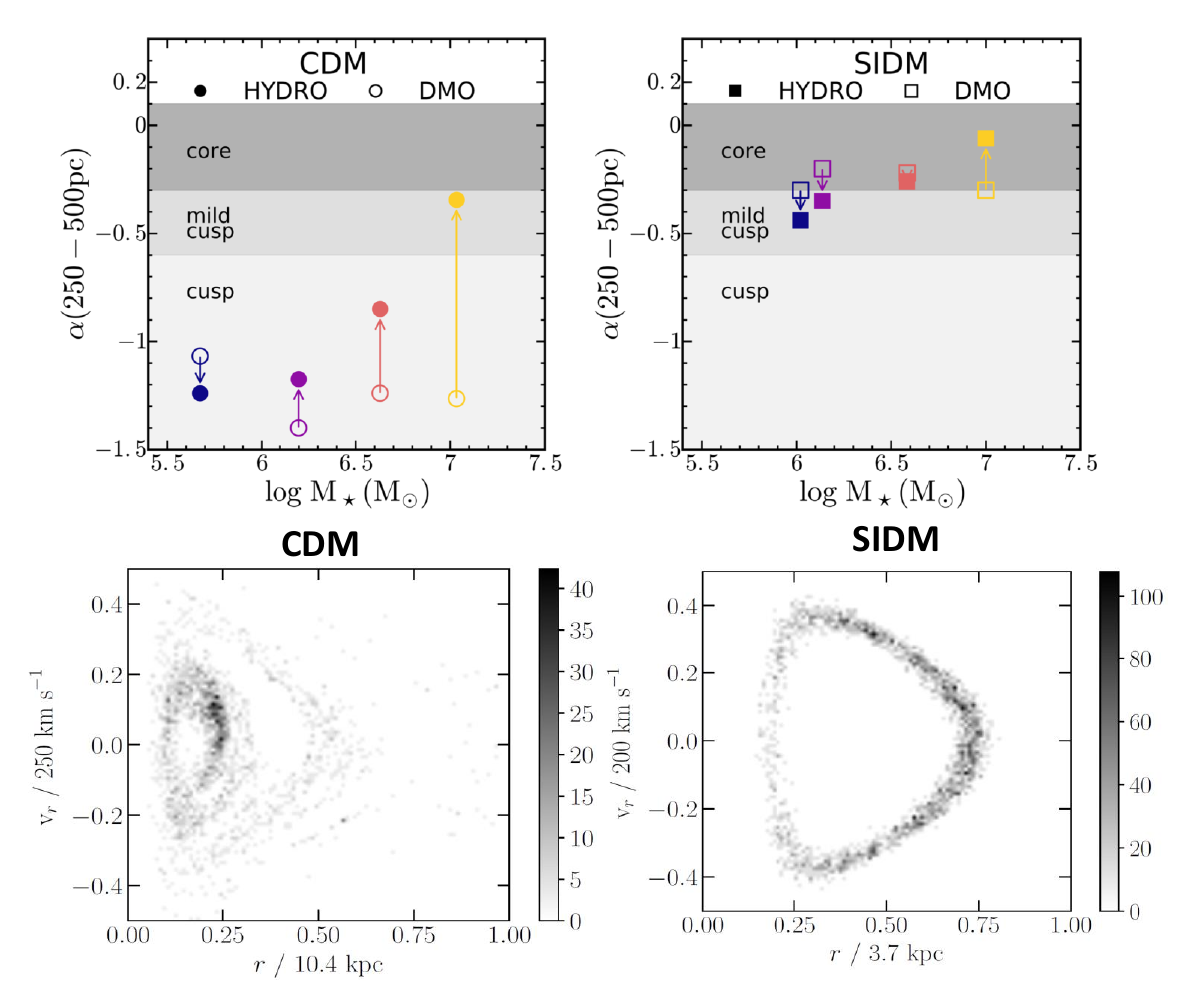}
 \caption{
 Two potential ways to disentangle CDM + supernova feedback from SIDM as mechanisms of cusp-core transformations in dwarf galaxies. The upper panels from Ref. \cite{Robles2017} show the slope $\alpha$ (within $0.5-1\%$ of the virial radius) of the halo density profile as a function of stellar mass for a full baryonic physics implementation (solid symbols; FIRE implementation, see Ref.~\cite{Hopkins2018}) and the corresponding DM-only simulations (open symbols). CDM in the upper left and SIDM ($\sigma/m_\text{DM}=1$ cm$^2/$g) in the upper right. \red{Note that this assumes that all the subhalos are in the core expansion phase, which seems to be disfavored currently (see text). For SIDM models where some of the subhalos are in the core collapse phase, we would expect greater diversity in the slopes, possibly spanning the whole range from slopes from 0 to -2. } For small stellar-to-halo mass ratios, the supernova feedback mechanism is (energetically) less efficient in producing sizeable cores, while the SIDM mechanism (constant cross section $\sigma/m_\text{DM} \lesssim1$ cm$^2/$g) remains efficient at all masses. The bottom panels are from the isolated controlled DM-only simulations of a dwarf-size halo from Ref.~\cite{Burger2019}. The halos in both panels result in a $\sim0.5$~kpc core after $t\sim4$~Gyr, with the simulation on the \add{left} having a toy supernova feedback model, while the one on the right has a standard SIDM implementation ($\sigma/m_\text{DM}=2$cm$^2/$g). The plots show the radial 2D phase space density of 2000 tracers originally set as an orbital family (narrow energy and angular momentum distribution) at the end of the simulation. Supernova feedback is an impulsive mechanism that irreversible perturbs the orbital family creating distinct shells in phase space, while the SIDM case is an adiabatic mechanism, which leaves the orbital family of traces largely unperturbed.}
\label{fig:cdm_vs_sidm}
\end{figure*}

On the theory side, a key avenue of progress is to identify observables in dwarf galaxies that are truly distinct between CDM and SIDM (see also Sec.~\ref{sec:unique}). The most pressing issue is to distinguish between dark-matter halo cores driven by baryons or by physics in the dark matter sector.  Cores formed due to DM self-interactions and due to supernova feedback have been shown to be widely degenerate (when it comes to the shape of the density profile, see Ref.\cite{Burger2021b}). However, the supernova-driven gas blowouts that are responsible for expanding the orbits of DM particles to create cores need to be impulsive (relative to the orbital timescales), while self-interactions are effectively an adiabatic process of energy redistribution leading to the formation of cores \cite{PG,Burger2019,Burger2021}. This intrinsically different nature leads to distinct signatures in the orbits of stars: shells in phase space are formed in the impulsive SN-feedback scenario, while they are not present in the SIDM case \cite{Burger2019}; see bottom panel of Fig.~\ref{fig:cdm_vs_sidm}. The presence (or absence) of such shells in the stellar distribution of dwarf galaxies with confirmed cores would strengthen (weaken) the supernova feedback scenario as the main mechanism of core formation.

Furthermore, to distinguish between SIDM and CDM, it will be important to determine the relative efficacy of baryons and dark matter in driving core formation as a function of halo mass.  The efficiency of supernova feedback as a cusp-core transformation mechanism is strongly dependent of the energy released relative to the change of potential energy in the halo from being cuspy to cored \cite{Read2005,Penarrubia:2012bb,Katz2018,Burger2021}. The efficiency peaks at around $M_h\sim10^{11}$M$_\odot$, just around the upper mass end of dwarf galaxies, and drops sharply towards higher and lower masses \cite{DiCintio2014,Chan2015}. For smaller dwarf galaxies, the stellar-to-halo mass ratio is expected to be too low to be efficient at forming large cores \cite{Robles2017} (although see Ref.\cite{Read2016}). On the other hand, in the SIDM scenario with $\sigma/m_\text{DM} \sim1$ cm$^2/$g, cores of a size comparable to the scale radius of the halo are expected across all masses \cite{Rocha13} \red{in the core expansion phase. In the core collapse phase, we would expect much greater diversity in the values of the slopes as discussed in this section. For the lowest concentration halos or halos with larger $V_{\rm max}$ where the cross section is small enough, we could have cored halos. For smaller mass halos or for halos with high concentrations or more tidally-stripped halos, one would expect slopes between 0 and -2 depending on the size of the core and the radius at which we are able to measure the slopes.} Unambiguously identifying and characterizing sizeable cores for stellar masses $\lesssim10^6$M$_\odot$ is a promising avenue to constrain physical mechanisms of core formation (see top panel of Fig.~\ref{fig:cdm_vs_sidm}). A key challenge for this program is to quantify how the various hydrodynamic recipes for star formation and feedback affect predictions on halo density profiles \cite{azartash2024}.

This line of theory work and new observational facilities also offer tests of SIDM in dwarf galaxies outside the MW.  It is important to measure the central densities (and slopes, if possible) of dwarf halos in different environments because the tidal effects in dense environments are expected to accelerate SIDM halo evolution relative to the field \cite{nishikawa2019,Yang:2022mxl,Zeng:2023fnj}.  With the proposed Maunakea Spectroscopic Explorer \cite{Li:2019nud} or 30-m-class telescopes \cite{simon2019BAAS}, it should be possible to measure central dark matter densities of dwarf galaxies within the Local Group.

We close with a potentially radically different approach to halo profile measurement: weak lensing.  Weak lensing plays an essential role in quantifying the galaxy-halo connection \cite{Wechsler:2018pic} and density profiles \cite{schulz2010} for massive galaxies.  Recently, new methods to obtain well-characterized dwarf-scale lens samples have been proposed to extend weak lensing to the dwarf regime \cite{thornton2023,luo2024}.  The first measurement, by \cite{thornton2023}, shows a galaxy-halo connection consistent with abundance-matching-based methods down to galaxies as small as the Small Magellanic Cloud.  It is expected that measurements will become sensitive to much smaller dwarf galaxies, and to deeper in the halos, in the Rubin, Roman, and DESI-II eras \cite{leauthaud2020PDU}.

\subsection{Rotation curves of spiral galaxies}
\label{sec:RCs}

\red{Problems associated with explaining the DM halo density profiles persist in higher mass galaxies, where they manifest themselves in several different but related forms. Unlike the situation with dwarf spheroidals where it has proven difficult to measure unambiguously the presence of constant density cores, the situation in rotation-supported galaxies is clearer. The  evidence for cores in many spiral galaxies is strong, particularly those of low surface brightness (e.g. \cite{deBlok_1, Donato,Naray_LSB,Relatores2019,mcquinn2022}). This evidence is sufficiently ubiquitous to lead some to suggest a universal cored density profile with a well-defined relation between core density and size \cite{Salucci_core, Karukes, Gentile}. NFW fits to the rotation curves of individual galaxies of a wide range of type often perform poorly or require concentrations significantly outside the $\Lambda$CDM mass--concentration relation scatter  (e.g. \cite{McGaugh_NFW, Katz}).}

\red{Fits to individual galaxies can suffer from systematic issues related to inclination angle estimation, non-circular motion and presence of bars~\cite{Rhee:2003vw,Valenzuela:2005dh,2016MNRAS.462.3628R,Oman:2017vkl}. The results discussed in the previous paragraph can be extended in a statistical sense across the galaxy population by employing  dynamical scaling relations between the masses, sizes and characteristic velocities of galaxies. Empirical methods for assigning galaxies to the $\Lambda$CDM halo population from $N$-body simulations (e.g. subhalo abundance matching \cite{Conroy, Moster}) generate a larger ratio of dark to luminous matter than observed across the rotation curves of galaxies. This is manifest in the fact that the $\Lambda$CDM halo mass function with NFW (or similar) profile normalises both the Tully--Fisher and Faber--Jackson relations too high \cite{McGaugh_RV, Desmond_TFR, Desmond_FJR}, and overestimates velocities in the low-acceleration outer regions of galaxies \cite{Desmond_MDAR}. This may be seen as extending the Too Big To Fail problem \cite{TBTF} to higher masses. There is evidence that the overprediction of dynamical mass in $\Lambda$CDM is in fact more general than implied solely by these relations \cite{Karachentsev}.  On the other hand, relaxing assumptions about the density profile of the dark-matter halos, allowing for cored halos, leads to better consistency between the halo masses preferred by rotation curves and those implied by abundance matching \cite{Katz:2018wao,Li:2019zvm,mcquinn2022}.}

\red{Several other aspects of galaxy phenomenology have proven challenging to $\Lambda$CDM and may hold clues to the nature of DM. Scaling relations like the baryonic Tully--Fisher \cite{McGaugh_BTFR} and mass discrepancy--acceleration \cite{McGaugh_MDAR1, McGaugh_MDAR2} (or radial acceleration; \cite{RAR}) relations and Fundamental Plane appear to have smaller scatter than the $\Lambda$CDM halo mass--concentration relation and galaxy--halo connection (e.g. $M_*-M_\text{vir}$), from which they should receive contributions \cite{Borriello, Desmond_BTFR, Desmond_MDAR}. The diversity of rotation curve shapes has also been seen as problematic for CDM and may point towards DM interactions \cite{Oman, Kamada, Creasey} or strong feedback~\cite{Santos-Santos}. However, recent work has argued that models with cuspy dark matter halos (i.e., $\Lambda$CDM halos without strong feedback) are likely to be more consistent with the observed rotation curves if the gas disks in late-type spirals are like those simulated in EAGLE simulations~\cite{Santos-Santos:2019vrw,roper2022}. The origin of this conclusion lies in the discrepancy between recovered and actual rotation curves caused by systematic effects related to non-circular motion, thickness of gas disk and departures from dynamical equilibrium in a systematic way that makes cuspy halos more likely to look like cored halos. If this conclusion holds true, this may be an indication that core collapse due to self interactions is more common in dwarf halos than discussed in the SIDM investigations thus far. }

\red{Other observations that are potentially surprising in the context of $\Lambda$CDM include spatially and kinematically correlated planes of satellites around the MW and nearby galaxies \cite{Pawlowski},
a dearth of DM in the central regions of high-mass ellipticals \cite{Romanowsky}, and the strong point-by-point correlation of kinematics and baryonic surface density in spirals \cite{Famaey, RAR}.}

\begin{figure}
\centering
\includegraphics[width=0.6\columnwidth]{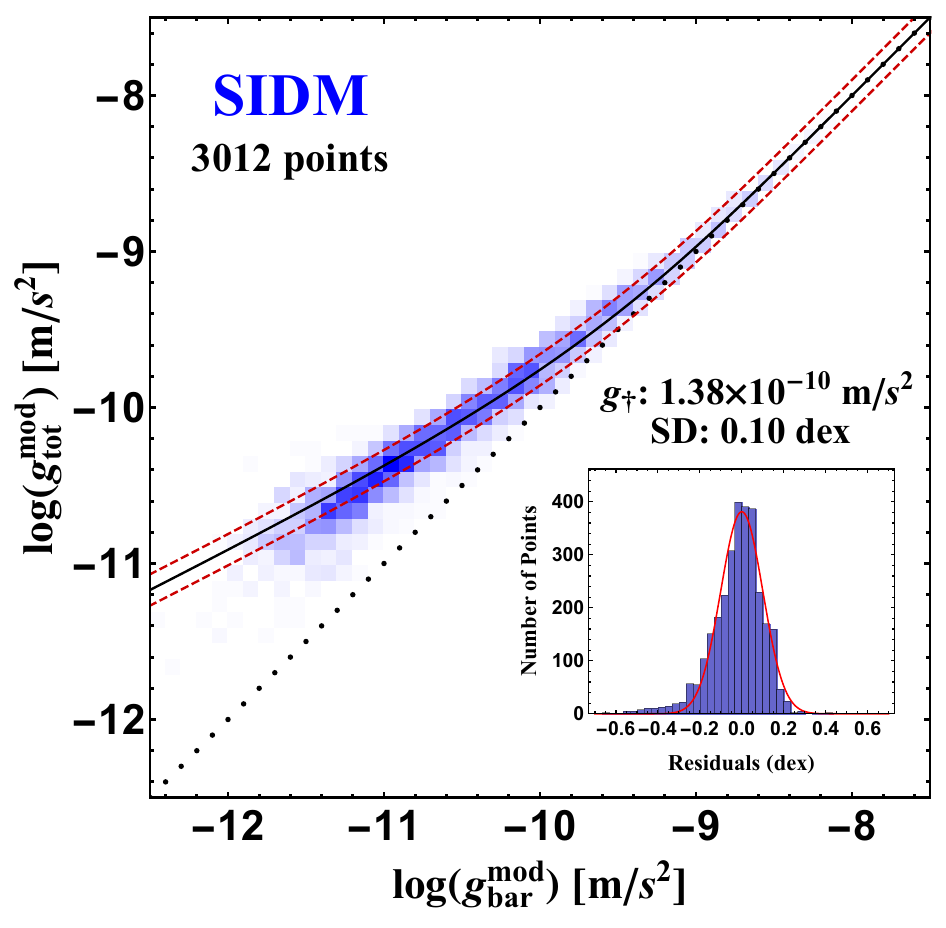}
\caption{The radial acceleration relation from SIDM fits to low-redshift galaxies with $\sigma/m_\text{DM} = 3$ cm$^2$/g. The black solid and red dashed lines show an empirical fitting function (and its $1\sigma$ scatter) that describes the observational data well. The inset shows residuals with the fit function removed. From \cite{Ren:2018jpt}.} \label{fig:RAR}
\end{figure}

\red{DM self-interactions may offer a solution to some of these problems.
To explain the smallest core densities in the most extreme LSBs, the cross section needs to be $\gtrsim 3\text{cm}^2/\text{g}$~\cite{Ren:2018jpt}. Such large cross sections are disfavored from cluster measurements as discussed in Sec.~\ref{sec:lens-main-halo}. These arguments therefore directly motivate the velocity-dependent cross section model, which we have argued is generic in the context of dark sector models. The velocity dependence allows for large effects on galaxy scales while maintaining consistency with constraints from cluster lensing studies. Small cross sections (well below $1 \rm cm^2/g$) may also work but will likely not be distinguishable from CDM.}
\add{The mass discrepancy--acceleration (or radial acceleration) relation} has also been explicitly studied in the context of SIDM \cite{Ren:2018jpt} (see Fig.~\ref{fig:RAR}); the good agreement with observations provides support for the redistribution of DM mass effected by self-interactions.

Currently, the majority of detailed studies of kinematics in low mass galaxies are limited to the Local Volume, however there are a few cases where spectroscopic measurements of strongly lensed galaxies up to z$\simeq$2 have been made (e.g. \cite{SL1, SL2, SL3, SL4}). It is also possible to measure the Tully--Fisher relation as a function of redshift ($z\lesssim2$) to study the relative evolution of galaxies' stellar and DM components. Evidence for variation in the Tully--Fisher normalisation is however controversial, with some authors finding no significant trend (e.g. \cite{Miller, Molina, Pelliccia}) and other a trend towards relatively more DM at high $z$ (e.g. \cite{Lorenzo++09, Cresci, Tiley, Price})
\red{The uncertainties in each of these methods are large due to the difficulty of reliably measuring the kinematic tracers and small sample sizes but they hold promise for testing the physics being implemented in hydrodynamic simulations.}

Advancing this field will require a synergy of observational and modelling techniques. On the one hand, empirical knowledge will benefit both from the compilation of homogeneous datasets with high quality photometry and kinematics (e.g. \cite{sparc}), and from the increase in statistics and precision afforded by current and next-generation IFU studies such as MaNGA \cite{manga}. Problems primarily based on photometry will benefit from DES, LSST, Euclid and Nancy Grace Roman Space Telescope \cite{WFIRST}\footnote{\url{https://roman.gsfc.nasa.gov/}} data. Large future datasets will enable stringent cuts on quantities like inclination and the degree of pressure support, reducing their confounding effects. On the other hand, advances in modelling will sharpen the likelihood function for dynamical observables for given theoretical assumptions. This requires not only further modelling of a range of systems in SIDM, but also increased resolution in $\Lambda$CDM hydrodynamic simulations---to push more of the relevant physics above the grid scale and hence reduce uncertainty in the effect of feedback---and an increase in the sophistication with which models and data are compared. \add{In addition it is important to bear in mind that solutions to astrophysical puzzles based on baryonic feedback, modified gravity and other DM physics are also possible however. For example, baryonic feedback may address the diversity problem due to the varying rates at which SNe explode and impact galaxies' dark matter distributions, although feedback strong enough to produce the large cores seen in LSBs do not produce galaxies with higher stellar and DM densities and therefore have trouble explaining the kinematics of HSBs \cite{Kaplinghat:2019dhn}. See Sec.~\ref{sec:unique} for further discussion of these points.}
High-redshift studies of galaxy kinematics (especially at low mass) may provide a way to disentangle SIDM from the effects of baryons by measuring the time scale of core formation.

The CDM simulations without strong baryonic feedback effects fail to explain the large dark matter cores seen in low surface brightness galaxies. On the other hand, with strong feedback, CDM simulations do not produce galaxy analogs with high stellar and dark matter densities, and therefore they have trouble in explaining the rotation curves of high surface brightness galaxies.

\subsection{Ultra-diffuse galaxies}
\label{sec:UDGs}

Long suspected of dominating the low-luminosity galaxy population \cite{Disney1976}, LSBs with central surface brightness fainter than \add{$\mu = 23~ \hbox{mag}/\text{arcsec}^2$}
are experiencing a second
\red{wave} of discovery after the initial discoveries in the 1990's enabled by CCDs \cite{McGaugh1995,Dalcanton1997,Leisman2017,Greco2018,Tanoglidis2021}.  With new telescope and survey designs, LSBs with surface brightnesses as faint as \add{$\mu = 31-32 ~\hbox{mag}/\text{arcsec}^2$}
can be detected either as resolved stellar populations \cite{Torrealba2019} or in diffuse light \cite{Abraham2014}.  This population is heterogeneous.  Many are ``ordinary" dwarf galaxies, which are low surface brightness simply by following the same size-luminosity relation as higher mass galaxies \cite{Munoz2015,Somerville2018,Jiang2019size}.  \add{However, a number of LSBs are outliers in the size--luminosity and/or stellar mass--halo mass relations.}

\add{Ultra-diffuse galaxies (UDGs) lie at the extreme low surface brightness end of the LSB population, with sizes of $r_e>1.5$ kpc.} Since 47 of these galaxies were found by the Dragonfly telescope in the Coma cluster \cite{vanDokkum:2014cea}, large populations have been observed in well-studied clusters, including Coma, Virgo, and Fornax \cite{2015ApJ...813L..15M,vandenBurg2017, 2017ApJ...838L..21K, Koda:2015gwa, 2015ApJ...813L..15M,Zaritsky2019}. In recent years, UDGs have been observed in a diversity of environments, including the field \cite{Fliri2016,Leisman2017,Roman2017,Greco2018,Tanoglidis2021,Scott2021}.  There are differences in the properties of UDGs in field and cluster environments--field galaxies tend to be gas-rich, blue, and star-forming, while those in clusters tend to be redder and quenched \cite{Leisman2017,2017ApJ...838L..21K,Zaritsky2019,karunakaran2020,Gault2021}.  Moreover, a number of cluster UDGs have unusual properties, such as being surrounded by anomalously many globular clusters given their stellar mass, and inhabiting halos far larger than indicated by the mean stellar mass--halo mass relation \cite{vandokkum2015,vandokkum2016,vandokkum2017,Beasley2016,forbes2020,Lim2020,Somalwar2020}.  Because of these differences in field and cluster environments, and because of the odd properties of cluster UDGs, an important question to answer is if these galaxies are true outliers in galaxy formation, or if they are part of a continuum of LSBs.  It is important to understand if there are different formation mechanisms for the field and cluster populations \cite{roman2017b}, and
how those mechanisms arise in the context of CDM and SIDM galaxy formation models.

In the context of CDM simulations, separate formation pathways are identified for field and cluster UDGs, and SIDM may additionally play a role in both.  In the field, although high spin was originally posited as the driver of UDG formation \cite{Amorisco2016}, the current conventional wisdom is that flybys and minor mergers drive the large size of $M_*\sim 10^8 M_\odot$ UDGs \cite{Wright2021,Jackson2021}.
Vigorous outflows can play a role too \cite{DiCintio2017}, and  enhance the effect of flybys and mergers. \red{However, the presence of a large reservoir of gas in the field UDGs makes the fly-by and concomitant tidal stripping unlikely explanations for field UDGs~\cite{Gault2021,PinaMancera:2021wpc}. In addition, the rotation curves of these field UDGs present challenges for all models of galaxy formation (whether based on CDM or SIDM) making this one of the most interesting avenues to pursue in the coming years~\cite{Kong:2022oyk}. Self-interactions of DM may play a role in UDG formation, by enhancing the distribution of galaxy sizes and rotation curve shapes at fixed stellar mass, as hinted at for the broader class of LSBs \cite{Kamada,Ren:2018jpt}, but more work is clearly needed to flesh out the predictions of SIDM models for these outliers.} \add{A potential solution is provided by a velocity-dependent SIDM model with $50 \lesssim \sigma/m \lesssim 100$ cm$^2/$g on UDG scales~\cite{Nadler:2023nrd}.}

Cluster UDGs are shaped by their environment.  CDM simulations show that although some cluster UDGs were previously UDGs in the field, many became UDGs after infall \cite{Carleton2019,Jiang2019LSB,Tremmel2020,moreno2022}.  These simulations show that ``typical'' dwarf galaxies can become UDGs if they fall into clusters early and experience quenching \add{(shutdown of star formation)} by ram-pressure stripping, fading of light as the stellar population ages \cite{roman2017b}, and tide-driven expansion of the stellar population. \red{Intriguingly, Refs. \cite{Ogiya:2018jww,Carleton2019} argued that galaxies in DM halos with large cores more easily become UDGs.  Idealized simulation including DM self-interactions in Ref.~\cite{Yang:2020iya} have shown that it is possible to reproduce the properties of NGC1052-DF2 and DF4 in the SIDM models. }

\begin{figure*}
\centering
\includegraphics[width=0.48\textwidth]{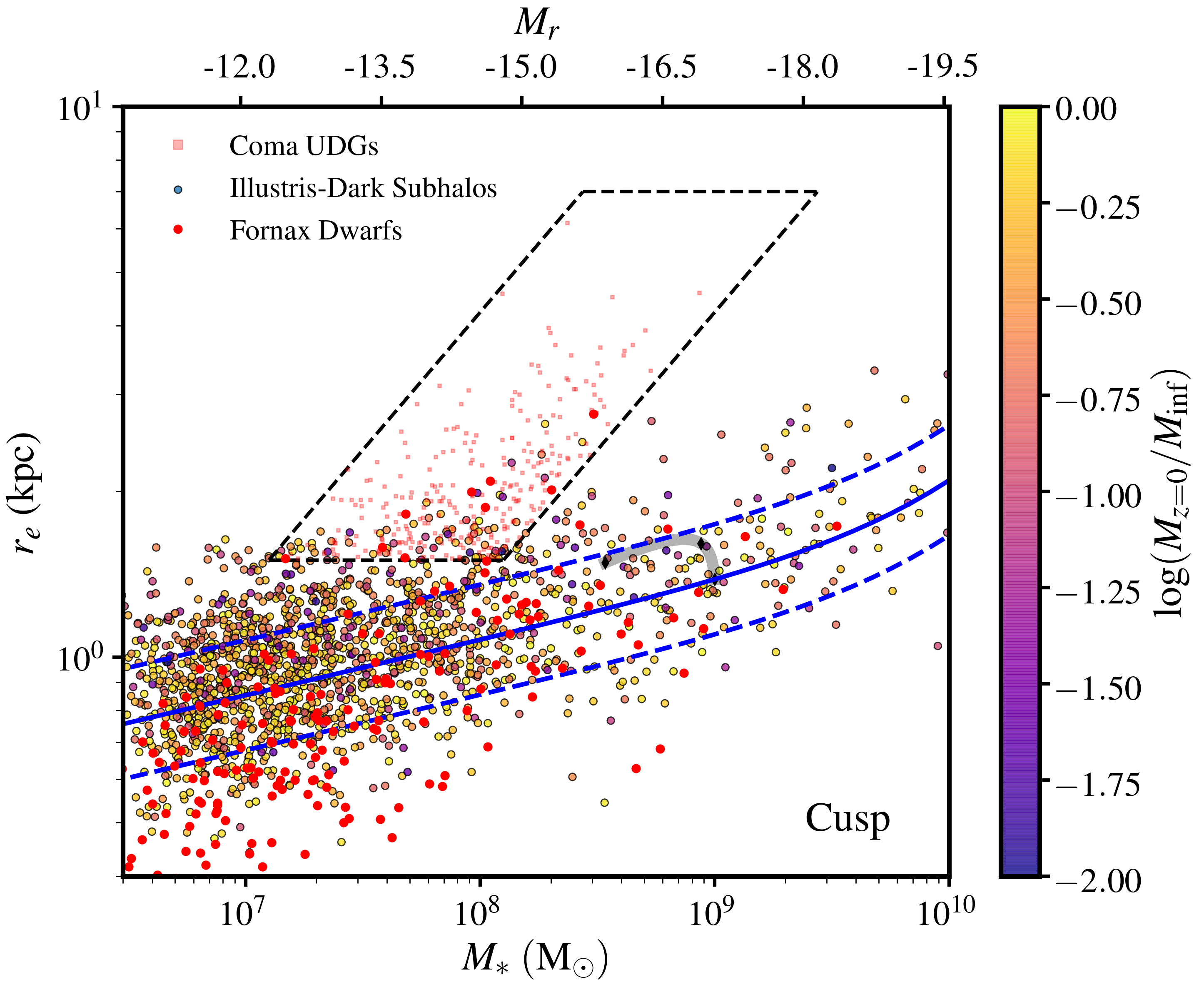}
\includegraphics[width=0.48\textwidth]{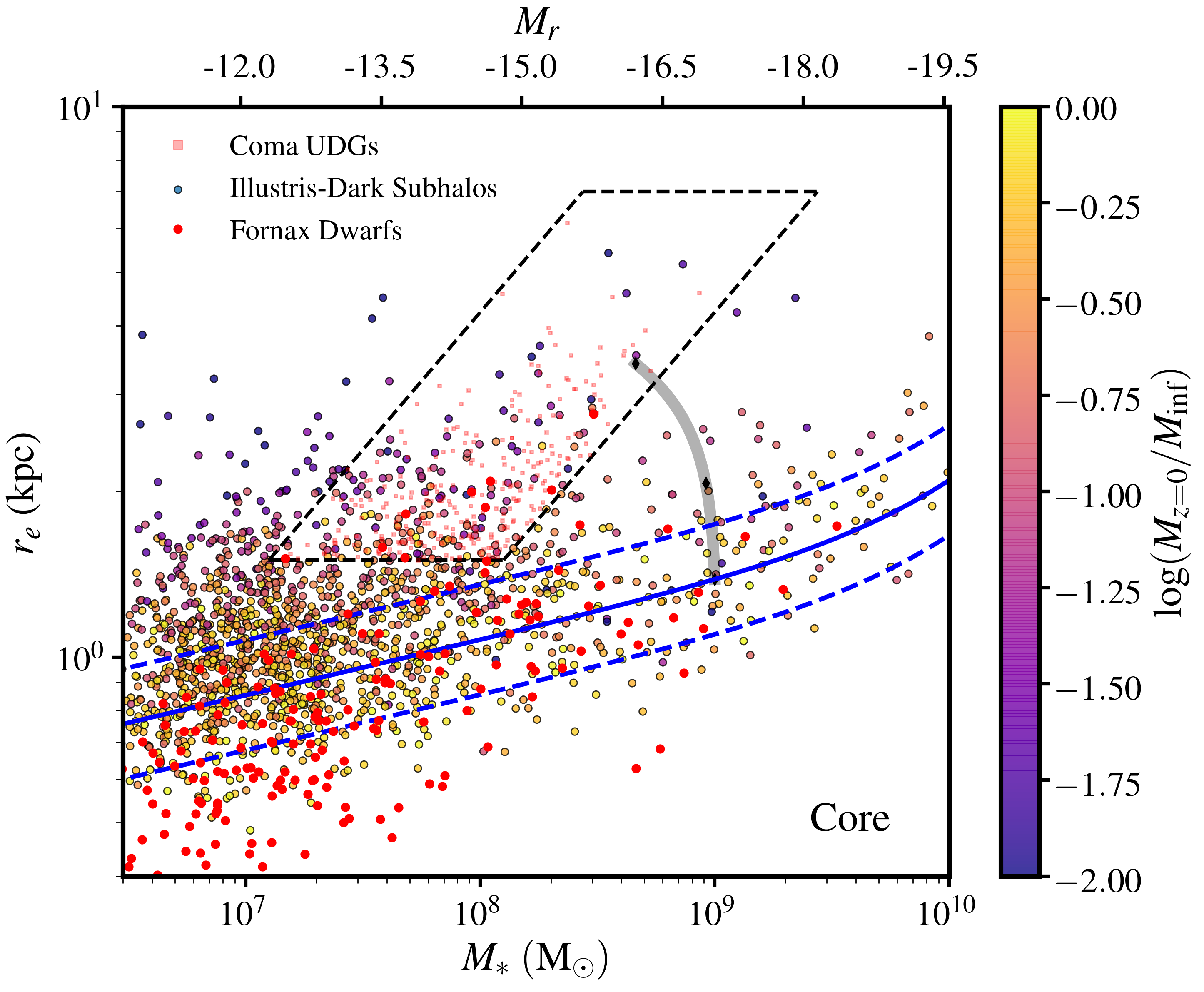}
\caption{A comparison of observed and simulated stellar masses -- half light radii relation of ultra diffuse galaxies (Figure reproduced from \cite{Carleton2019}.) within a massive cluster. The pink squares corresponds to observations from Coma cluster and the colored circles correspond to simulated objects. The color scale shows the degree of tidal mass loss from each UDG halo in the simulations. Also shown are the dwarfs within the Fornax cluster as red solid circles. The left panel corresponds to UDG halos with a cuspy profile and the right panel corresponds to halos that have cored profiles. In both cases the mass of the host cluster is $2.4\times 10^{14} M_odot$}
\label{fig:Carleton_fig_2}
\end{figure*}

The stellar component in dark-matter-dominated galaxies expands faster in cored halos rather than cusped halos in response to tides, which has been shown in the context of CDM \cite{Penarrubia:2010jk,Errani2015} as well as SIDM \cite{Dooley:2016ajo}.  In the semi-analytic model of Ref. \cite{Carleton2019}, cored halos play the essential role in populating the tail encompassing large UDG sizes, as shown in Fig. \ref{fig:Carleton_fig_2}.
\red{This model with cored halos at infall was able to reproduce the observed sizes and stellar masses of UDGs, as well as their abundance as a function of host halo mass. However, the assertion that cored halos are necessary has been questioned in Ref.~\cite{Sales:2019iwl} who argue that it is possible for a large number of UDGs to form in the field as LSBs and then be accreted into the cluster. In this scenario, there are still be some UDGs created close to the center of the cluster due to extreme tidal interactions in cuspy halos. More work on discriminating the two scenarios is warranted, as are cosmological CDM and SIDM simulations tailored to this problem.}

\subsection{Warping of stellar disks}
\label{sec:warps}

\begin{figure*}
\centering
\hspace{-1cm}
\includegraphics[height=6.3cm]{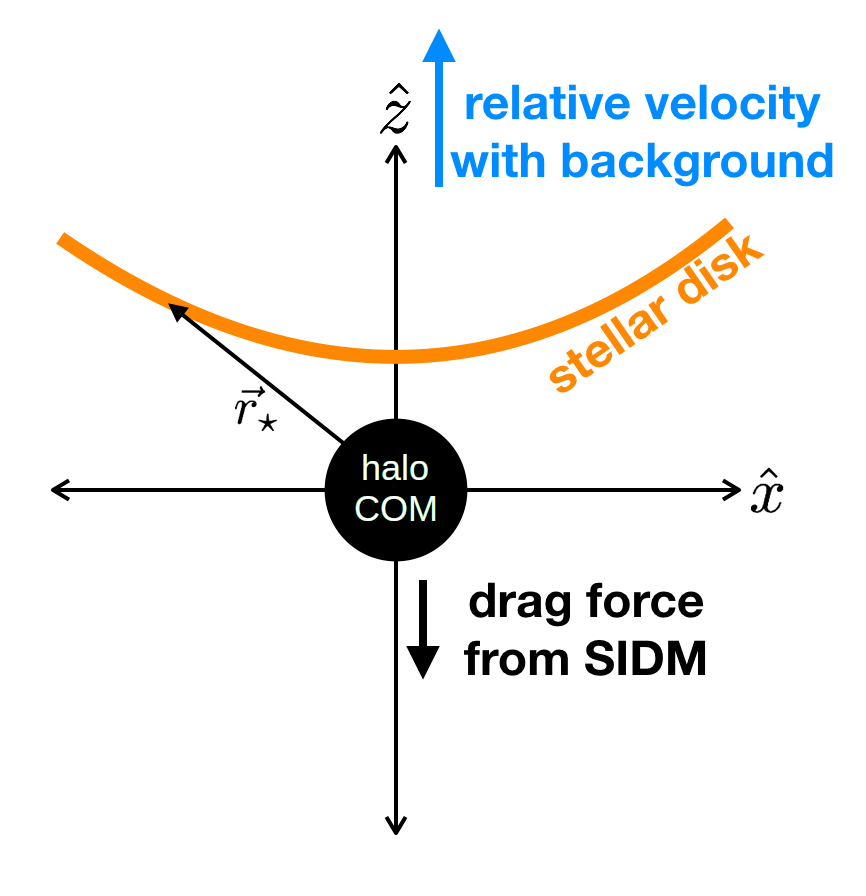}
\hspace{.2cm}
\includegraphics[height=6.3cm]{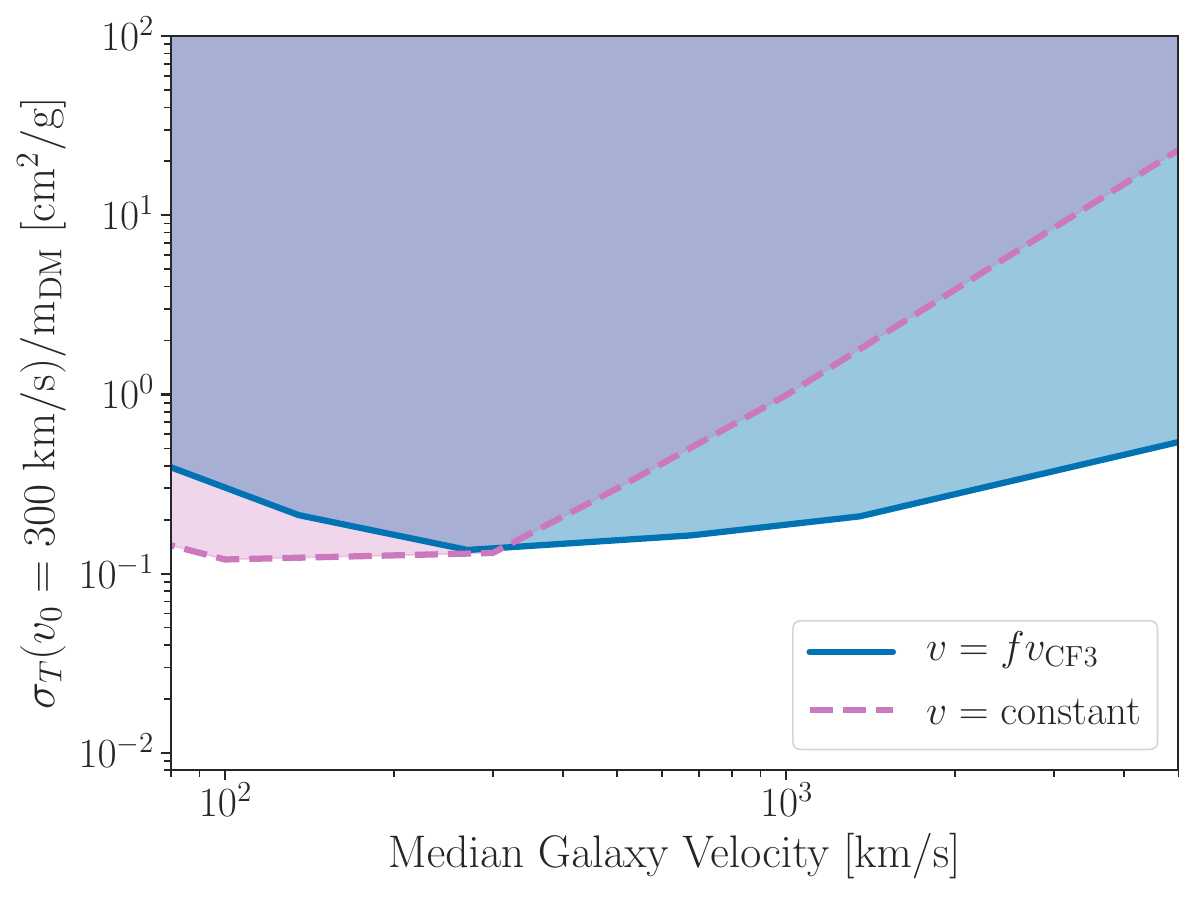}
 \caption{\emph{Left:} Cartoon of the formation of a U-shaped stellar disk warp in SIDM. \emph{Right:} Constraints on the momentum transfer cross section at 300 km/s for long-range (Rutherford) scattering from 3,213 measured warps in the nearby Universe, as a function of the median galaxy velocity relative to the background. Adapted from Ref.~\cite{Pardo}.}
\label{fig:warp}
\end{figure*}

We described in Sec.~\ref{sec:drag} how the drag force experienced in SIDM by halos falling through a medium of DM can cause a separation between the centroids of stars and DM in a galaxy. Both numerical~\cite{Secco} and simple analytic~\cite{Pardo} models show that the potential gradient that this displacement establishes across the stellar disk induces a U-shaped warp in the disk facing the direction of infall (Fig.~\ref{fig:warp}, left), and a longer-lasting disk thickening. \add{The shape of the warp can be understood from the requirement that in equilibrium all parts of the stellar must have the same acceleration as the halo centre in the direction of the large-scale gravitational field. This means that parts of the disk further from the centre, which have a lower restoring force due to the dark matter, must have a greater component of that force pointing in the direction of the halo's acceleration and hence bend up more strongly.} While S-shaped warps formed by tidal distortions of the stellar light profile are abundantly observed in cluster environments, indicating that they are readily produced by physics unrelated to DM self-interactions, such physics is unlikely to generate prominent U-shaped warps. These warps are only formed by a differential force on the disk and its halo, e.g. due to SIDM drag.

The simulations of Ref.~\cite{Secco} indicate that the warping effect may be observable on a galaxy-by-galaxy basis in current and next-generation photometric data when SIDM \add{momentum-transfer} cross sections are $\gtrsim0.5-1 \cmg$. \add(Note that this forecast corresponds to the infall velocity scale of a Milky Way like disk-galaxy into a host halo). This has inspired work aimed at constraining SIDM by means of this signal \cite{Pardo}. Here, the $r$-band images of $\sim$3000 edge-on disk field galaxies with $M_*>10^9 M_\odot$ and $D < 250$ Mpc from the \textit{Nasa Sloan Atlas} are reduced to yield estimates of their degree of U-shaped warping, and a forward-model is developed to predict this in SIDM as a function of interaction cross section and range, halo properties, and DM environment. This is then combined with a simple model for astrophysical and measurement noise to create a galaxy-by-galaxy likelihood for the warp statistic, and the noise and other nuisance parameters are marginalised over to obtain constraints on the SIDM cross section. Due to the requirement that the system satisfies the fluid approximation and that drag rather than evaporation is the dominant result of the self-interactions, only long-range (light mediator) models can be constrained in this way, although a sample of galaxies in higher-density multi-streaming environments (e.g. clusters) may in the future be used to constrain shorter-range forces.

No correlation of the warp strength or direction with the SIDM expectation is found in the analysis, leading to the constraints shown in Fig.~\ref{fig:warp} (right) on the momentum transfer cross section at $300$ km/s. These are plotted as a function of the median relative velocity of the galaxies and the background medium because this is considerably uncertain. The two models shown correspond to \textit{a)} a constant velocity over the entire sample and \textit{b)} velocities proportional to those from the CosmicFlows-3 catalogue, respectively. \add{The large-scale velocity of the background is estimated by applying the continuity equation to a reconstruction of the density field of the local universe constrained by galaxy number densities in voxels. The constraints become weaker at high velocity due to the reduction in cross-section of the long-range interaction model.} At face value these bounds are an order of magnitude tighter than those from other SIDM effects, e.g. dwarf galaxy evaporation. The model does however make a number of simplifying assumptions to render the inference tractable, for example that the halos are fully spherical, thermalised and equilibrated, and that all parts of the halo move together. Simulations of halos in environments similar to those of galaxies with measurable morphology will be needed to validate and refine this novel method for testing SIDM.

\subsection{Large-scale structure constraints}
\label{sec:lss_constraints}

\textcolor{black}{
As we have discussed in Sec.~\ref{sect:LSS}, there are several particle physics models that can modify the power spectrum of density fluctuations, thereby leaving signatures on the temperature and polarization of the CMB, the clustering of galaxies and the halo mass function. A necessary feature in these models is the presence of light mediators.
As discussed in Section \ref{sect:LSS} dark matter models that tightly couple to dark radiation in the early universe can lead to dark acoustic oscillations and diffusion damping in dark matter (like the the BAO and Silk damping), modifying the linear power spectrum. In Ref.~\cite{Cyr-Racine:2013fsa},  cosmological data from the lensing of the CMB, the clustering of galaxies on linear scale and the BAO were analyzed to derive constraints on the fraction of dark matter that can couple to dark radiation and the strength of the coupling. Their results show that a fraction of partially interacting dark matter, $f_{\rm int}>5\%$ is in severe tension with current observations of large-scale structure.}

\textcolor{black}{The damping scale of the power spectrum due to interactions between dark matter and radiation is set by the horizon at the epoch of kinetic decoupling. For a kinetic decoupling temperature, $T_{\rm kd}$, the suppression in the power spectrum occurs approximately for modes $k > 10 (T_{\rm kd}/\mathrm{keV})\ \rm Mpc^{-1}$ \cite{Green:2005fa,Loeb:2005pm,Bertschinger:2006nq,Feng:2009mn,CyrRacine:2012fz}.
SIDM coupling to a dark radiation component like dark photons or sterile neutrinos can significantly delay the decoupling of dark matter from the thermal bath leading to damping on scales of order $0.1$ Mpc/h or smaller. An effective framework to describe these interactions and the associated phenomenology has been developed by the ETHOS collaboration~\cite{Vogelsberger:2015gpr,Cyr-Racine:2015ihg}. One of the central predictions is a potentially observable cut-off in the halo mass function~\cite{Aarssen:2012fx,Bringmann:2013vra,Chu:2014lja,Bringmann:2016ilk,Archidiacono:2017slj,Huo:2017vef} arising from the damping of the power spectrum.}

\textcolor{black}{One of the ways to probe these scales is through the absorption spectrum of  Lyman-$\alpha$ forest, which measures the density of neutral hydrogen on large scales.  In Ref.~\cite{Huo:2017vef} the Lyman-$\alpha$ forest power spectrum measurements was used to put strong constraints on the damping scale (by recasting the warm dark matter constraints) to set an upper limit on the decoupling temperature of $T_{\rm kd}> 1$ keV and a minimum halo mass of $M_{\rm min} < 10^8 M_\odot/h$. The origin of this constraint is rooted firmly in the physics of the dark sector, including the production mechanism for the dark matter. The massive gauge boson that mediates the self-interaction in simple SIDM models has to decay in order to avoid over-closing the universe and in Ref.~\cite{Huo:2017vef} decay of the gauge boson to light DM particles were assumed. Decays to SM particles have much stronger direct detection constraints~\cite{PandaX-II:2021lap} and indirect search constraints~\cite{Bringmann:2016din}.
In the future, we expect the combined measurements of the Lyman-$\alpha$ power spectrum and the halo mass function from faint galaxies will strongly constrain the space of allowed parameter space for the models with dark radiation.
Prospects for the future in this direction are bright, with CMB-S4 \cite{S4}\footnote{\url{https://cmb-s4.org/}} combined with LSST likely to improve constraints on $\Delta N_{\rm eff}$ in these models by an order of magnitude~\cite{Krall:2017xcw}. Another promising avenue to constrain simple SIDM models and those with additional light degrees of freedom and interactions is to probe the linear and quasi-linear power spectrum to higher $k$ values using 21-cm data. Future measurements offer the possibility to constrain the power spectrum amplitude at the 10\% level integrated over $k=40-80\ \rm Mpc^{-1}$~\cite{Munoz:2019hjh}, which can probe decoupling temperatures well above a keV.
}

\subsection{Summary of constraints on the self-interaction cross section}
\label{sec:summary}

\red{The various observational probes for SIDM can be thought of as a mapping between the velocity dependence of the cross section and observable processes at various energy scales within halos of different masses during hierarchical structure formation.
On the galaxy cluster scales, where the internal velocity dispersions are of the order of a few 1000s of km/s, the most stringent constraints currently come from measured core densities from strong lensing \cite{Kaplinghat:2015aga}. The small sizes of the cores and the high central densities in these clusters constrain cross section to $\sigma/m_\text{DM}<0.13 {~\rm cm}^2$/g at $95\%$ confidence in one study~\cite{Andrade:2020lqq} and $\sigma/m_\text{DM}<0.35 {~\rm cm}^2$/g at $95\%$ confidence in another~\cite{Sagunski:2020spe}, as discussed in more detail in Sec.~\ref{sec:lens-main-halo}. Even though there is some evidence of a non-zero cross section in these analyses, it is not possible to conclude that we have a detection. In the future, detailed simulation work combined with joint analyses of strong lensing, weak lensing, X-ray and velocity dispersion data for a large number of clusters could provide a concrete path to measuring a non-zero cross section at relative velocities of about $1500 \ \rm cm^2/g$. Similar analysis at the group scale gives an upper limit of $1.1 \ \rm cm^2/g$, with a mild preference for a non-zero cross section ($0.5\pm 0.2 \ \rm cm^2/g$;~\cite{Sagunski:2020spe}).
A variety of other probes like the displacement between galaxies and dark matter in merging systems, the steepening of the profiles beyond the dark matter core and the existence of displaced BCGs will further help establish consistency between different observations once the measurement systematics and the interplay between baryonic and dark matter evolution are better constrained and understood. Among these, constraining BCG wobbles, which are correlated with the size of the constant density core, using offset BCGs seems highly promising with current constraints at $\sigma/m_\text{DM} < 0.39 \ \rm cm^2/g$ at the 95\% C.L.~\cite{Harvey:2018uwf}.}

At lower velocities, a wider range of cross sections is in principle still allowed as long as the model is consistent with the constraints at high velocities. As baryonic effects are more relevant for lower mass halos, the combination of baryons and SIDM leads to a wide range of halo mass profiles inside a region of about the NFW scale radius. In particular, for galaxies like the MW and large elliptical galaxies, the large concentration of baryons in the inner regions makes the density profiles of SIDM and CDM models indistinguishable~\cite{Kaplinghat:2013xca,Robles:2019mfq,2021MNRAS.507..720S,Vargya:2021qza}. We can expect constraints from evaporation of DM in satellites of the MW (and similar galaxies) due to scattering events between subhalo and host halo DM particles when the cross section at $200~\rm km/s$ exceeds $5-10 \rm cm^2/g$  (e.g.,~\cite{Nadler2020SIDM,Kahlhoefer_diversity,Jiang:2021foz}) but the constraints have not been robustly inferred. \add{This is made more difficult by the fact that standard halo finders may fail to identify subhalos in N-body simulations~\cite{Diemer}.} At lower velocities, fitting the rotation curves in SPARC database has shown that $\sigma/m_\text{DM} > 3 \ \rm cm^2/g$ at about $100 \rm km/s$ is needed to adequately explain the low-surface brightness galaxies with the lowest central DM densities~\cite{Ren:2018jpt}. Note that the predictions for the cored profiles of SIDM halos do not change dramatically as the cross section is increased, so cross sections even an order of magnitude larger cannot be easily distinguished based on galaxy rotation curve data alone~\cite{Elbert:2014bma}. If the cross sections are $\gtrsim 100\ \rm cm^2/g$, then core collapse will set in for median concentrations and it may become difficult to explain the low density DM cores of low-surface brightness galaxies.

\begin{table}
\vspace{0.1cm}
\centering
\begin{tabular}{|c|c|c|c|}
\hline
		Halo Mass [$M_\odot$] & Velocity [km/s] & Observational Probes & $\sigma/m$ [cm$^2/$g]  \\ \hline

            Clusters & $v>2000$ & Strong Lensing  & $< 0.28$ \cite{Sagunski:2020spe}  \\
            $M \gtrsim 3\times 10^{14} $ & & BCG offsets & $< 0.39$ \cite{Harvey:2018uwf}\\
            & & Merging clusters & $< 1.25$ \cite{Markevitch}\\
            \hline
            Groups  & $ 500 < v < 2000$ & Strong lensing  & $< 0.9$ \cite{Sagunski:2020spe} \\
              & &   & $< 0.13$ \cite{Andrade:2020lqq} \\
            $10^{12.5} \lesssim M \lesssim 10^{14}$ & & Weak Lensing  & $< 1$ \cite{Adhikari:2024aff}\\
            \hline
            Galaxy Halos  & $100 < v < 500$ & Rotation Curves & $> 3$ \cite{Ren:2018jpt} \\
            $10^{11} \lesssim M \lesssim 10^{12.5}$ & &  Satellite counts &  $<5-10$\\
             &  &   & (\citet{Nadler2020SIDM,Kahlhoefer_diversity}\\
             &  &   & \citet{Jiang:2021foz})\\
            \hline
            Dwarf Halos  & $v<100$ & Density profiles & $\geq3$\\
            & &  & (\citet{Valli2018, Correa2021}\\
            & &  & \citet{Jiang:2021foz,Silverman:2022bhs}\\
            & &  & \citet{Nadler:2023nrd})\\

            $M\lesssim10^{11}$ & &  & \\
            \hline

	\end{tabular}

\caption{\add{Summary of constraints on SIDM from astrophysical probes. Constraints on the interaction cross-sections from the clusters to the galaxy scales (first three rows) correspond to the $95\%$ confidence interval. The values quoted for dwarf halos are based on simulations exploring the $\sigma/m$ space. They should be interpreted as limits on $\sigma/m$ beyond which the simulations do not agree with the data (as opposed to statistical constraints from parameter estimation studies). Note that the results presented here are for isotropic and elastic scattering.}}
\label{tab:summary_constraints}

\end{table}

\red{Within the local universe, a self-consistent explanation for the distribution and inner structure of the satellites of the Milky way is not yet fully known. While the ``missing satellites problem'' is largely thought to be resolved given the current census of satellites, the exact nature of the galaxy--subhalo connection is still unclear and currently requires populating galaxies in subhalos with $V_{\rm peak} \sim~7 $ km/s that are below the collisional cooling limit. Recent work has shown that the minimum $V_{\rm peak}$ will have to be smaller for moderate cross sections $\sigma/\rm{m}_{\rm dm} \lesssim 5 \ \rm cm^2/g$~\cite{Silverman:2022bhs}. While the ``Too Big To Fail'' problem is resolved by moderate SIDM cross sections of $\sim$1 cm$^2/$g~\cite{Vogelsberger2013,Robles:2019mfq}, this underpredicts the central densities of ultra-faint satellites~\cite{Kim:2021zzw,Zavala:2019sjk,Kahlhoefer_diversity}. These conclusions are highly sensitive to the presence of the MW disk, but recent work including the disk potential has demonstrated the validity of these arguments~\cite{Silverman:2022bhs}. The issue is not restricted to the ultra-faints dwarfs---using analytic methods it has been found that if all the bright Milky Way satellites are in the core expansion phase, it is difficult to fit their measured stellar dispersions with a single cross section~\cite{Valli:2017ktb}.  A velocity-dependent cross section with $\sigma/m_\text{DM}>10 \ \rm cm^2/g$ at velocities relevant for the internal dynamics of satellites ($5-20 \ \rm km/s$) that drives the faint satellites to core-collapse is perhaps the best bet to explain these observations~\cite{nishikawa2019,Kahlhoefer_diversity,Zavala:2019sjk,Correa2021,Jiang:2021foz,Silverman:2022bhs}.}

\red{Putting together these constraints suggests that a significant velocity dependence is necessary to explain data constraining DM mass profiles from dwarf galaxy to clusters of galaxies. Compared to earlier work~\cite{Kaplinghat:2015aga}, the motivations for a strong velocity dependence have increased over time. An exciting possibility within models that feature large cross sections at low velocities is that many of the halos and subhalos could be in the core collapse phase of their evolution, with predictions for observables that are different from CDM as discussed in this review. Examples of such sharp velocity dependence are discussed in Refs.~\cite{Correa2021,Turner:2020vlf,Jiang:2021foz}, where analytic methods and N-body simulations are used to predict the central densities of Milky Way satellites in the core expansion and core collapse phases. More comprehensive analytic and numerical treatments including the effect of the MW disk are required to make further progress.  However, it seems clear that the satellites of the MW can provide a definite test of the SIDM models that aim to explain DM density profiles across the mass range from dwarf galaxies to galaxy clusters. }

\section{Beyond the simple SIDM framework}\label{sec:particle}
\label{sec:extensions}

In the basic setup of SIDM, a single species of DM particles interact through the exchange of a light (or massless) scalar or vector mediator with a cross sections per unit mass of $\mathcal{O}(1~\mathrm{cm}^2/\mathrm{g})$, as discussed in Sec.~\ref{sec:theory}.
We have, however, largely neglected the implications of a hidden sector beyond the phenomenology of the DM particle itself.
There are also more complex hidden sectors that predict interesting phenomenological signatures, not only in astrophysical systems, but in cosmology and potentially in laboratory experiments as well.
While there are model-dependent constraints on any specific particle physics realization of SIDM, many models may produce similar phenomenological signatures and are thus subject to similar restrictions from data.

In this section, we review the consequences of non-minimal SIDM models and of the hidden sectors themselves.
We first discuss two broad generalizations of the simple SIDM scenario that we have assumed thus far.
In Sec.~\ref{sec:extensions-dissipation}, we relax the assumption that self interactions are elastic and allow a mechanism for SIDM to dissipate energy.
In Sec.~\ref{sec:extensions-subcomponent}, we discuss the possibility that SIDM as a subcomponent of the overall DM content, allowing it to have very strong self interactions.
The hidden sector contains light mediators that allow DM particles to self interact, and we discuss the consequences of the light mediator in Sec.~\ref{sec:extensions-mediators}.
Finally, in Sec.~\ref{sec:extensions-SM}, we mention various complementary constraints on DM sectors that interact with the SM.

\subsection{SIDM with dissipation}
\label{sec:extensions-dissipation}

If SIDM is not a simple single-state particle, it is possible for SIDM to undergo inelastic processes between its ground state and an excited state.
The presence of exicited states can arise in models of composite SIDM, such as atomic-like DM~\cite{CyrRacine:2012fz,Boddy:2016bbu} and nuclear-like DM~\cite{Boddy:2014yra}, in which fundamental dark-sector particles are bound under a dark-sector force.
The composite nature of DM permits a spectrum of internal energy levels, along with a naturally large self-interaction cross section.
Excited states also arise if DM is comprised of different fundamental particles with a mass splitting and an off-diagonal interaction~\cite{ArkaniHamed:2008qn,Schutz:2014nka,Blennow:2016gde,Das:2017fyl}.
In either case, collisions may induce a transition to the excited state, which may then relax back to the ground state through collisional de-excitation or decay via the emission of some form of radiation.
This radiation may be a massless force carrier, such as a dark photon, or a light mediating particle, with a mass smaller than the energy splitting between the ground and excited states.

Assuming there is sufficient kinetic energy available, collisions within DM halos can excite DM particles.
If the decay width of the excited state is large and if the halo is optically thin to the emitted radiation, the halo loses energy and could eventually collapse~\cite{Boddy:2016bbu}.
As discussed in Sec.~\ref{sec:core_collapse}, the gravothermal collapse process with elastic SIDM interactions can be accelerated by environmental effects~\cite{nishikawa2019}; for an isolated halo, the collapse process can be accelerated through energy dissipation~\cite{Essig:2018pzq, Huo:2019yhk}, assuming DM particles are not ejected from the halo~\cite{Vogelsberger:2018bok}.
The observation of uncollapsed dwarf galaxies and LSBs can place restrictions on the acceptable amount of halo cooling in those systems~\cite{Essig:2018pzq}.
On the other hand, the gravothermal collapse of DM halos with dissipation may be able to form the seeds for the observed population of supermassive black holes~\cite{Xiao:2021ftk}.

\subsection{Subcomponent SIDM}
\label{sec:extensions-subcomponent}

Some models of DM postulate that while most of the DM in the universe may be (approximately) collisionless, there may be a small component that has very large self-interactions, including dissipative interactions.
This small component could have a large impact on the observational predictions.
For example, even a very small amount of ultra-strong SIDM can cause a halo to experience gravothermal collapse, as discussed in Sec.~\ref{sec:core_collapse}, and form seed black holes~\cite{Boddy:2014yra,Pollack:2014rja}.
Alternatively, some models also propose that that up to 15\% of the DM of the Universe exhibit a behavior similar to baryons (i.e., experience energy dissipation), including disk collapse~\cite{Fan:2013yva,Fan:2013tia} and plasma instabilities~\cite{Heikinheimo:2015kra,Sepp:2016tfs,Heikinheimo:2017meg}.
These effects have for example been studied in the context of the formation of a dark core in major mergers such as Abell 520~\cite{Jee:2012sr,Jee:2014hja}.
Dissipation could also lead to qualitatively new predictions for the small-scale structure of DM, including asymmetric dark stars~\cite{Chang:2018bgx}.

\subsection{Phenomenology of light mediators}
\label{sec:extensions-mediators}

{In many models of SIDM, the self interaction arises from the exchange of a mediator that is light compared to the DM particle. For the case of thermal DM with a mass in the GeV--TeV range, such a light mediator could for example be a 1--100 MeV scalar or vector particle with a Yukawa coupling to DM~\cite{Tulin:2013teo}.}
The mediating particles themselves can impact a wide range of observations, allowing for additional probes of SIDM hidden sectors.

The production mechanism of SIDM in the early Universe can play an important role throughout cosmic evolution (see, e.g., Ref.~\cite{Bernal:2015ova}).
A common assumption is that DM particles were at some point in thermal equilibrium with the bath of CMB photons, through a common reheating mechanism.
In a completely isolated hidden sector, however, DM may be produced in its own thermal bath through a separate reheating mechanism, resulting in isocurvature perturbations that are constrained by CMB observations~\cite{Akrami:2018odb}, and thus place constraints on self interactions~\cite{Heikinheimo:2016yds}.

With a nontrivial dark sector, the SIDM relic abundance can be set through mechanisms analogous to the standard thermal freeze-out scenario.
For instance, SIDM may undergo dark-sector freeze-out (in which SIDM particles annihilate into lighter dark-sector particles) or the SIMP mechanism (in which SIDM particles experience number-changing processes).

The mediator constitutes the thermal bath for DM and contributes an additional $\Delta N_\textrm{eff}$ to the standard number of relativistic degrees of freedom in the early Universe.
Measurements of the CMB and of nuclei abundances restrict the amount of $\Delta N_\textrm{eff}$ permitted at the time of recombination and Big Bang Nucleosynthesis (BBN), respectively~\cite{Aghanim:2018eyx}.
Moreover, for models of light DM (with masses in the MeV range), the DM itself can contribute to $\Delta N_\textrm{eff}$ during BBN~\cite{Hufnagel:2017dgo}.
Since no interaction with the SM is required in a minimal model of SIDM, the hidden-sector bath can be slightly colder than the CMB and thus easily evade $\Delta N_\textrm{eff}$ constraints~\cite{Feng:2008mu,Boddy:2014yra}.
However, if the hidden sector does communicate with the SM, $\Delta N_\textrm{eff}$ will set constraints on such models.

The presence of a hidden thermal bath for SIDM also leads to the suppression of small-scale fluctuations in the CMB, as described in Sec.~\ref{sec:pheno} for the case of dark photon mediators.
\textcolor{black}{Such signatures can be constrained by the CMB, as well as large-scale structure observables, such as the Lyman-$\alpha$ forest~\cite{Bringmann:2016ilk,Huo:2017vef,Krall:2017xcw} as discussed previously in Sec.~\ref{sec:lss_constraints}.
If models predict a strong cutoff of the matter power spectrum, then additional powerful constraints at small scales arise from the existence of the smallest DM halo observed~\cite{Nadler:2019zrb}.
}

\subsection{Connections to the Standard Model}
\label{sec:extensions-SM}

\textcolor{black}{
While not strictly required, there are good reasons to consider non-gravitational interactions of SIDM with SM particles. These interactions allow SIDM to be produced from the SM thermal bath, which can constrain the model space  stringently~\cite{Bringmann:2016din} and also provide simple viable models~\cite{Bernal:2015ova,Hambye:2018dpi,Hambye:2019tjt}.}
These interactions can impact the composition and cosmological evolution of the Universe and may give rise to observational signatures in laboratory experiments and astrophysical systems (see e.g., Refs.~\cite{Gluscevic:2019yal,Boddy:2022knd,Bechtol:2022koa}).

While there are many different ways in which SIDM particles can couple to the SM, most attention has been paid to two benchmark scenarios, in which the couplings of SIDM particles are proportional to the electric charge and mass of the various SM particles, respectively~\cite{Evans:2017kti}. The former case arises from the (kinetic) mixing between a new vector boson and the photon of the SM~\cite{HOLDOM198665}, while the latter case arises from the mixing between a new scalar boson and the Higgs boson of the SM~\cite{Lebedev:2012zw}. The new exchange particle (often referred to as the mediator) is typically considered to also be responsible for the generation of strong self-interactions.

It may be possible to understand the cosmological abundance of SIDM particles in terms of their annihilation and scattering rates~\cite{Bernal:2015ova}. If these rates are large enough to bring SIDM particles into thermal equilibrium with the SM plasma, the relic abundance can be set through the standard WIMP freeze-out mechanism. Even if the interaction rates do not bring the two sectors into equilibrium, the relic abundance may be set, for example, via the freeze-in mechanism~\cite{Hall:2009bx}. Moreover, many SIDM models require the presence of additional (lighter) states in the dark sector, as discussed in Sec.~\ref{sec:extensions-mediators}. In order to ensure that these additional particles do not constitute a large fraction of DM or dark radiation, they typically need to be unstable, such that they decay into SM particles. In order to evade constraints on $\Delta N_\textrm{eff}$ from BBN and the CMB and in order not to modify the predictions of BBN through the photodisintegration of light elements, these decays should occur sufficiently early in cosmological history, placing a lower bound on the interaction strength between SIDM particles and SM states~\cite{Hufnagel:2018bjp}.

On the other hand, cosmological considerations also place an upper bound on the interaction strength for DM annihilation into and DM scattering with SM particles.
DM annihilation can inject energy into the SM plasma, affecting the optical depth of CMB photons post-recombination~\cite{Padmanabhan:2005es,Mapelli:2006ej,Zhang:2007zzh,Galli:2009zc,Slatyer:2009yq,Finkbeiner:2011dx,Bringmann:2016din,Cirelli:2016rnw,Binder:2017lkj,Baldes:2017gzu}.
CMB constraints on DM annihilation are strong for $s$-wave annihilation, but significantly weaken for $p$-wave etc.\ due to the velocity suppression of cold DM.
Additionally, there are constraints on DM scattering with ordinary matter, which can induce spectral distortions~\cite{Ali-Haimoud:2015pwa,Ali-Haimoud:2021lka} and a suppression of anisotropies in the CMB~\cite{Chen:2002yh,Sigurdson:2004zp,Dvorkin:2013cea,Gluscevic:2017ywp,Boddy:2018kfv,Xu:2018efh,Slatyer:2018aqg,Boddy:2018wzy,Nguyen:2021cnb,Dvorkin:2022bsc,Boddy:2022tyt} and the matter fluctuations~\cite{Nadler:2019zrb,DES:2020fxi,Maamari:2020aqz,Rogers:2021byl}.
These constraints are complementary to the more traditional search methods of direct and indirect detection, described below.

Direct detection experiments place strong constraints on the interactions of DM particles with the SM and are known to exclude many DM models that exploit the freeze-out mechanism to explain the DM relic abundance. \textcolor{black}{
Another concrete way in which the direct detection of a SIDM particle can be motivated is the requirement in some models that the mediators decay before the beginning of BBN~\cite{Kaplinghat:2013yxa}. These experiments are particularly sensitive if the mediator of the interaction is light, such that scattering with small momentum transfer is enhanced~\cite{DelNobile:2015uua,Kahlhoefer:2017ddj,PandaX-II:2021lap}.
In many models of SIDM, the DM particle mass is below 1 GeV, for which the most promising strategy is to search for DM-electron interactions~\cite{Battaglieri:2017aum}.}

A second source for observable signals of SIDM particles are annihilation processes occurring in regions of enhanced DM density, such as MW satellites or the Galactic Centre. The resulting fluxes of photons, neutrinos and/or changed anti-particles can be searched for with indirect detection experiments. These constraints are particularly strong if the annihilation cross section in enhanced for small relative velocities (so-called Sommerfeld enhancement~\cite{Ackerman:mha,Pospelov:2008jd,Buckley:2009in,Essig:2010em,Feng:2010zp,Loeb:2010gj}). We note, however, that annihilation signals may also be strongly suppressed or absent, for example if the annihilation cross section is suppressed in the non-relativistic limit, or if SIDM particles have a particle-antiparticle asymmetry similar to the one of baryons~\cite{Baldes:2017gzu}.

Finally we note that SIDM particles may also be produced at accelerators, although the resulting signatures are expected to look very similar to other DM models, such as WIMPs. A notable exception occurs in the case of the SIDM particles produced form a bound state (called WIMPonium or Darkonium), which subsequently decays into SM particles~\cite{Tsai:2015ugz,An:2015pva,Elor:2018xku,Krovi:2018fdr,Tsai:2018vjv,Aboubrahim:2020lnr, Laha_2, Laha_3}.

\section{Future Prospects}\label{sec:future}

\subsection{Observational probes}

Upcoming galaxy surveys like the LSST and Dark Energy Spectroscopic Instrument (DESI; \cite{DESI})\footnote{\url{http://desi.lbl.gov}} will allow us to observe the faint satellite galaxies that inhabit the DM halo of the MW, allowing a follow-up similar to the discovery of satellites and streams that DES, among other surveys, has already made possible. Apart from observing the satellites of the MW, there are several surveys that target the faint galaxies in the nearby universe, among them the SAGA survey \cite{Geha:2017iqy, Mao:2020rga} which searches for satellites around MW analogs, ELVES \cite{elves} which provides a volume-limited sample of classical satellites around MW analogs with confirmed distances, LBT-SONG \cite{2021MNRAS.500.3854D} which searches for satellites around intermediate-mass hosts beyond the Local Group, and MADCASH \cite{madcash} which observes Magellanic analogs in the Local Volume. Apart from focussing on satellite systems, surveys like DELVE \cite{2021arXiv210307476D}---a deep, multi-component survey that uses the Dark Energy Camera to study low-surface brightness, DM dominated galaxies in the local universe---will observe field dwarf galaxies. Similarly the Merian survey \cite{Merian}\footnote{\url{https://merian.sites.ucsc.edu/}} uses the HSC to measure weak lensing profiles of Small and Large Magellanic Cloud (LMC) analogs in the local universe. Spectroscopic follow-up of these galaxies will be enabled by Keck, the Thirty Meter Telescope and proposed experiments like the Maunakea Spectroscopic Explorer (MSE; \cite{MSE}) allowing us to characterize the detailed properties of these faint objects. Furthermore, surveys like the Widefield ASKAP L-band Legacy All-sky Blind surveY (WALLABY) \cite{Koribalski:2020ngl}---an HI survey in the local universe---will illuminate the baryonic component of faint galaxies in the nearby universe.

Within the next decade, the Rubin Observatory and Euclid are expected to discover thousands of new galaxy-scale strong gravitational lenses \cite{Oguri:2010ns}. In order to exploit the future data to the maximum possible extent, new methods have been developed to accelerate detections of such systems as well as perturbers in the images \cite{DiazRivero:2019hxf, Brehmer:2019jyt,Alexander:2019puy,Varma:2020kbq,Ostdiek:2020cqz, Ostdiek:2020mvo}. The vast increase in sample sizes will provide much stronger constraints on DM scenarios. Similar machine learning-based methods may also facilitate the analysis of other data from upcoming surveys.

A dedicated follow-up effort for a well-chosen subsample of these lens systems to obtain spectroscopic redshifts and (for Rubin-discovered systems) high-resolution imaging will enable improved constraints on SIDM. On the one hand, gravitational imaging analyses of extended arcs and rings are likely to lead to the detection of several DM perturbers (either substructure or line-of-sight halos) \cite{Vegetti_2010_2,Vegetti_2012,2014MNRAS.442.2017V,Ritondale:2018cvp}. Detailed modeling of the density profiles of these perturbers \cite{Minor:2020hic,Minor:2020bmp} will provide constraints on the SIDM cross section at small velocities. On the other hands, a joint analysis of a large population of quasar lenses \cite{Gilman:2019bdm,Gilman:2021sdr} could also be used to constrain the SIDM cross section. In parallel, follow-up ALMA observations of lensed dusty star-forming galaxies discovered in South Pole Telescope (SPT; \cite{SPT})\footnote{\url{https://pole.uchicago.edu/public/Home.html}} maps \cite{Hezaveh:2012ai,Hezaveh:2016ltk} and Very Long Baseline Interferometry (VLBI) observations \cite{Spingola:2018tup,Spingola:2019ojd} of lensed radio sources will also be used to discover small DM halos and subhalos. In both cases, detailed studies of the density profile of these perturbers will lead to constraints on SIDM. \red{As we have discussed, the possibility of the detection of a number of perturbers that are denser (or more compact) than expected in CDM can be a clean path to the discovery of dark sector DM. } Finally, high-resolution imaging of galaxy-scale lenses could be used to measure the small-scale power spectrum \cite{Bayer:2018vhy,Chatterjee2018,Cyr-Racine:2018htu,Rivero:2017mao,Rivero:2018bcd,Sengul:2020yya} independent of how of the density profile of line-of-sight halos and subhalos are modeled, hence providing complementary constraints on SIDM.

\red{The upcoming galaxy surveys across a wide range of halo masses, including both satellite and field galaxies, will help constrain the cross section on multiple velocity scales giving us a rich understanding  of the varied phenomenology that derives from the interplay of hierarchical structure formation and the particle properties of DM. While observations discussed above will help chart out the nearby Universe, we note also that data from the James Webb Space Telescope \cite{JWST}\footnote{\url{https://www.jwst.nasa.gov/}} and SphereX \cite{spherex}\footnote{\url{https://www.jpl.nasa.gov/missions/spherex}} in combination with the Rubin Observatory data will give us novel constraints on the nature of DM from the first galaxies at high redshifts from the epoch of reionization.}

In addition, constraints on SIDM cross sections from massive halos, cluster and group profiles inferred via weak lensing \cite{Banerjee:2019bjp, Adhikari:2024aff} and core sizes \cite{Andrade:2019wzn,Sagunski:2020spe,Andrade:2020lqq} determined via strong lensing and stellar velocity are expected to tighten significantly given the number of such massive clusters ($M\gtrsim 1\times 10^{14} M_\odot/h$) that will be detected by a combination of optical surveys---such as Rubin and Euclid, as well as Y6 data from DES---and CMB surveys such as the SPT, Atacama Cosmology Telescope (ACT; \cite{ACT})\footnote{\url{https://act.princeton.edu/}}, Simons Observatory \cite{SO}\footnote{\url{https://simonsobservatory.org/}} and CMB-S4. Given its sky coverage, Rubin alone is expected to detect more that $50,000$ clusters in this mass range, up from $\sim 5000$ clusters detected in the DES Y1 data. At slightly lower mass scales ($5 \times 10^{13}M_\odot/h \lesssim M \lesssim 10^{14} M_\odot/h$), the eROSITA mission \cite{eROSITA}\footnote{\url{https://www.mpe.mpg.de/eROSITA}} is expected to map out $\sim 100,000$ low-mass clusters and galaxy groups. On those parts of the sky where eROSITA overlaps with optical surveys, weak lensing measurements around these objects will improve constraints on SIDM cross sections at the relevant velocity scale. Moreover, with an increasing sample of clusters and precise weak lensing maps from telescopes like Roman and Euclid cross section constraints from offsets between BCGs and potential centers of galaxy clusters are also expected to improve by a factor of two \cite{Harvey:2018uwf}. Experiments like MSE will allow complementary studies of positional offsets with velocity offsets. \red{A key advance in the future will be robustly combining weak lensing, strong lensing, X-ray and stellar velocity dispersion data with a dedicated analysis that includes a self-consistent halo model for SIDM. It is possible that we may be able to measure cross sections of order $0.1 \ \rm cm^2/g$ with the next generation of observations and analysis tools. A necessary requirement for this to become a reality is the availability of cosmological simulations of galaxy clusters using cross sections of $0.1 \ \rm cm^2/g$ and smaller. We discuss simulations in more detail in the next subsection.}

\textcolor{black}{Finally, the upcoming surveys, several of which are optimized to answering questions about the large-scale structure and evolution of the universe, will provide  unprecedented data sets to test models that predict modifications to the power spectrum (Secs.~\ref{sect:LSS}, \ref{sec:lss_constraints} and \ref{sec:extensions-mediators}). Conservatively, CMB-S4 for example is expected to measure the energy density in weakly-coupled light particles, parametrized by $\Delta N_{\rm eff}$, to a precision of $\sigma(N_{\rm eff})=0.02-0.03$ \cite{S4}. As we have discussed, $\Delta N_{\rm eff}$ is a critical discriminant of SIDM models and dark sector physics in general. These experiments will also be contemporaneous with galaxy surveys from the Rubin Observatory, Euclid, DESI and the Roman Space Telescope, which will measure cosmological parameters from the clustering of galaxies, gravitational lensing and the abundance of clusters. Simons Observatory and CMB-S4 will also improve CMB lensing measurements by an order of magnitude compared to Planck~\cite{S4}. Combined with measurements of Lyman-$\alpha$ forests from spectrographs like Prime Focus Spectrograph on the Subaru telescope \cite{PFS},  WEAVE on the  William Herschel telescope \cite{pieri2016sf2a},  and in the future MOSAIC  on the Extremely Large Telescope \cite{2019A&A...632A..94J}, WFOS on TMT~\cite{TMT:2015pvw} and MSE~\cite{MSE}, we will have an opportunity to combine several different measurements to constrain these models of dark matter with light mediators. Moreover, 21cm surveys like HERA, SKA and LOFAR along with the James Webb Space Telescope will allow us to push to the epoch of reionization and measure the small-scale power spectrum~\cite{Munoz:2019hjh}. Combined with the small-scale power spectrum, the mass function of dark matter halos and progress in understanding galaxy formation over a wide range of redshifts,  we can be optimistic about making significant strides in pinning down the microphysical nature of dark matter in the near future.}

\subsection{Simulations and modeling}

With the wealth of data that will be available in the coming years, there is a need for simultaneous advancement in simulation and modeling methods of alternate DM models in order to use this data to constrain properties of DM. For SIDM models of the type considered in this work, much of the progress has been based on DM-only (i.e. without baryons) modified $N$-body simulations. Different groups involved in the simulation efforts broadly agree on the types of phenomena produced by the presence of elastic self-interactions, as well as the size of these effects as a function of interaction cross section. On the other hand, a number of powerful probes of SIDM presented in this work involve modeling of small-scale structure, especially the evolution of substructure inside a larger virialized object, where the different simulations techniques have not been rigorously tested against each other. For collisionless $\Lambda$CDM models, different implementations of $N$-body simulations have been stringently calibrated against each other over the years, ensuring that numerical artifacts from any specific implementation can be detected and corrected \cite{Heitmann2005,Heitmann2008} (also see the AGORA code comparison project for hydrodynamic simulation algorithms \cite{Agora_paper1,Agora_paper2, Agora_paper3}). A code comparison project of this type for different SIDM implementations will be hugely beneficial toward the goal of predicting the effects of SIDM on small-scale structure robustly (cf. \cite{CodeComp_SIDM}). Furthermore, since the small-scale structure of interest often lie close to the resolution limits of these simulations, a thorough understanding of whether the resolution affects the predictions of SIDM effects, even within individual implementations, is needed.

In recent years, there has been a growing focus on the mechanism of disruption of subhalos in collisionless CDM $N$-body simulations (see, e.g., \cite{Penarrubia2008, Vandenbosch2018a, Vandenbosch2018b,Fattahi2018,Errani2022}). It has been pointed out that much finer force resolution may be needed to faithfully follow the tidal stripping of subhalos inside a more massive host halo than what was estimated previously, and has been used in much of the existing literature. Lack of sufficient force resolution results in larger amounts of tidal stripping than is expected physically. A related issue is that of subhalos being stripped to below the mass resolution of the applied halo-finder and disappearing from the substructure catalog, even though the central cusp retains its identity, and can continue to host a luminous galaxy. While some of these issues can be overcome by additional modeling, e.g. including an ``orphan model" for galaxies living in the substructure that fall below the halo finder detection threshold, these numerical effects can have major implications on predictions of strong lensing, satellite weak lensing signals and the phase space distributions of satellites in the nearby Universe. An initial exploration of this effect was done in \cite{2021arXiv210608292B}, showing that accounting for disrupted orphans is important to model satellite weak lensing in clusters and to understand the overall distribution of subhalos. Additionally, it should be noted that tracking orphans in SIDM simulations is non-trivial as interactions can often scatter particles near the core of a subhalo, making methods like core-tracking inherently difficult. A detailed understanding of the effects of artificial disruption in SIDM $N$-body simulations is essential, especially as they can be degenerate with certain physical signatures of DM self-interactions. Ram pressure stripping of subhalos in SIDM simulations represent an additional mass-loss mechanism, and the presence of a central core in SIDM subhalos instead of a cusp can mean that the substructure can be completely destroyed.

{Several recent simulation studies have begun to explore the core collapse regime. Typically as the interaction cross-section is allowed to be large at lower velocity scales, the halos that are most likely to core-collapse are low mass objects that live as subhalos in tidal environments of their hosts.   \cite{Zeng:2021ldo}  have used a hybrid approach combining semi-analytic methods and N-body simulations to study in detail the gravothermal evolution low mass subhalos, focusing  specifically on the interplay of tidal evolution, evaporation and core-collapse. The host halo potential is treated analytically while the subhalo is treated as an N-body system. This allows one to study the subhalo evolution with high resolution, simulated with a large number of particles, without the need to simulate the complete N-body system with a large dynamic range that is usually computationally very expensive. \cite{Yang:2022mxl} and \cite{Nadler:2023nrd} on the other hand have run full cosmological zoom-in simulations of Milky Way mass dark matter halos in the core-collapse regime. This allows exploration of the full distribution of subhalo properties in the context of hierarchical structure formation.  These studies follow the implementation of velocity-dependent interactions based on \cite{Robertson:2017mgj}.In \cite{Yang:2022mxl}  they find that nearly $20\%$ of low mass subhalos are core-collapsed and in addition nearly $10\%$ of isolated systems are collapsed for a velocity dependent cross-section which is $\sim 1 \cmg$ at the LMC mass scale. Naturally, both these works imply that a large particle number and small time-step is required to achieve convergence of subhalo properties.
\\
With regard to N-body simulations in the core-collapse regime however, several recent studies have pointed to some challenges that require further detailed exploration. The challenges of simulating core-collapse arises primarily from the unique numerical errors associated with N-body simulations that include scattering events \cite{Mace:2024uze}. For example, core-collapse typically occurs at high interaction cross-sections that drive particles to scatter several times within one dynamical time. Most algorithms that are currently employed to implement SIDM in $N$-body simulations are designed explicitly to work in the limit of rare self-interactions, i.e. when an individual simulation particle interacts via at most one other particle through self-interactions in a given time-step. This condition is imposed to ensure accurate post-scatter velocity and energy distributions and to avoid the complications that arise from choices of ordering interactions. This  typically also corresponds to the regime where most particles do not scatter in any particular time-step. On the other hand, core-collpase by design requires frequent self-interactions. In this then it becomes important to make the timestepping of the simulations significantly small to meet the simulation conditions required to avoid numerical errors. Moreover, since we’re typically studying cores of very low mass subhalos large number of particles are required to capture the density profile accurately. All of these issues make the core-collapse regime significantly expensive to simulate. Detailed exploration of the various numerical issues are presented in \cite{Mace:2024uze, Fischer:2024eaz, Palubski:2024ibb}. All these studies point out issues related to energy conservation, convergence that arise in current N-body treatments of SIDM and suggest improvements with regard to time-stepping, force softening and resolution conditions. Additionally, an earlier work \cite{Zhong:2023yzk} that studies the simulation of an LMC-mass isolated NFW-halo, with a central baryonic potential, also found that convergence of the profiles in the core-collapse phase can be inaccurate.  While the algorithms do reasonably well with the prescription provided in \cite{Yang:2022mxl}, deep in the core-collapse phase artificial heating can stall the evolution. In this context, it is also worth investigating whether $N$-body techniques are indeed the most appropriate tools to model this regime, or whether gravothermal fluid techniques \cite{Essig:2018pzq, nishikawa2019} in these regions can perform better. \cite{Huo:2019yhk} also explore deterministic approaches (rather than the current probabilistic techniques) to simulate SIDM in the presence of dissipation, where similarly collapse is expected to occur.}


DM-only simulations, of course, make predictions for the evolution of DM halos and subhalos. On the other hand, cosmological surveys generally measure the positions, and in some cases, velocities of luminous galaxies. The connection between the observed galaxies and the DM halos in which they live needs to be effectively modeled in order for the simulations to be used in the interpretation and analysis of observational data. In the context of CDM, the galaxy--halo connection has been studied extensively (see \cite{Wechsler:2018pic} for a recent review), and various effective parameterizations have been developed to marginalize over the unknown galaxy formation physics, and its potential impact back-reacting onto the dynamics and evolution of DM substructure. Such studies are also needed in the context of SIDM to promote results from DM-only SIDM simulations to analysis tools. An important question to answer in this context is whether the same parameterizations developed in the context of CDM can be applied to the SIDM context, or whether the galaxy--halo connection in SIDM is qualitatively different, and needs a completely different approach.

Along with a detailed scrutiny of cosmological simulations of DM it is also essential to understand the varied aspects of the interplay between baryonic physics and SIDM over a wide range of halo mass scales. Ideally one would like to simultaneously explore the parameter space in SIDM along with varying the feedback mechanisms in hydrodynamic simulations.
At the very least, when using simulations to investigate how observable quantities are affected by the DM model one should compare this with the differences seen with a fixed DM model when different implementations of baryonic physics are used.

Another avenue for future code development is to include a larger number of possible DM interactions. To date, the bulk of work on SIDM has considered elastic scattering, typically with an isotropic differential cross section. There has been a small amount of work that considered anisotropic scattering cross sections \cite{Robertson:2017mgj, Banerjee:2019bjp}, or inelastic scattering \cite{Vogelsberger:2018bok}, but these scenarios have not been exhaustively explored and do not cover the full range of phenomenology that DM with self-interactions could possess.

\subsection{Promising tests of simple SIDM models}

\red{We have described a variety of tests of SIDM models in this review, ranging from the least massive dark subhalos that could perturb lensed images to galaxy clusters and  large-scale structure. It is fascinating that the simple modification of including elastic self-interactions to the CDM model, which is  well-motivated from the particle physics perspective, could impact structure formation in so many different ways. It is also sobering to realize that despite these striking predictions, the remarkable progress in simulating galaxies, and the plethora of data that already exists, there are no definitive findings in favor of or against SIDM. The reasons  include the fact that CDM models are an excellent fit to the majority of data, and where it may have trouble either star-formation physics (with large modeling uncertainties) is important or the data is not yet decisive. SIDM models  predict deviations from the CDM model predictions predominantly in the central regions of galaxies, where baryonic physics is important. This is a feature of simple SIDM models and not a fine-tuning, as we have reviewed in Sec.~\ref{sec:pheno}.
}

\textcolor{black}{
We have identified multiple avenues to test SIDM through astrophysical observables. A key prediction of all DM models is the mass function of dark subhalos and their internal density profiles. Strong lensing studies aimed at testing these predictions will mature in the coming years, as we have discussed. Observables associated with resolved and unresolved subhalos and their power spectrum should provide a definitive test of a wide swathe of SIDM models that predict halos and subhalos are mostly in the core-expansion phase of their evolution. We have also discussed the recent progress in fleshing out the predictions of SIDM models that predict some subhalos could be in the core-collapse phase. These subhalos with their steep density profiles in the inner regions (radii smaller than the NFW scale radius) provide a striking target for future observations of strongly-lensed systems and stellar streams. Much more work remains to be done to work out the predictions for these models and the critical correlations between subhalo's density profile, survival and  mass loss due to stripping and evaporation.
}

We also highlighted the possibility of throwing the full power of lensing (strong and weak) reconstructions, X-ray observations, BCG offsets, mergers (major and minor) and stellar kinematics in groups and clusters of galaxies to put definitive constraints on the cross sections at relative velocities of $1000-2000 \ \rm km/s$.
It is only possible to put upper limits currently but the massive influx of data in the coming years, along with the possibility of improvements in theory and modeling, makes this an exciting avenue to search for a positive detection.

The most concrete astrophysical motivations for SIDM have  been the core-cusp and diversity issues of rotation curves, and the too-big-to-fail problem of the Milky Way satellites. We have reviewed these problems in detail along with the constraints imposed by the growing census of Milky Way satellites. In the near future, there will be an explosion of data on galaxies and satellites at low redshift, which will provide critical tests of SIDM models. In this context we have highlighted the unique opportunities provided by the ultra-faint dwarf spheroidals of the Milky Way, other dwarfs in the local volume and the ultra-diffuse galaxies in the field and clusters. It is safe to say that there is not currently a concrete understanding of the dark matter or stellar distribution of these objects in the context of either CDM or SIDM models. As with the dark subhalos, the prevalence of core expansion vs core collapse phases in the satellites will be critical. It seems clear that SIDM models where interactions are not strong enough to push satellites into the core collapse phase will be definitively tested in the near future.
SIDM models provide a viable, predictive and versatile framework for interpreting astrophysical observables on galactic and sub-galactic scales.

The predictions of these models need to be developed further if we are to maximize the science returns from the amazing data expected this decade and next. SIDM models provide a unique and compelling opportunity to detect the particle nature of DM through astrophysical observables, as we have highlighted throughout this review. They also provide a well-defined and predictive foil to the dominant CDM paradigm; with predictions for simple SIDM models in hand, the tests of the CDM model become clearer. However, progress will only be possible if there is a serious investment of resources in simulating SIDM models. We have discussed many aspects of the required simulations throughout this review. This investment in simulations is critical given the incomplete progress in mapping out the small-scale structure of even the simplest (single-component,  elastic-scattering-only) SIDM models. It is also vital because progress in understanding galaxy formation and the phenomenology of DM self-interaction are intertwined. We need a concerted effort on the theory side to take full advantage of the incredible data that we will have in the near future and develop the astrophysical observables we have reviewed into decisive probes of dark sector physics.

\section*{Acknowledgements}

Marisa March helped get this review and the Novel Probes Project off the ground. We are very grateful for her energy, vision, and support in the early stages of this work. We thank Alex Drlica-Wagner,
Matthew Walker and Matthew Buckley for early discussions. \newline

KB acknowledges support from the NSF under Grant No.~PHY-2112884.
HD was supported by a Junior Research Fellowship at St John's College, Oxford, a McWilliams Fellowship at Carnegie Mellon University and a Royal Society University Research Fellowship (grant no. 211046).
CD was partially supported by the Department of Energy (DOE) Grant No. DE-SC0020223.
BJ is supported in part by the US Department of Energy Grant No. DE- SC0007901.
FK was funded by the Deutsche Forschungsgemeinschaft (DFG) through the Emmy Noether Grant No. KA 4662/1-1.
MK acknowledges support from the NSF under Grant No.~PHY-1915005.
AHGP and FYCR were supported in part by the National Aeronautics and Space Administration under award number 80NSSC18K1014.
JZ acknowledges support by a Grant of Excellence from the Icelandic Research fund (grant number 206930).

\bibliographystyle{apsrmp4-1}
\bibliography{references}

\end{document}